\documentclass[a4paper,11pt]{article}
\usepackage[margin=1in,footskip=0.25in, a4paper]{geometry}

\usepackage[english]{babel}
\usepackage{amsmath,amssymb, enumitem}
\usepackage{bbm, dsfont}
\usepackage{tikz}
\usepackage{mytikzstyles}
 
\usepackage{natbib}

\usepackage[utf8]{inputenc}
\usepackage{hyperref}
\usepackage{graphicx}
\usepackage{ntheorem}

\usepackage{authblk}

\newtheorem{theorem}{Theorem}
\newtheorem{proposition}{Proposition}
\newtheorem{lemma}{Lemma}

\newtheorem{corollary}{Corollary}
\newtheorem{definition}{Definition}
\newtheorem{example}{Example}
\newtheorem{remark}{Remark}

\theoremstyle{break}

\def\g{\boldsymbol}
\def\ci{\perp\!\!\!\perp}
\def\perpe{\perp_e}
\DeclareMathOperator{\argmin}{arg\,min}

\DeclareMathOperator{\ph}{ph}

\DeclareMathOperator{\var}{Var}

\DeclareMathOperator{\MST}{mst}

\newcommand{\R}{\mathbbm{R}}
\newcommand{\dist}{\rho}
\newcommand{\weak}{\leadsto}
\DeclareMathOperator{\Var}{Var}

\usepackage[utf8]{inputenc} 
\newcommand{\E}{\mathbb{E}}
\newcommand{\bP}{\mathbb{P}}
\newcommand{\eps}{\varepsilon}

\newcommand{\einsfun}{\g 1} 

\title{Structure learning for extremal tree models}

\author[1]{Sebastian Engelke}
\author[2]{Stanislav Volgushev}

\affil[1]{ Research Center for Statistics, University of Geneva, Boulevard du Pont d’Arve 40, 1205 Geneva, Switzerland.}

\affil[2]{Department of Statistical Sciences, University of Toronto, 700 University Ave., Toronto, ON M5G 1X6, Canada.}

\date{}

\begin{document}
\maketitle

\begin{abstract}
    Extremal graphical models are sparse statistical models for multivariate extreme events. The underlying graph encodes conditional independencies and enables a visual interpretation of the complex extremal dependence structure. For the important case of tree models, we develop a data-driven methodology for learning the graphical structure. We show that sample versions of the extremal correlation and a new summary statistic, which we call the extremal variogram, can be used as weights for a minimum spanning tree to consistently recover the true underlying tree. Remarkably, this implies that extremal tree models can be learned in a completely non-parametric fashion by using simple summary statistics and without the need to assume discrete distributions, existence of densities, or parametric models for bivariate distributions.   
\end{abstract}

\section{Introduction}\label{sec:intro}

Extreme value theory provides essential statistical tools to quantify the risk of rare events such as floods, heatwaves or financial crises \citep[e.g.][]{kat2002, poo2004, eng2017a}. The univariate case is well understood and the generalized extreme value and Pareto distributions describe the distributional tail with only few parameters. In dimension $d \geq 2$, the dependence between large values in the different components of a random vector $\g X=(X_1,\ldots,X_d)$ can become very complex. Estimating this dependence in higher dimensions is particularly challenging because the number of extreme observations $k_n$ is by definition much smaller than the number $n$ of all samples in a data set. Constructing sparse models for the multivariate dependence between marginal extremes is therefore crucial for obtaining tractable and reliable methods in multivariate extremes; see \cite{eng2021} for a review of recent advances.

One line of research aims at exploiting conditional independence structures \citep{daw1979} and corresponding graphical models. In the setting of max-stable distributions, {which arise as limits of component-wise block maxima of independent copies of $\g X$, \cite{gis2018} and \cite{klu2019} study max-linear models on directed acyclic graphs. The distributions considered in there do not have densities, and a general result by \cite{papastathopoulos2016conditional} shows that there exist no non-trivial density factorization of max-stable distributions on graphical structures.

A different perspective on multivariate extremes is given by threshold exceedances and the resulting class of multivariate Pareto distributions. Such distributions are the only possible limits that can arise from the conditional distribution of $\g X$ given that it exceeds a high threshold \citep{roo2006,roo2018}. For a $d$-dimensional random vector $\g Y$ that follows a multivariate Pareto distribution, \cite{eng2018} introduce suitable notions of conditional independence and extremal graphical models with respect to a graph $G$. They further show that these notions are natural as they imply the factorization of the density of $\g Y$ through a Hammersley--Clifford type theorem. Extremal graphical models are also related to limits of regularly varying Markov trees studied in \cite{seg2019} and \cite{ase2020}.

In most of the above work, the graphical structure $G$ is assumed to be known \textit{a priori}. It is either based on expert knowledge in the domain of application or it might be identified with an existing graph, as for instance a river network for discharge measurements. However, often no or insufficient domain knowledge on a prior candidate for a graphical structure is available, and a data-driven approach should be followed in order to detect conditional independence relations and to estimate a sensible graph structure. In this work we discuss structural learning for extreme observations.

An important sub-class of general graphs for which structure learning for extremes turns out to be possible in great generality is given by trees. A tree $T=(V,E)$ with nodes $V$ and edge set $E$ is a connected undirected graph without cycles. Most structure learning approaches for trees are based on the notion of the minimum spanning tree. For a set of symmetric weights $\dist_{ij} > 0$ associated with any pair of nodes $i,j\in V$, $i\neq j$, the latter is defined as the tree structure
\begin{align}\label{Tmin}
  T_{\MST} = \argmin_{T = (V,E)} \sum_{(i,j)\in E} \dist_{ij},
\end{align}
that minimizes the sum of distances on that tree. Given the set of weights, there exist greedy algorithms that constructively solve this minimization problem \citep{kruskal1956shortest, pri1957}.

The crucial ingredient for this algorithm are the weights $\dist_{ij}$ between the nodes, and for statistical inference it is generally desirable to choose them in such a way that $T_{\MST}$ recovers the true underlying tree structure that represents the conditional independence relations.
A common approach in graphical modelling is to use the Chow--Liu tree \citep{cho1968}, which is the conditional independence structure that maximizes the likelihood for a given parametric model \citep[cf.,][Chapter 11]{cow2006}. This method uses the negative mutual information as edge weights $\rho_{ij}$ in \eqref{Tmin}, and in general this requires formulating parametric models for the bivariate marginal distributions. In the Gaussian case the weights then simplify to $\dist_{ij}  = \log (1-r_{ij}^2)/2$, where $r_{ij}$ are the correlation coefficients \citep[cf.,][]{drt2017}.

For a multivariate Pareto distribution $\g Y$ that is an extremal graphical model on a tree $T$,
\cite{eng2018} proposed to use the negative maximized bivariate log-likelihoods as edge weights.
This approach has two disadvantages. First, in higher dimensions $d$ it may become prohibitively costly to compute $d^2$ censored likelihood optimizations, and second, a set of parametric bivariate models has to be chosen in advance.

In this paper we study structure learning for extremal tree models in much larger generality. We show that a function of the extremal correlation $\chi_{ij}$, a widely used coefficient to summarize the strength of extremal dependence between marginals $i,j\in V$ \citep[e.g.,][]{col1999}, can be used as weights $\rho_{ij}$ in \eqref{Tmin} to retrieve the underlying tree structure $T$ as the minimum spanning tree under mild non-parametric assumptions. We further introduce a new summary coefficient for extremal dependence, the extremal variogram $\Gamma_{ij}$, which turns out to take a similar role in multivariate extremes as covariances in Gaussian models. More precisely, the extremal variogram of $\g Y$ is shown to be an additive tree metric on the tree $T$ and, as a consequence, it can be used as well as weights $\rho_{ij}$ of the minimum spanning tree to recover the true tree structure. Surprisingly, these results are stronger than for non-extremal tree structures, since we do not require any further parametric assumptions or the existence of densities. This phenomenon originates from the homogeneity of multivariate Pareto distributions and particularly nice stochastic representations of extremal tree models.

In practice, we usually observe $n$ samples of $\g X$ in the domain of attraction of $\g Y$, that is, the conditional distribution of $\g X$ given $\g X$ exceeds a high threshold converges to the distribution of $\g Y$ after proper scaling; see Section~\ref{sec:mpd} for a formal definition. We then rely on estimators of the quantities $\chi_{ij}$ and $\Gamma_{ij}$ to plug into \eqref{Tmin}. To take into account that $\g X$ is only in the domain of attraction of $\g Y$, typical estimators in extreme value theory use only the most extreme observations. We use an existing estimator $\hat \chi_{ij}$ of extremal correlation and a new empirical estimator of the extremal variogram to show that the extremal tree structure can be estimated consistently in a non-parametric way, even when the dimension increases with the sample size.

The remaining paper is organized as follows. In Section~\ref{graph_models} we revisit the notion of extremal graphical models and extend existing representations to the case where densities may not exist. The extremal variogram is introduced in Section~\ref{sec:EV} and its properties are discussed in detail. In Section~\ref{sec:learning} we prove the main results on the consistent recovery of extremal tree structures based on extremal correlations and extremal variograms, both on the population level and using empirical estimates. The simulation studies in Section~\ref{sec:simu} illustrate the finite sample behavior of our structure estimators and show that extremal variogram based methods typically outperform methods working with the extremal correlation. We apply the new tools in Section~\ref{sec:application} to a financial data set of foreign exchange rates. The Appendix and the Supplementary Material contain the proofs and additional illustrations. The methods of this paper are implemented in the R package \texttt{graphicalExtremes} \citep{graphicalExtremes}.

\section{Extremal graphical models}
\label{graph_models}

\subsection{Multivariate Pareto distributions}\label{sec:mpd}

Let $\g X = (X_i)_{i \in V}$ be a random vector with eventually continuous marginal distribution functions $F_i$. Extreme value theory studies marginal and joint tail properties of $\g X$. Univariate extreme value theory focuses on the behavior of marginal components $X_i$, see e.g. \cite{emb1997} and \cite{col1999}. Multivariate extreme value theory is concerned with the dependence structure among different components of extreme observations from $\g X$; see \citet[][Chapter~5]{res2008}, \citet{deh2006a}, \citet{ber2004} or \cite{eng2021} for an introduction. 

One way to describe such tail properties is based on threshold exceedances; here only observations that land above a high threshold are considered. Multivariate Pareto distributions arise as the limits of such high threshold exceedances and are thus natural models for extreme events \citep{roo2006}.
To formally define threshold exceedances in dimension $d > 1$, we need to specify the notion of a high threshold in a multivariate setting. Throughout the paper, we consider multivariate exceedances of the random vector $\g X$ as those realizations where at least one component of $\g X$ exceeds a high marginal quantile. In order to guarantee the existence of the limit of the exceedance distribution, a regularity condition called multivariate regular variation \citep[][Chapter~5]{res2008} is needed. Intuitively, this assumption ensures that the dependence between different components of this conditional distribution stabilizes if the marginal quantile is sufficiently large. More formally, this means that there exists a random vector $\g Y$ supported on $\mathcal L = \{\g x \geq \g 0: \|\g x\|_\infty > 1\}$ such that for all continuity points $\g x \in \mathcal L$ of the distribution function of $\g Y$ we have
\begin{align}\label{mpd_limit}
\mathbb P(\g Y \le \g x) = \lim_{q \to 0} \mathbb P(F(\g X) \le 1 - q/\g x \mid F(\g X) \not \le 1 - q),
\end{align}
where we define $F(\g x) = (F_1(x_1),\dots,F_d(x_d))$. Note that the condition $\{F(\g X) \not \le 1 - q \}$ states that at least one component of $\g X$ exceeds its marginal $1-q$ quantile, explaining the terminology of threshold exceedances. Distributions of random vectors $\g Y$ that can arise in the above limit are called multivariate Pareto distributions. We say that the random vector $\g X$ is in the max-domain of attraction of the multivariate Pareto distribution $\g Y = (Y_i)_{i\in V}$.

The class of multivariate Pareto distributions is very general and contains many different parametric sub-families. Nevertheless, since the random vector $\g Y$ arises as a limiting distribution, it has an important structural property called homogeneity:
\begin{align}\label{MPD}
\mathbb P( \g Y \in t A) = t^{-1}  \mathbb P( \g Y \in A), \qquad t \geq 1,
\end{align}
where for any Borel subset $A \subset \mathcal L$ we define $t A = \{t\g x : \g x\in A\}$.
This explains the name multivariate Pareto distribution since it implies that for any $i \in V$ we have $\mathbb P(Y_i \leq  x \mid Y_i > 1) = 1- 1 /x$ for $x\geq 1$, that is, $Y_i\mid Y_i > 1$ follows a standard Pareto distribution. Moreover, since {$F(\g X)$} has identically distributed margins, it follows from~\eqref{mpd_limit} that $\mathbb P(Y_1 > 1)= \dots = \mathbb P(Y_d > 1)$. Conversely, if the latter holds and $\g Y$ is homogeneous as in~\eqref{MPD}, then $\g Y$ is a multivariate Pareto distribution; for a proof of this equivalence and the relationship to limits appearing in~\cite{seg2019} see Supplementary Material~\ref{sec:prop1}.

\subsection{Extremal Markov structures}

Since the support $\mathcal L$ of multivariate Pareto distributions is not a product space, the definition of conditional independence is non-standard and relies on auxiliary random vectors derived from $\g Y$. For any $m\in V$,
we consider the random vector $\g Y^m$ defined as $\g Y$ conditioned on the event that $\{Y_m > 1\}$, which has support on the space $\mathcal L^m = \{ \g x \in \mathcal L: x_m > 1\}$. {For general random vectors $\g X \in \R^d$ and ordered sets $A \subset \{1,\dots,d\}$, let $\g X_A$ denote the subvector of $\g X$ with indices in $A$. The notation $\g Y^m_A$ will be used to denote the subvector of $\g Y^m$ with indices in $A$.}

With this notation we can state a definition of conditional independence for
multivariate Pareto distributions that is more general than the one in \cite{eng2018},
since we do not assume existence of densities.
\begin{definition}\label{CI_def}
  For disjoint subsets $A,B,C \subset V = \{1,\dots, d\}$,
  we say that $\g Y_A$ is conditionally independent of $\g Y_C$ given $\g Y_B$ 
\begin{align}\label{eq:citail3}
  \forall m\in  \{1,\dots, d\}: \quad \g Y^m_A\ci \g Y^m_C \mid \g Y^m_B.
\end{align}
In this case we write $\g Y_A\perpe \g Y_C\mid \g Y_B$.
\end{definition}
{The subscript $e$ in $\perpe$ indicates that this conditional independence notion is defined for extreme observations, which are described by the multivariate Pareto distribution $\g Y$ according to~\eqref{mpd_limit}.}

We view the index set $V$ as a set of nodes of a graph $G = (V, E)$,
with connections given by a set of edges $E \subset V\times V$ of pairs of distinct
nodes. The graph is called undirected if for two nodes $i,j\in V$, $(i,j)\in E$ if and only if $(j,i)\in E$. For notational convenience, for undirected graphs we sometimes represent edges as unordered pairs $\{i,j\}\in E$. When counting the number
of edges, we count $\{i,j\}\in E$ such that each edge is considered only once.
For disjoint subsets $A,B, C \subset V$, $B$ is said to separate $A$ and $C$ in $G$ if every path from $A$ to $C$ contains as least one node in $B$.
{For an illustration of these definitions see Supplementary Material~\ref{sm:graph}.}

The notion of an extremal graphical model is then naturally defined as a multivariate Pareto
distribution that satisfies the global Markov property on the graph $G$ with respect to the conditional independence
relation $\perpe$, that is, for any disjoint subsets $A,B,C \subset V$ such that $B$ separates $A$ from $C$ in $G$,
\begin{align}\label{egm}
  \g Y_A \perpe \g Y_C \mid \g Y_B.
\end{align}
{In line with the definition in the graphical models literature \citep[][Chapter 3]{Lauritzen}, the definition allows for additional conditional independence relations that are not encoded by graph separation. This means that there are typically several graphs $G$ that are consistent with the distribution of $\g Y$; for instance, any multivariate Pareto distribution is an extremal graphical model on the fully connected graph.}

In the case of a decomposable graph $G$ and if $\g Y$ possesses a positive and continuous density $f_{\g Y}$, \cite{eng2018} show that this density factorizes into lower-dimensional densities, and that the graph $G$ is necessarily connected. If $\g Y$ does not have a density, then the extremal graph can be disconnected and the connected components are mutually independent of each other \citep[see Kirstin Strokorb's discussion contribution]{eng2018}. Note that we require the global Markov property in the definition of extremal graphical models as opposed to the pairwise Markov property used in \cite{eng2018}. Both properties are equivalent in the case of positive, continuous densities, but in general, the former implies the latter but not the other way around \citep[see][Chapter 3]{Lauritzen}.

\subsection{Extremal tree models}

An important example of a sparse graph structure is a tree. A tree $T=(V,E)$ is a connected undirected graph without cycles and thus $|E| = |V| - 1$. Equivalently, a tree is a graph with a unique path between any two nodes. If $\g Y$ is an extremal graphical model satisfying the global Markov property~\eqref{egm} with respect to a tree $T$, we obtain a simple stochastic representation of $\g Y^m$. This stochastic representation will be the crucial building block for the results on tree learning given in the next section. 

To this end we need to introduce the concept of extremal functions. Define the extremal function relative to coordinate $m$ as the $d$-dimensional, non-negative random vector $\g W^m$ with $W^m_m = 1$ almost surely that satisfies the stochastic representation
\begin{align}
\label{extr_fct}
\g Y^m \stackrel{(d)}{=} P\g W^m,
\end{align}
where $P$ is a standard Pareto random variable, $\mathbb P( P \leq x) = 1 - 1/x$, $x \geq 1$, which is independent of $\g W^m$, {and $\stackrel{(d)}{=}$ stands for equality in distribution}. Such a representation is possible by homogeneity \eqref{MPD} of $\g Y$, which is inherited by $\g Y^m$. Indeed, given homogeneity of $\g Y^m$ we see that $Y^m_m$ follows a standard Pareto distribution. Moreover, writing $\g W^m := \g Y^m/Y_m^m$, homogeneity of $\g Y^m$ and a simple calculation implies that $\g W^m$ and $Y_m^m$ are independent, resulting in the representation~\eqref{extr_fct}.

The representation~\eqref{extr_fct} is an alternative way of describing the distribution of $\g Y$, and indeed, the set of the $d$ extremal functions $\g W^1,\dots,\g W^d$ uniquely defines the multivariate Pareto distribution. We refer to \cite{dom2013} and \cite{dom2016} for additional technical background on extremal functions.

\begin{example}\label{ex_biv}
	In the case $d=2$, due to homogeneity, the bivariate Pareto distribution $\g Y = (Y_1, Y_2)$ can essentially be characterized by a univariate distribution. Indeed, for any non-negative random variable $W^1_2$ with $\mathbb E W^1_2 \leq 1$, the random vector $\g W^1 = (1,W^1_2)$ is the extremal function relative to the first coordinate of a unique bivariate Pareto distribution~$\g Y$. The extremal function relative to the second coordinate $\g W^2 = (W^2_1, 1)$ is obtained through a change of measure 
	\begin{align}\label{biv_change}
	\mathbb P( W^2_1 \leq z, W^2_1 > 0) &= \mathbb E( \einsfun\{1/W^1_2 \leq z\} W^1_2),\quad  z > 0,
	\end{align}
	which implies that $\mathbb E(W^1_2) = 1 - \mathbb P(W^2_1 = 0)\leq 1$.  
\end{example}
An elementary proof of~\eqref{biv_change} can be found in Appendix~\ref{sec:proofchange}.

We now proceed to a stochastic representation for $\g Y^m$ that involves {only the univariate random variables $W_i^j$}. Define a new, directed tree $T^m = (V,E^m)$ rooted at an arbitrary but fixed node $m\in V$. The edge set $E^m$ consist of all edges $e\in E$ of the tree $T$ pointing away from node $m$. For the resulting directed tree we define a set $\{W_e: e\in E^m\}$ of independent random variables, where for $e = (i,j)$, the distribution of $W_e = W^i_j$ is {$j$th coordinate of the extremal function of $\g Y$ relative to coordinate $i$.}

The following result generalizes Proposition 2 in \cite{eng2018} to extremal tree models with arbitrary edge distributions. 

\begin{proposition}\label{prop_tree}
Let $\g Y$ be a multivariate Pareto distribution that is an extremal graphical model on the tree $T = (V,E)$. Let $P$ be a standard Pareto distribution, independent of $\{W_e: e \in E^m \}$. Then we have the joint stochastic representation for $\g Y^m$ on $\mathcal L^m$
\begin{align}\label{tree_rep}
  Y^m_i \, \stackrel{(d)}{=} 
  \begin{cases}
     P, & \text{for } i = m,\\
     P \times \prod_{e\in \ph(mi; T^m)} W_e, &\text{for } i\in V\setminus \{m\},
  \end{cases}
\end{align}
where $\ph(mi; T^m)$ denotes the set of edges on the unique path from node $m$ to node $i$ on the tree~$T^m$; see Figure \ref{dir_tree} for an example with $m=2$.

Conversely, for any set of independent random variables $\{W_i^j, W_j^i; \{i,j\} \in E\}$, where $W_i^j$ and $W_j^i$ satisfy the duality~\eqref{biv_change}, the construction~\eqref{tree_rep} defines a consistent family of extremal functions $\g W^1, \dots, \g W^d$ that correspond to a unique $d$-dimensional Pareto distribution $\g Y$ that is an extremal graphical model on $T$.
\end{proposition}
The above result formally establishes the link of the conditional independence in Definition~\ref{CI_def} to the limiting tail trees in \cite{seg2019}; see also Proposition~\ref{prop_equiv} in the Supplementary Material for details on this link. In this sense, the first part of Proposition~\ref{prop_tree} can be deduced from Theorem~1 in \cite{seg2019}.

{Note that $\g Y$ defined as in Proposition~\ref{prop_tree} can also be an extremal graphical model on a disconnected graph $G$; {see paragraph after~\eqref{egm}}. The representation~\eqref{tree_rep} then remains true and some of the $W_e$ are almost surely equal to zero.}

\begin{figure}[ht]
  \centering
\begin{tikzpicture}
  [scale=1,auto=left]
  \node[nodeObs] (n1) at (0,0) {$Y_1$};
  \node[nodeObs] (n2) at (2,1)  {$Y_2$};
  \node[nodeObs] (n5) at (4,0)  {$Y_5$};
  \node[nodeObs] (n4) at (4,2)  {$Y_4$};
  \node[nodeObs] (n3) at (2,-1)  {$Y_3$};
  \draw[edgeObs, ->] (n2) -- node[above] {$W^2_1$} (n1);
  \draw[edgeObs, ->] (n2) -- node[above] {$W^2_4$} (n4);
  \draw[edgeObs, ->] (n2) -- node[below] {$W^2_5$} (n5);
  \draw[edgeObs, ->] (n1) -- node[below] {$W^1_3$} (n3);
\end{tikzpicture}
\caption{A tree $T^2$ rooted at node $m=2$ with the extremal functions on the edges.}
\label{dir_tree}
\end{figure}
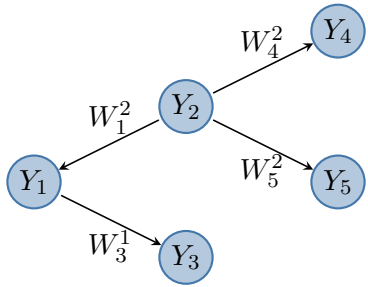

\begin{remark}
  It is remarkable that for an extremal tree model $\g Y$, the distribution of its extremal functions, and therefore also of the multivariate Pareto distribution itself, is characterized by the set of univariate random variables $\{W_i^j, W_j^i; \{i,j\} \in E\}$. This indicates that the probabilistic structure is simpler than in the non-extremal case, where in general both univariate and bivariate distributions are needed to describe a tree graphical model.
\end{remark}

\section{The extremal variogram}
\label{sec:EV}
Covariance matrices play a central role in structure learning for Gaussian graphical models
due to their connection to conditional independence properties.
In multivariate extreme value theory, several summary statistics have been developed
to measure the strength of dependence between the extremes of different variables.
The most popular one is the extremal correlation, which for $i,j\in V$ is defined as
\begin{align}\label{EC}
  \chi_{ij} := \lim_{q\to 0} \chi_{ij}(q) := \lim_{q\to 0} \mathbb P\left\{F_i(X_i) > 1- q\mid  F_j(X_j) >1-q \right\},
\end{align}
whenever the limit exists. It ranges between $0$ and $1$ where the boundary cases are asymptotic independence and complete extremal dependence, respectively \citep[cf.,][]{col1999, sch2003}. In particular, if $\g X$ is in the max-domain of attraction of the multivariate Pareto distribution $\g Y$, then the extremal correlation always exists and $ \chi_{ij} =  \mathbb P(Y_i > 1 \mid Y_j > 1)$.
There are many other coefficients for extremal dependence in the literature, including the madogram \citep{coo2006} and a coefficient defined on
the spectral measure introduced in \cite{lar2012} and used for dimension reduction in \cite{coo2018} and \cite{for2020}.

While designed as summaries for extremal dependence, none of these coefficients has an obvious relation to conditional independence for multivariate Pareto distributions
or density factorization in extremal graphical models of \cite{eng2018}. In this section
we define a new coefficient that will turn out to take a similar role in multivariate
extremes as covariances in non-extremal models.

\subsection{Limiting extremal variogram}

The variogram is a well-known object in geostatistics
that measures the degree of spatial dependence of a random field \citep[cf.,][]{chi2012, wac2013}.
It is similar to a covariance function, but instead of positive definiteness, a variogram
is conditionally negative definite; for details, see for instance \citet[][Appendix B]{eng2018}.
For Brown--Resnick processes, the seminal work of \cite{kab2009} has shown that negative
definite functions play a crucial role in  spatial extreme value theory.
We define a variogram for general multivariate Pareto distributions.

\begin{definition}\label{EV}
  For a multivariate Pareto distribution $\g Y$ we define
  the extremal variogram rooted at node $m\in V$ as the matrix $\Gamma^{(m)}$ with entries
  \begin{align}\label{vario}
    \Gamma_{ij}^{(m)} &= \var \left\{ \log Y^m_{i} - \log Y^m_j \right\}, \quad i,j\in V,
  \end{align}
  {whenever the right-hand side exists and is finite. }
\end{definition}
 We can interpret the $\Gamma^{(m)}_{ij}$ as a distance between the variables $Y^m_i$ and $Y^m_j$ that is large if {their extremal dependence is weak} and \emph{vice versa}.

 \begin{proposition}
   \label{prop_vario}
  Let $\g Y$ be a multivariate Pareto distribution.
  \begin{itemize}
  \item[(i)]
    For $m\in V$, we can express the extremal variogram in terms of the extremal function relative to coordinate $m$,
    \begin{align*}
    \Gamma_{ij}^{(m)} &= \var \left\{ \log W_i^m - \log W_j^m \right\}, \quad i,j\in V.
  \end{align*}      
  \item[(ii)]
    For $m\in V$, the matrix $\Gamma^{(m)}$ is a variogram matrix, that is, it is conditionally negative definite.    
  \item[(iii)]
    Let $\g Y_n$ be a sequence of multivariate Pareto distributions
    with extremal coefficients {$\chi_{n,im}$ between the $i$th and $m$th coordinate of $\g Y_n$} satisfying $\chi_{n,im} \to 0$ as $n\to \infty$ for
    some $i,m\in V$. Then the {corresponding} extremal variograms satisfy $\Gamma^{(m)}_{n,im} \to \infty$ {as $n\to \infty$}.
  \end{itemize}
\end{proposition}

Part (iii) in the above proposition underlines the interpretation of the extremal variogram.
When the variables become asymptotically independent, then the extremal variogram grows
and eventually diverges to $+\infty$. Note that the inverse statement is not true in general, since
there are cases where $\Gamma_{im}^{(m)} = \infty$ but $\chi_{im} > 0$.
We proceed with several examples where the extremal variogram can be computed explicitly. Figure~\ref{gamma_chi} shows the extremal variogram values for these models as a function of the corresponding extremal correlation.

\begin{example}\label{ex_logistic}
  The extremal logistic distribution with parameter $\theta\in (0,1)$ can be defined through its extremal functions \citep[see][]{dom2016}
  $$\g W^m = \left( {U_1}/{U_m}, \dots, {U_d}/{U_m}\right),$$  
where $U_1,\dots, U_d$ are independent and $U_i$, $i\neq m$ follow $\text{Fr\'echet}(1/\theta, G(1-\theta)^{-1})$ distributions, and $(G(1-\theta) U_m)^{-1/\theta}$
  follows a $\text{Gamma}(1-\theta,1)$ distribution; here $G(x)$ is the Gamma function evaluated at $x \geq 0$. It turns out that for the logistic model we have
\[
\Gamma_{ij}^{(m)}  = \begin{cases}
{\pi^2\theta^2}/{3},  &\text{ if } i,j \neq m,\\
       \theta^2 \{\psi^{(1)}(1-\theta) + \pi^2/6\}, & \text{ if } i = m, j\neq m,
  \end{cases}
\]
where $\psi^{(1)}$ is the trigamma function defined as the second derivative of the logarithm of the gamma function.

{The corresponding extremal correlations have the form $\chi_{ij} = 2 - 2^\theta$, $i,j\in V$.}
\end{example}

The proof of this representation of the extremal variogram in the logistic model can be found in the Supplementary Material~\ref{proof_log}.

\begin{example}\label{ex_dirichlet}
  The extremal Dirichlet distributions with parameters $\alpha_1, \dots, \alpha_d$ \citep[cf.,][]{coles1991modelling} has extremal functions
  $$ \g W^m = (U_1/U_m, \dots, U_d/U_m),$$
  where $U_1,\dots, U_d$ are independent and $U_i$, $i\neq m$ follow $\text{Gamma}(\alpha_i, 1/\alpha_i)$ distributions, and $U_m$ follows a $\text{Gamma}(\alpha_m+ 1, 1/\alpha_m)$ distribution.
  By straight-forward calculations,
  $$\Gamma_{ij}^{(m)}  = \begin{cases}
      \psi^{(1)}(\alpha_i) + \psi^{(1)}(\alpha_j),  &\text{ if } i,j \neq m,\\
    \psi^{(1)}(\alpha_m + 1) + \psi^{(1)}(\alpha_j), & \text{ if } i = m, j\neq m,
  \end{cases}
  $$
  with $\psi^{(1)}$ denoting the trigamma function as in Example~\ref{ex_logistic}.

  {The corresponding extremal correlations do have have a closed form but can be calculated numerically.}
\end{example}

For the class of H\"usler--Reiss distributions the extremal variogram turns out to be very natural.

\begin{example}\label{ex_HR}
  The H\"usler--Reiss distribution is parameterized by a variogram matrix $\Gamma \in \mathbb R^{d\times d}$; see \cite[][]{eng2018} for details. For any $d$-variate centered normal random vector $\g U$ with variogram matrix $\Gamma$, the extremal function relative to coordinate $m\in V$ has representation 
  \begin{align}\label{Yk}
     \g W^m  =  \exp\{ \g U - U_m  - \Gamma_{\cdot m} / 2\},
  \end{align}
  see  \citet[][Prop.~4]{dom2016}. The extremal variogram $\Gamma^{(m)}$ for any $m\in V$ is then equal to the variogram matrix $\Gamma$ from the definition of the  
    H\"usler--Reiss distributions, and, in particular, it is independent of the root node,
    $$\Gamma_{ij} = \Gamma_{ij}^{(1)} = \dots = \Gamma_{ij}^{(d)}, \quad i,j \in V.$$
    {The corresponding extremal correlations have the form $\chi_{ij} = 2 - 2 \Phi\big(\sqrt{\Gamma_{ij}}/2\big)$, where $\Phi$ is the standard normal distribution function.}
\end{example}

\begin{figure}
\centering
\includegraphics[height=.38\textwidth]{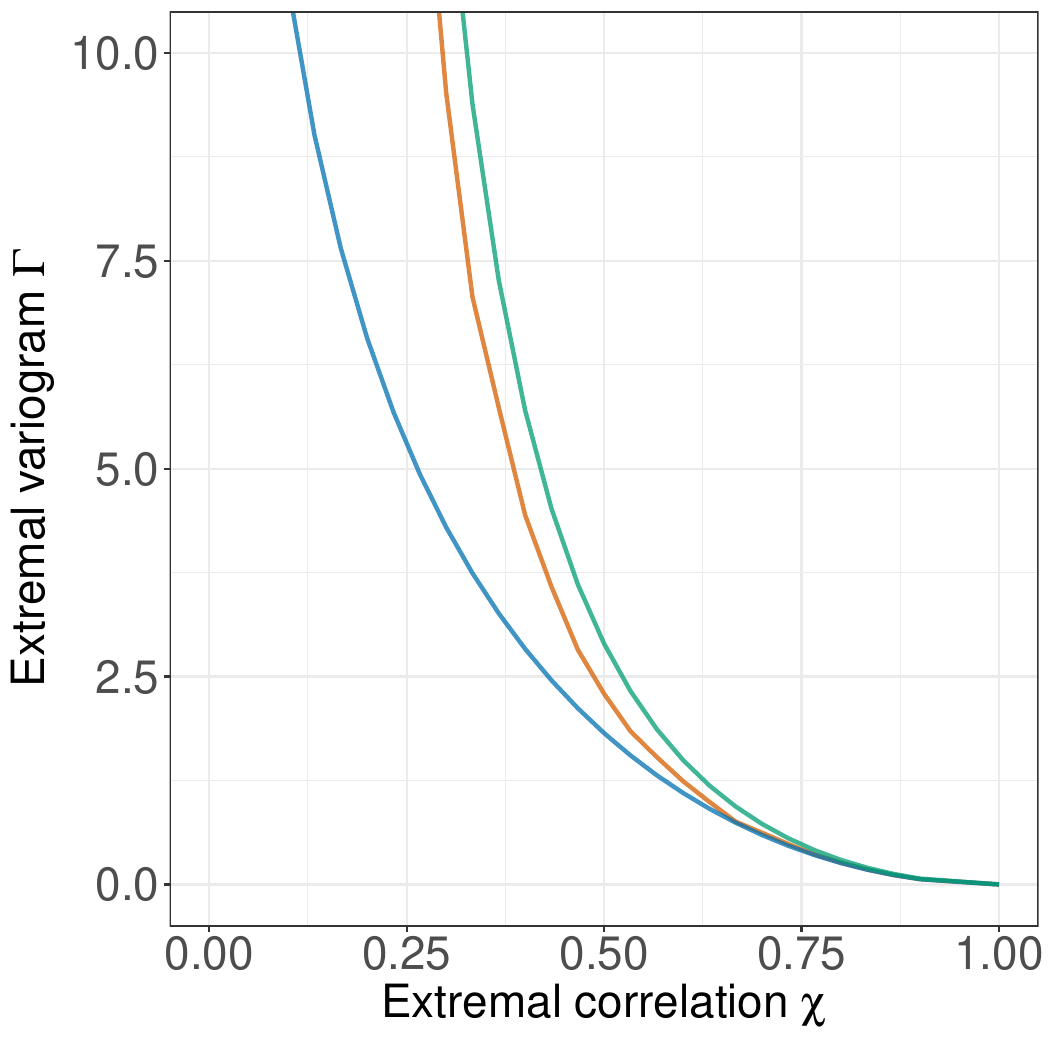}%
\caption{Values of the extremal variogram $\Gamma_{12}^{(1)}$ as a function of the extremal correlation $\chi_{12}$ for the bivariate H\"usler--Reiss (blue), symmetric Dirichlet (orange) and logistic (green) models. Note that in all three cases we have that $W_1^2 \stackrel{(d)}{=} W_2^1$ and therefore $\Gamma_{12}^{(1)} = \Gamma_{12}^{(2)}$.}
\label{gamma_chi}
\end{figure}

\subsection{Pre-asymptotic extremal variogram}
\label{sec:preasym}

Similar to the extremal correlation in \eqref{EC} we can define the extremal
variogram as the limit of pre-asymptotic versions. 
\begin{definition} \label{def:variopre}
	For a multivariate distribution $\g X$ with continuous marginal distributions we define
	the pre-asymptotic extremal variogram at level $q\in(0,1)$ rooted at node $m\in V$ as the matrix $\Gamma^{(m)}(q)$ with entries
	\begin{align*}
	\Gamma_{ij}^{(m)}(q) &= \var \left[\log\{1 -  F_i(X_{i})\}- \log\{1 - F_j(X_j)\} \mid F_m(X_m) > 1-q\right], \quad i,j\in V,
	\end{align*}
	whenever right-hand side exists and is finite. 
\end{definition}
{Note that for $q$ close to zero the conditional distribution of the terms $-\log\{1 -  F_i(X_{i})\}$ given $F_m(X_m) > 1-q$ is approximately that of $\log Y^m_i$, $i \in V$.}

Next we provide conditions which ensure the convergence $\Gamma_{ij}^{(m)}(q) \to \Gamma_{ij}^{(m)}$ as $q \to 0$.
We introduce the following notation: for a vector $\g x \in \R^d$ and $I \subset \{1,\dots,d\}$, let $\g x_I$ denote a vector in $\R^{|I|}$ with entries $x_{j}, j \in I$. For a distribution function $F$ of a $d$-dimensional random vector $\g X$ define $F_I$ as the distribution function of the corresponding random vector $\g X_I$ and {let $\g Y_{(I)}$ denote the limit obtained in relation~\eqref{mpd_limit} when $F, \g X, \g x$ are replaced by $F_I, \g X_I, \g x_I$. Note that $\g Y_{(I)}$ is not the same as $\g Y_I$, the subvector of $\g Y$ with entries in $I$, because the latter is not supported on $\mathcal{L}_I = \{\g x \geq \g 0: \|\g x\|_\infty > 1\} \subset \R^{|I|}$. {The distribution of $\g Y_{(I)}$ can} be obtained from that of $\g Y_I$ by conditioning.}
\begin{enumerate}
\item[(B)] {There exist constants} $\xi > 0, K_B < \infty$ {such that for} any $I \subset V$ with $|I| \in\{2,3\}$ and all $q \in (0,1)$ 
\begin{align}\label{eq:MRVsecor}
\sup_{\g x_I \in [1,\infty]^{|I|}} \Big| \mathbb P\left( F_I(\g X_I) \leq  1 - q/\g x_I \mid F_I(\g X_I) \nleq  1 - q\right) - \mathbb P({\g Y_{(I)}} \leq \g x_I)\Big| \leq K_B q^\xi.
\end{align}

\item[(T)] There exists a $\gamma > 0$ such that for any $i,m \in V$ the extremal function satisfies
\begin{align}
\label{gamma_moment}
\mathbb E \left(W^{m}_i\right)^{-\gamma} \leq K_W < \infty.
\end{align} 
\end{enumerate}

Assumption (B) is a strengthening of~\eqref{mpd_limit} for bivariate and trivariate distributions as it imposes that convergence to the limit should take place uniformly and at a certain rate. It is closely related to typical second order conditions on the stable tail dependence function that are fairly standard in the literature; see for instance \cite{einmahl2012} and \cite{fougeres2015} among many others. Additional details on this matter are given in the Supplementary Material~\ref{seq:equivalence}. Condition (T) is a mild assumption on the extremal functions $W^{m}_i$, which holds for all examples considered in the previous section. This condition prevents the distribution of $W_i^m$ from putting too much mass close to zero. 

\begin{proposition}\label{prop:consvar}
	Under conditions (B), (T) we have for any $m,i,j \in V$
	\begin{align*}
	\Gamma^{(m)}_{ij}(q) \to  \Gamma^{(m)}_{ij}, \qquad {\text{as }} q \to 0.
	\end{align*}
\end{proposition}

We note that condition (T) already implies that $\Gamma^{(m)}_{ij} \in [0,\infty)$ for any $i,j$, so the convergence above is always to a finite limit. 

\section{Structure learning for extremal tree models}
\label{sec:learning}

\subsection{Extremal tree models}\label{sec:population}

Extremal graphical models where the underlying graph is a tree were considered as
a sparse statistical model in \cite{eng2018}. As explained in the introduction, their approach of
using a censored maximum-likelihood tree becomes prohibitively costly in higher dimension $d$
and requires parametric assumptions on the bivariate distributions of the tree.

Ideally, one would like to have summary statistics, similar to the correlation coefficients $r_{ij}$ in the Gaussian case, that can be estimated empirically and that guarantee to recover the true underlying tree
structure when used as edge weights. The extremal variogram defined in Section~\ref{sec:EV} turns out to be a so-called tree metric, and as such a natural quantity to infer the conditional independence structure in extremal tree models. We underline that the extremal variogram $\Gamma^{(m)}$ is defined for arbitrary multivariate Pareto distributions and in the case of the H\"usler--Reiss distribution it coincides with the parameter matrix.

\begin{proposition}\label{prop_tree_gamma}
  Let $\g Y$ be an extremal graphical model with respect to the tree $T = (V,E)$ and suppose that the extremal variogram matrix $\Gamma^{(m)}$ exists for all $m\in V$. Then we have that
  \begin{align}\label{gamma_additivity}
    \Gamma_{ij}^{(m)} = \sum_{(s,t) \in \ph(ij; T)} \Gamma^{(m)}_{st}.
  \end{align}
  In other words,  for any $m\in V$, the extremal variogram matrix $\Gamma^{(m)}$ defines an additive tree metric.
\end{proposition}

\begin{corollary}\label{cor1}
  Let $\g Y$ be an extremal graphical model with respect to the tree $T = (V,E)$. Suppose that the extremal variogram matrix $\Gamma^{(m)}$ exists {and is finite} for all $m\in V$ and that $\bP(Y_i \neq Y_j) > 0$ for all $i,j\in V$, $i\neq j$ (or equivalently, $\Gamma^{(m)}_{ij} > 0$).
  For any $m\in V$, the minimum spanning tree with $\dist_{ij} = \Gamma^{(m)}_{ij}$ is unique and satisfies 
  $$T_{\MST} = T.$$    
\end{corollary}

For extremal tree models, Corollary~\ref{cor1} shows that independently of any distributional assumption, the extremal variogram contains the conditional independence structure of the tree $T$. This result is quite surprising, since it is stronger than what is known in the classical, non-extremal theory of trees. Indeed, as discussed in the introduction, for Gaussian graphical models, a analogous result holds for a minimum spanning tree with weights $\dist_{ij}  = \log (1-r_{ij}^2)/2$ for $r_{ij}$ denoting the correlation between the $i$th and $j$th component of the Gaussian random vector under consideration. The assumption of Gaussianity is crucial and the result no longer holds outside this specific parametric class. 

Beyond the world of Gaussian graphical models, there exists some literature on the non-parametric estimation of graphical models on tree structures, see \cite{cho1968} for an early contribution and \citet[][Section 3.1]{drt2017} for an overview. However, one either needs to assume discrete distributions \citep{cho1968} or the existence of densities \citep{LXGGLW2011, laf2012}, and non-parametric density estimation is required in the latter case. To the best of our knowledge, multivariate Pareto distributions are the first example for a non-parametric sub-class of multivariate distributions where tree dependence structures can be learned using simple moment-based summary statistics without additional parametric assumptions. It is also remarkable that there is no need to assume the existence of densities and that the distributions we consider can simultaneously have continuous and discrete components.

The reason why such a strong result can hold can be explained by
the homogeneity of the multivariate Pareto distribution $\g Y$. For trees, all cliques
contain two nodes and therefore the density $f_{\g Y}$ factorizes into bivariate Pareto
densities. Because of the homogeneity, such a bivariate density can be decomposed into
independent radial and angular parts; see Example \ref{ex_biv}.
Bivariate Pareto distributions only differ in terms of the angular distribution, whose support
is {the subset of a one-dimensional sphere with all coordinates positive}. Consequently, an extremal tree model in $d$ dimensions can essentially be reduced to $d-1$ univariate angular distributions; see also Proposition~\ref{prop_tree}. This provides an intuitive explanation why the result in Corollary~\ref{cor1} can hold.

We can go further and show that a linear combination of the matrices $\Gamma^{(m)}$, $m\in V$, which are possibly different from each other, still induces the true tree as the minimum spanning tree.

\begin{corollary}\label{cor2}
Under the same assumptions as in Corollary \ref{cor1}, the minimum spanning tree with distances
\[
\dist_{ij} = \sum_{m=1}^d w_m\Gamma^{(m)}_{ij}
\]
given by a linear combination of the extremal variograms rooted at different nodes with coefficients $w_m\geq 0$, $m\in V$, $\max_{m\in V} w_m >0$, is unique and satisfies $T_{\MST} = T.$
\end{corollary}

The extremal correlation coefficients $\chi_{ij}$ do not form a tree metric, that is, they are not additive according to the tree structure as the extremal variogram in \eqref{mst_diff}. It is therefore a non-trivial question whether these coefficients can also be used as weights in a minimum spanning tree to infer the underlying conditional independence structure. Interestingly, the next result gives {a partially affirmative answer.}

\begin{proposition} \label{prop:chimst}
  Let $\g Y$ be {an extremal graphical model on the tree $T = (V,E)$. 
Then the extremal correlation coefficients satisfy for any $h,l \in V$ with $h\neq l$ that
\begin{equation}\label{eq:orderchi2}
  \chi_{hl} \leq \chi_{ij} \quad \forall (i,j) \in \ph(hl; T).
\end{equation}
Under the additional assumption that this inequality is strict as soon as $(i,j) \neq (h,l)$,} the minimum spanning tree corresponding to distances $\dist_{ij} = -\log (\chi_{ij})$ is unique and satisfies 
  $$T_{\MST} = T.$$    
\end{proposition}
{The assumption that $\chi_{hl} < \chi_{ij}$ for any $(i,j) \in \ph(hl; T)$ with $(i,j)\neq (h,l)$ is not satisfied for all tree models. Indeed, a counterexample ({J. Segers, personal communication, 2022-07-07}) with index set $V=\{1,2,3\}$ and edges $E=\{(1,2), (2,3)\}$ is the following. Let the extremal function $W^1_2\sim \text{Unif}([1/2, 3/2])$ and let $W^2_3$ have a discrete distribution $\mathbb P(W^2_3 = 1/4) = 4/5$ and $\mathbb P(W^2_3 = 4) = 1/5$; both are valid extremal functions as in Example~\ref{ex_biv}. In this case $\chi_{13} = \chi_{23} {= 2/5}$ and the set of minimum spanning trees is not unique. We can then only guarantee that the true underlying tree $T$ lies in the set of all possible minimum spanning trees; {this follows from a close inspection of the proof of Proposition~\ref{prop:chimst}}.

There are simple conditions to ensure that inequality~\eqref{eq:orderchi2} is strict for all $(i,j)\neq (h,l)$. For instance, a sufficient condition for this to hold is that all extremal functions $W^i_j$ for $(i,j) \in E$ have support equal to the whole space $[0,\infty)$; see Lemma~\ref{lemma_sufficient} in Appendix~\ref{app:sufficient}. This covers many relevant examples such as the H\"usler--Reiss, the extremal logistic and the extremal Dirichlet distributions in Examples~\ref{ex_logistic},~\ref{ex_dirichlet} and~\ref{ex_HR}, respectively. A weaker condition for strict inequality was recently obtained by \cite{hu2022}.
}

\begin{remark}
 Both the extremal variogram $\Gamma_{ij}^{(m)}$ and the extremal correlation $\chi_{ij}$ contain  information on conditional independence structure for extremal tree models. {The extremal correlation is defined for any model but needs additional assumptions to correctly recover the tree. The extremal variogram does not exist if $\g Y$ has mass on lower-dimensional sub-faces of $\mathcal L$ but is guaranteed to recover the underlying tree whenever all extremal variograms exist.} When their sample versions are used (see Section~\ref{sec:estimation}), the probability of correctly identifying the underlying tree may differ {even when both approaches work on population level}; see Section~\ref{sec:simu}. 
\end{remark}

\subsection{Estimation}\label{sec:estimation}

Throughout this section assume that we observe independent copies $\g X_1,\dots,\g X_n$ of the $d$-dimensional random vector $\g X$, which is in the max-domain of attraction of a multivariate Pareto distribution $\g Y$, an extremal graphical model on the tree $T$ according to~\eqref{egm}. Our aim is to estimate $T$ from the observations $\g X_1,\dots,\g X_n$.
Motivated by Proposition~\ref{prop:chimst} and Corollaries~\ref{cor1},~\ref{cor2} we propose to achieve this through a two-step procedure. We first construct estimators for the quantities $\chi_{ij}$ and $\Gamma^{(m)}_{ij}$, and then compute the  minimal spanning trees corresponding to those estimators.

The empirical estimator for $\chi_{ij}$ is defined as
\[
\hat \chi_{ij} := \frac{n}{k} \sum_{t = 1}^n \einsfun\{\tilde F_i(X_{ti}) >1- k/n , \tilde F_{j}(X_{tj}) > 1 - k/n\},
\]
where $k = k_n$ is an intermediate sequence and $\tilde F_i$ denotes the empirical distribution function of $X_{1i},\dots,X_{ni}$. Standard arguments imply that under~\eqref{mpd_limit} and {provided that $k \to \infty$ and $k/n \to q \in [0,1]$ as $n \to \infty$,} we have for any $i,j \in V$ 
\begin{equation}\label{eq:chicons}
\hat \chi_{ij} = \chi_{ij}(q) + o_\bP(1),\qquad {\text{as } n \to \infty},
\end{equation}
where $\chi_{ij}(q)$ is defined in~\eqref{EC} and $\chi_{ij}(0) := \chi_{ij}$.
In particular, if $q=0$ then $\hat \chi_{ij}$ is a consistent estimator of $\chi_{ij}$.

The extremal variogram matrix $\Gamma^{(m)}$ for the sample $\g X_t$, $t=1,\dots, n$, is estimated by
\begin{align*}
\hat \Gamma_{ij}^{(m)} &:= \widehat{\Var}\Big(\log (1 - \tilde F_i(X_{ti})) - \log(1 - \tilde F_j(X_{tj})) : \tilde F_m(X_{tm}) \geq 1 - k/n  \Big),
\end{align*}
where $\widehat{\Var}$ denotes the sample variance. Under the assumption $k/n \to q \in [0,1]$ {as $n\to\infty$} and mild conditions on the underlying data generation, this estimator can be shown to be consistent for the pre-asymptotic version $\Gamma_{ij}^{(m)}(q)$ as introduced in Definition~\ref{def:variopre}.

\begin{theorem} \label{th:Gammacons}
Let assumptions (B), (T) hold and assume that $k \geq n^\theta$ for some $\theta > 0$ and that $k/n \to q \in [0,1]$ {as $n\to \infty$}. Then we have for any $m, i, j \in V$
\[
\hat \Gamma^{(m)}_{ij} = \Gamma^{(m)}_{ij}(q) + o_\bP(1),\qquad {\text{as } n \to \infty,}
\]
where $\Gamma^{(m)}_{ij}(0) := \Gamma^{(m)}_{ij}$.
\end{theorem}

The proof of this result turns out to be surprisingly technical, details are given in the Supplementary Material~\ref{sec:prGammacons}. The main challenge arises from the fact that in the definition of $\Gamma_{ij}^{(m)}$ only the observations in component $m$ are extreme while observations in other components may also be non-extreme. This is different from the setting that is typically considered in asymptotically dependent extreme value theory.

\begin{remark}
  By choosing $q=0$, the above theorem implies consistency of the empirical extremal variogram $\hat \Gamma^{(m)}$. This result is of independent interest, since it is the first proof of consistency of the moment estimators
  \[
  {\hat \Sigma^{(m)}_{ij} = \frac{1}{2} \{\hat \Gamma_{im}^{(m)}+ \hat \Gamma_{jm}^{(m)}- \hat \Gamma_{ij}^{(m)}\}, \quad i,j\neq m},
  \]
  which were introduced in \cite{Engelke2015} as estimators for the parameters of the H\"usler--Reiss distribution. 
\end{remark}

\begin{remark}
The assumption that the data $\g X_1,\dots,\g X_n$ are independent was only made to keep the presentation simple. The consistency result in Theorem~\ref{th:Gammacons} continues to hold under a high-level assumption that allows for temporal dependence and is spelled out in detail at the beginning of Supplementary Material~\ref{sec:prGammacons}. 
\end{remark}

Now we have all results that are needed for consistent estimation of the underlying tree structure. Given a general distance $\dist$ with estimator $\hat \dist$ on pairs $(i,j) \in V\times V$, we consider plug-in procedures of the form
\begin{align}\label{T_est}
\hat T_\dist := \arg \min_{ T = (V,E)} \sum_{(i,j) \in E} \hat \dist_{ij},
\end{align}
with three cases of particular interest given by
\begin{align*}
  \hat \dist_{ij} &= -\log (\hat \chi_{ij}), \qquad \hat \dist_{ij} = \hat \Gamma_{ij}^{(m)},\qquad \hat \dist_{ij} = \sum_{m=1}^d w_m \hat \Gamma_{ij}^{(m)},
\end{align*}
resulting in the estimators $\hat T_\chi, \hat T_\Gamma^{(m)}, \hat T_\Gamma^{w}$, respectively. The special case of $w_1=\dots = w_d = 1/d$ is denoted by $\hat T_\Gamma$. We solve the minimum spanning tree problem~\eqref{T_est} by Prim's algorithm, which is guaranteed to find a global optimizer of problem~\eqref{T_est} that is unique if the distances $\hat \dist_{ij}$ are distinct for all pairs \citep{pri1957}.

\begin{theorem}\label{th:consMST}
Assume that {$\g Y$ is an extremal graphical model on the tree $T$}. {Assume~\eqref{mpd_limit} holds and that the inequality in~\eqref{eq:orderchi2} is strict whenever $(i,j) \neq (h,l)$}.  If {$k \to \infty$ as $n \to \infty$}  then there exists $q^* > 0$ such that under the additional assumption $k/n \to q \in [0,q^*]$ {as $n \to \infty$,} 
\[
\bP( \hat T_\chi = T) \to 1 ,\qquad {\text{as } n \to \infty}.
\]
If {instead of~\eqref{mpd_limit} and strict inequality in~\eqref{eq:orderchi2}} assumptions (B), (T) hold, {$\bP(Y_i \neq Y_j) > 0$ for all $i,j\in V$, $i\neq j$ (or equivalently, $\Gamma^{(m)}_{ij} > 0$),} and if $k \geq n^\theta$ for some $\theta >0$ then for any $m \in V$ there exists $q^*_m > 0$ such that for $k/n \to q \in [0,q^*_m]$  {as $n \to \infty$,} we have
\[
\bP( \hat T_\Gamma^{(m)} = T) \to 1 ,\qquad {\text{as } n \to \infty}.
\]
The same is true for $\hat T_\Gamma^w$ provided the weights $w_m$ satisfy $w_m \geq 0, \max_m w_m > 0$. 
\end{theorem}

\begin{remark}
As pointed out by a referee, it would be of interest to find weights that maximize (asymptotically) the probability of correct tree structure recovery by $\hat T_\Gamma^w$. This would require precise information on the joint asymptotic distribution of $\hat \Gamma^{(m)}$ for different $m$, which is currently an open question. 
\end{remark}

\begin{remark}\label{rem:kn}
At first glance it might seem surprising that the tree structure can be estimated consistently even when $k_n/n$ does not converge to zero. The latter would be a classical minimal assumption in extreme value theory and would be required for consistent estimation of $\chi_{ij}$ or $\Gamma_{ij}^{(m)}$. We explain the intuition behind this result for the extremal correlation, the arguments for the extremal variogram are exactly the same. {Assume that the inequality in~\eqref{eq:orderchi2} is strict whenever $(i,j) \neq (h,l)$, making the minimal spanning tree with respect to $- \log \chi_{ij}$ unique.} The key insight is that even biased estimators of $\chi_{ij}(q)$ can lead to the correct minimal spanning tree since all we need is 
\[
\sum_{(i,j)\in E'} -\log \chi_{ij}(q) > \sum_{(i,j)\in E} -\log \chi_{ij}(q)
\] 
for all trees $T' = (V, E') \neq T$, where $T$ denotes the true underlying tree. Multivariate regular variation~\eqref{mpd_limit} implies that $\chi_{ij}(q) \to \chi_{ij}$ as $q \to 0$ for all $i,j$, so {there exists $q_0 > 0$ such that the above inequality is satisfied for all $q < q_0$ }. Since in addition $\hat \chi_{ij} = \chi_{ij}(k/n) + o_\bP(1)$ as $n \to \infty$ under the assumption $k \to \infty$, consistency follows.  
\end{remark}

Theorem~\ref{th:consMST} shows that the proposed procedures are able to consistently recover the tree structure under rather weak assumptions on the sequence $k = k_n$. It is natural to wonder which choices of $k$ correspond to higher probabilities of recovering the tree structure consistently. Here we provide some indicative discussion of this issue for minimal spanning trees based on $\chi_{ij}$ without going into technical details. Standard results from empirical process theory show that under mild assumptions and for $k/n \to q \in [0,1]$  {as $n \to \infty$,} all $\sqrt{k}(\hat \chi_{ij} - \chi_{ij}(k/n))$ converge jointly to a multivariate normal distribution with covariance matrix $\Sigma_q$. The latter matrices satisfy $\Sigma_q \to \Sigma_0$ as $q \to 0$. Combined with the delta method this implies that for any tree $T' = (V,E') \neq T$ 
\[
\sum_{(i,j) \in E'} \hat \dist_{ij} - \sum_{(i,j) \in E} \hat \dist_{ij} = \Delta_{k,n} + \frac{1}{\sqrt{k}} Z_{k,n}  := \sum_{(i,j) \in E'} \dist_{ij}(k/n) - \sum_{(i,j) \in E} \dist_{ij}(k/n) + \frac{1}{\sqrt{k}} Z_{k,n} 
\]
where $\dist_{ij}(k/n) := - \log \chi_{ij}(k/n)$ and $\hat \dist_{ij} := - \log \hat \chi_{ij}$, and $Z_{k,n}$ is a weighted linear combination of differences $\sqrt{k}(\hat \chi_{ij} - \chi_{ij}(k/n))$ and thus approximately centered normal with variance $\sigma_q^2$. The probability that the sum over estimated distances on $T'$ is shorter than the sum over true tree $T$ is given by $\bP(- Z_{k,n} > \sqrt{k} \Delta_{k,n})$. Under the assumptions for asymptotic normality of $\hat \chi_{ij}$, $\Delta_{k,n}$ converges to $\Delta(q) := \sum_{(i,j) \in E'} \dist_{ij}(q) - \sum_{(i,j) \in E} \dist_{ij}(q)$. Combining all of the above approximations we find $\bP(-Z_{k,n} > \sqrt{k} \Delta_{k,n}) \approx \bP(\sigma_q \mathcal{N}(0,1) > \sqrt{n}\sqrt{q}\Delta(q))$. Since $\sigma_q \to \sigma_0 >0 $ and $\Delta(q) \to \Delta(0) > 0$ as $q \to 0$, it is easy to see that there exists $q_0 > 0$ such that $\sqrt{q}\Delta(q)/\sigma_q < \sqrt{q_0}\Delta(q_0)/\sigma_{q_0}$ for all $q < q_0$, and thus the probability of selecting $T'$ instead of the true tree $T$ starts to increase as the limit of $k/n$ decreases after $q_0$. This suggests that an optimal value for $k$ in terms of maximizing the probability of estimating the true tree would satisfy $k/n \to \tilde q$ {as $n \to \infty$} for some $\tilde q >0$. Turning the above arguments into a formal proof would require many technicalities which are beyond the scope of the present paper, but the intuition obtained here is also confirmed in the simulations in Section~\ref{sec:simu}.

\subsection{Estimation in growing dimensions}
\label{sec:high_dim} 
The consistency results in the previous section were derived for data of fixed dimension for sample size tending to infinity. Here we provide an extension of those results by adding non-asymptotic bounds on the probability of consistently estimating the true tree. {Throughout this section, the underlying tree can change with the sample size $n$.} 

We start with discussing results for $\hat T_\chi$. This requires the following additional notation. Assume that {$\g Y$ is an extremal graphical model on the tree $T$} and define the corresponding extremal correlation
$\chi_{ij}^Y := \mathbb P(Y_i > 1 \mid Y_j > 1)$. Let 
\begin{equation*}
\mu_\chi^Y := \min_{(h,l) \notin E} \min_{(i,j) \in \ph(hl;T)} (\chi_{ij}^Y - \chi_{hl}^Y ).  
\end{equation*}
To gain some intuition on the reason for this definition, {recall that in~\eqref{eq:orderchi2} in  Proposition~\ref{prop:chimst}} we show that 
\[
\chi_{hl}^Y \leq \chi_{ij}^Y \quad \forall (i,j) \in \ph(hl;T).
\]
{To ensure that the minimal spanning tree corresponding to $-\log \chi_{ij}^Y$ is unique, we need to rule out equality in the above statement whenever $(i,j) \neq (h,l)$, which follows from $\mu_\chi^Y > 0$.}
Thus the quantity $\mu_\chi^Y$ can be interpreted as a lower bound on the increase of the sum of distances on the edges if we move from the true tree $T$ to $T'\neq T$. This is formalized in the proof of Theorem~\ref{th:consMST-hd}. We are now ready to state the first main result.
   
\begin{theorem}\label{th:consMST-hd}
Assume that {$\g Y$ is an extremal graphical model on the tree $T$} and that $\g X_1,\dots,\g X_n$ are independent copies of $\g X$, a random vector with continuous marginal distributions. Let {$\chi_{ij}(k/n)$ denote the pre-asymptotic extremal coefficients corresponding to $\g X$ in the sense of~\eqref{EC} and define}
\[
\delta_{k/n} := \max_{i \neq j} |\chi_{ij}(k/n) - \chi_{ij}^Y|.
\]
Then there exists a universal constant $K>0$ such that
\begin{equation}\label{eq:chitail}
\bP\big(\hat T_\chi \neq T\big) \leq 5d^2 \exp\Big(-\frac{3k}{10}\Big\{\Big(\frac{\mu_\chi^Y - 2\delta_{k/n}}{2K}\Big)_+^2\wedge 1 \Big\} \Big).
\end{equation}	
\end{theorem}

Note that above we did not assume that $\g X$ is in the max-domain of attraction of $\g Y$. A link between $\g X$ and $\g Y$ is implicitly provided through $\delta_{k/n}$ which measures the distance between $\chi_{ij}(k/n)$ computed from $\g X$ and the extremal coefficients $\chi_{ij}^Y$ which correspond to $\g Y$. 

Some comments on the implications of the above result are in order. On a high level, larger dimensions $d$, smaller values of $\mu_\chi^Y$, and larger bias $\delta_{k/n}$ lead to a larger bound. The effects of dimension $d$ and bias $\delta_{k/n}$ are intuitive: larger dimensions or more bias make the tree recovery problem more difficult. The effect of $\mu_\chi^Y$ is also expected because smaller values of $\mu_\chi^Y$ imply that, on population level, there exist trees that are closer to the true tree and estimation becomes more difficult. 

For a more quantitative discussion assume that (B) in Section~\ref{sec:preasym} holds with constants $K_B, \xi$ independent of $n, d$. In this case $\delta_{k/n} \leq K_R (k/n)^\xi$ for a possibly different constant $K_R$ which is still independent of $n,d,\xi$; see~\eqref{eq:bounRnkR'} in Supplementary Material~\ref{sm_prop3}. Note that the exponent can be bounded by $-3n(k/n) \{[(\mu_\chi^Y - 2K_B(k/n)^{\xi})_+^2/(4K^2) ] \wedge 1\}$. Straightforward but tedious computations optimizing this rate over $k$ show that the largest achievable rate for this exponent is of order $n(\mu_\chi^Y)^{2+1/\xi}$ if we let $k = c n(\mu_\chi^Y)^{1/\xi} $ for a suitable constant $c \in (0,\infty)$ which depends on $K, K_B, \xi$ only. With this choice of $k$ consistent tree structure recovery is possible if $\log d = o(n(\mu_\chi^Y)^{2+1/\xi})$ {as $n \to \infty$}. If $\mu_\chi^Y$ stays bounded away from zero this simplifies to $\log d =o(n)$ {as $n \to \infty$}, which allows the dimension to grow exponentially in $n$. In contrast, if the dimension $d$ is fixed but we consider observations from a triangular array with the same tree but changing value of $\mu_\chi^Y$, we require $n(\mu_\chi^Y)^{2+1/\xi} \to \infty$  {as $n \to \infty$}, provided that $k$ is chosen as described above. This condition becomes more stringent if $\xi$ is smaller, which is intuitive since it corresponds to slower decaying bias.

We now discuss tree structure recovery with $\hat T_{\Gamma}^{(m)}$ and $\hat T_{\Gamma}^{w}$. A key result here are concentration bounds on $\hat \Gamma_{ij}^m$. Such bounds are established in~\cite{ELV2021} and reproduced in the proof of Theorem~\ref{th:consMSTGamma-hd} given in the Supplementary Material~\ref{sec:proofth:consMSTGamma-hd}. To state those bounds we need an additional assumption.

\begin{enumerate} \label{assum:r}
\item[(D)] For all $I \subset V$ with $|I| = 2$ the random variables $\g Y_{(I)}$ have densities $f_I$. There exists an $\eps > 0$ such that for all $\beta \in [-\eps,1-\eps]$ there is a constant $K(\beta)$ such that 
\[
f_I(x,y) \leq K(\beta) \frac{1}{y^{1+\beta}x^{2-\beta}} \quad x,y \in (1,\infty)^2.
\]

\end{enumerate}
This is equivalent to Assumption 2 in~\cite{ELV2021}; see the discussion around~\eqref{eq:densequiv} in Supplementary Material~\ref{sec:proofth:consMSTGamma-hd}. \cite{ELV2021} show that it holds for H\"usler--Reiss distributions, for instance. This condition is implied by the simpler but stronger condition $f_I(x,2-x) \leq K_r(x(2-x))^{1+\eps}$ for some $\eps > 0$ and all $x \in (0,2)$; this follows from elementary calculations involving the homogeneity of $f_I$ which is derived in~\eqref{eq:fIhom} in Supplementary Material~\ref{sec:proofth:consMSTGamma-hd}. 

\begin{theorem}\label{th:consMSTGamma-hd}
Assume that {$\g Y$ is an extremal graphical model with respect to the tree} $T$ and that $\g X_1,\dots,\g X_n$ are independent samples of $\g X$, a random vector with continuous marginal distributions. Assume that (B), (T), (D) hold and that $k \geq n^\theta$ for some $\theta >0 $. Then there exist constants $c,C,M >0 $ depending only on the constants from (B), (T), (D) and $\theta$ such that for all $k \geq 1$ and $b_{k/n} := (k/n)^\kappa(\log(n/k))^2$ where $\kappa := \gamma\xi/(1+\gamma+\xi)$
\begin{equation}\label{eq:gammatail}
\bP\big(\hat T_\Gamma^{(m)} \neq T\big) 
\leq  
M d^3\exp\Big(-ck\Big\{\Big(\frac{\min_{(i,j) \in E} \Gamma_{ij}^{(m)}}{2C}-b_{k/n}\Big)_+^2\wedge \frac{1}{(\log n)^8}\Big\} \Big).
\end{equation}
For $w_m \geq 0$ with $\sum_{m=1}^d w_m = 1$ the same bound holds for $\bP\big(\hat T_\Gamma^w \neq T\big)$  with $\min_{(i,j) \in E} \Gamma_{ij}^{(m)}$ replaced by $\min_{(i,j) \in E} \sum_{m=1}^d w_m \Gamma_{ij}^{(m)}$.
\end{theorem}

We note that Assumption (D) can be dropped at the cost of introducing an additional $1/(\log n)^4$ factor; details are provided in Supplementary Material~\ref{sec:proofth:consMSTGamma-hd}. Similarly to Theorem~\ref{th:consMST-hd} we do not explicitly assume that $\g X$ is in the domain of attraction of $\g Y$. Assumption (B) provides the link between $\g X$ and $\g Y$ in terms of their bivariate and trivariate distributions.

We briefly comment on the result in Theorem~\ref{th:consMSTGamma-hd}. Observe that the general structure of the bound is similar to the corresponding result for tree structure recovery based on $\chi$. The fact that $\wedge 1$ in~\eqref{eq:chitail} is replaced by $\wedge (\log n)^{-8}$ in~\eqref{eq:gammatail} is due to technical details in the derivation of tail bounds for $\hat \Gamma_{ij}^{(m)}$, which has a more complex structure than the simple estimator $\hat \chi_{ij}$. Similarly to $\mu_\chi^Y$ in ~\eqref{eq:chitail}, $\min_{(i,j) \in E} \Gamma_{ij}^{(m)}$ can be interpreted as measuring the minimal separation between the length of shortest and second-shortest minimal spanning tree. The quantity $b_{k/n}$ appearing in Theorem~\ref{th:consMSTGamma-hd} stems from bounds on bias terms in estimating $\Gamma_{ij}^{(m)}$ and plays a similar role as $\delta_{k/n}$ for $\chi_{ij}$. Comments on the fastest possible growth of the dimension $d$ and minimal separation conditions that still allow for consistent tree structure recovery follow along the same lines as in the discussion following Theorem~\ref{th:consMST-hd} and are omitted for the sake of brevity.

\section{Simulations}
\label{sec:simu}

The minimum spanning trees based on the empirical versions of the extremal variogram and extremal correlation both recover asymptotically the underlying extremal tree structure. In this section  we study the finite sample behavior of the different tree estimators on simulated data. The results and figures of Sections~\ref{sec:simu} and~\ref{sec:application} can be reproduced with the code at \url{https://github.com/sebastian-engelke/extremal\_tree\_learning}.

Let $T = (V, E)$ be a random tree structure that is generated by sampling uniformly $d-1$ edges and adding these to the empty graph under the constraint to avoid circles. Throughout the whole section, we simulate $n$ samples $\g X_1, \dots, \g X_n$ from a random vector $\g X$ in the domain of attraction of a multivariate Pareto distribution $\g Y$ that is an extremal graphical model on the tree $T$ in dimension $d = |V|$.  As random vector $\g X$ we take the corresponding max-stable distribution \citep[e.g.,][]{deh1984}, which is indeed in the domain of attraction of $\g Y$ in the sense of~\eqref{mpd_limit}. In order to perturb the samples, a common way is to add lighter tailed noise \citep[e.g.,][]{ein2016}. More precisely,
\begin{align}\label{sim_model}
  \g X_i = \g Z_i + \varepsilon_i, \quad \varepsilon_i \ci \g Z_i,\qquad i=1,\dots, n,
\end{align}
where $\g Z_i$ is a max-stable random vector with standard Fr\'echet margins associated to $\g Y$, and $\varepsilon_i$ is
a lighter-tailed noise vector which is independent of $\g Z_i$.
We consider two scenarios for the noise distribution, where in both cases the marginal distribution is transformed to a Fr\'echet distribution with $\mathbb P(\varepsilon_{ij} \leq x) = \exp(-1/x^2)$, $x \geq 0$, $j\in V$. 
\begin{enumerate}
\item[(N1)] The noise vector $\eps_i$ has independent entries.
\item[(N2)] The noise vector $\varepsilon_i$ in~\eqref{sim_model} is generated from an extremal tree model on a fixed tree $T_{\text{noise}}$ that is generally different from the true tree $T$.
\end{enumerate}	
Since the marginals of the noise vector have lighter tails, the limit of $\g X_i$ in~\eqref{mpd_limit} is not altered by $\varepsilon_i$. The main difference between the two noise mechanisms lies in the type of bias they introduce for large $k$, and we observe that this has an interesting impact on the recovery of the tree structure underlying $\g Y$.

We consider two different parametric classes of distributions for $\g Y$.
  \begin{enumerate}
    \item[(M1)]
      The H\"usler--Reiss tree model is a multivariate Pareto distribution that factorizes on $T = (V, E)$, where each bivariate distribution $(Y_i, Y_j)$ for $(i,j) \in E$ is H\"usler--Reiss with
      parameter $\Gamma_{ij} > 0$; see Example~\ref{ex_HR}. The joint distribution is then also H\"usler--Reiss with parameter matrix $\Gamma$ induced by the tree structure through~\eqref{gamma_additivity}. The coefficients on the edges are generated as
      $$\Gamma_{ij} \sim \text{Unif}([0.2,1]), \quad (i,j) \in E.$$
    \item[(M2)]
      For the second model we let each bivariate distribution be given by the family of asymmetric Dirichlet distributions; see Example~\ref{ex_dirichlet}. We generate the two parameters of the bivariate Dirichlet models independently as
      $$\alpha_1, \alpha_2 \sim \text{Unif}([1, 10]), \quad (i,j) \in E.$$
      Note that the resulting $d$-dimensional Pareto distribution is not in the family of Dirichlet distributions.
    \end{enumerate}

We compare four different estimators for the weights on the minimum spanning tree $\hat T_\rho = (V, \hat E_\rho)$ in~\eqref{T_est}:
  \begin{enumerate}
    \item[(i)]
      $\hat \rho_{ij} = -\log\hat \chi_{ij}$, where $\hat \chi_{ij}$ is the empirical extremal correlation;  
    \item[(ii)]
      $\hat \rho_{ij} = \hat \Gamma_{ij}^{(m)}$, the extremal variogram estimator for one fixed $m\in V$;
    \item[(iii)]  
      $\hat \rho_{ij} = \hat \Gamma_{ij}$, the combined extremal variogram estimator;
    \item[(iv)]
      $ \hat \rho_{ij}$ are the censored negative log-likelihoods of the bivariate H\"usler--Reiss model $(Y_i, Y_j)$, evaluated at the optimizer.
  \end{enumerate}
The estimators (i)--(iii) were introduced in Section~\ref{sec:estimation} and their consistency has been derived. The estimator in (iv) is the one used in \cite{eng2018} to learn the structure of H\"usler--Reiss tree models. Note that for this estimator, no theoretical justification is available.
As performance measures we choose the average proportion of wrongly estimated edges
\begin{align}\label{err} 
\mathbb E_{T=(V,E)}\mathbb E \left(1 - \frac{| \hat E_\rho \cap E|}{d-1}\right),
\end{align}
and the probability of not recovering the correct tree structure
\begin{align}\label{srr} 
\mathbb E_{T=(V,E)} \mathbb P( \hat T_\rho \neq T),
\end{align}
where the outer expectations signify that the tree $T$ is randomly generated in each repetition. {Each experiment is repeated 300 times in order to estimate these errors empirically.} We report only the results on the structure recovery rate error~\eqref{srr} and provide the corresponding results on the wrong edge rate~\eqref{err} in the Supplementary Material~\ref{sm:simu}.

We first investigate the choice of the intermediate sequence $k=k_n$ of the number of exceedances used for estimation. We simulate from the H\"usler--Reiss tree model (M1) in dimension $d=20$ and consider the minimum spanning trees $\hat T_\Gamma$ and $\hat T_{\text{CL}}$ based on the combined extremal variogram and the censored likelihoods, respectively. Figure~\ref{fig_HR_kn} shows the structure recovery rate error as a function of the exceedance probability $k/n$ for different samples sizes $n$. Interestingly, the two noise patterns lead to qualitatively different results: while consistent recovery of the limiting tree seems possible even when $k=n$ for noise model (N1), noise model (N2) with a dependence structure also introduces a bias in the corresponding minimal spanning tree and the true tree can not be recovered when the limit of $k/n$ is too large. 
It is interesting to observe that the optimal exceedance probability $k/n$ seems to converge to a positive value $q^*$, especially for noise (N1). This is consistent with the intuition given at the end of Section~\ref{sec:estimation} in the paragraph after Remark~\ref{rem:kn}. This is in contrast to classical asymptotic theory for consistent estimation in extremes where $k = o(n)$ is required to remove the approximation bias and therefore $q^*=0$.   
\begin{figure}
	\centering
	\includegraphics[height=.33\textwidth]{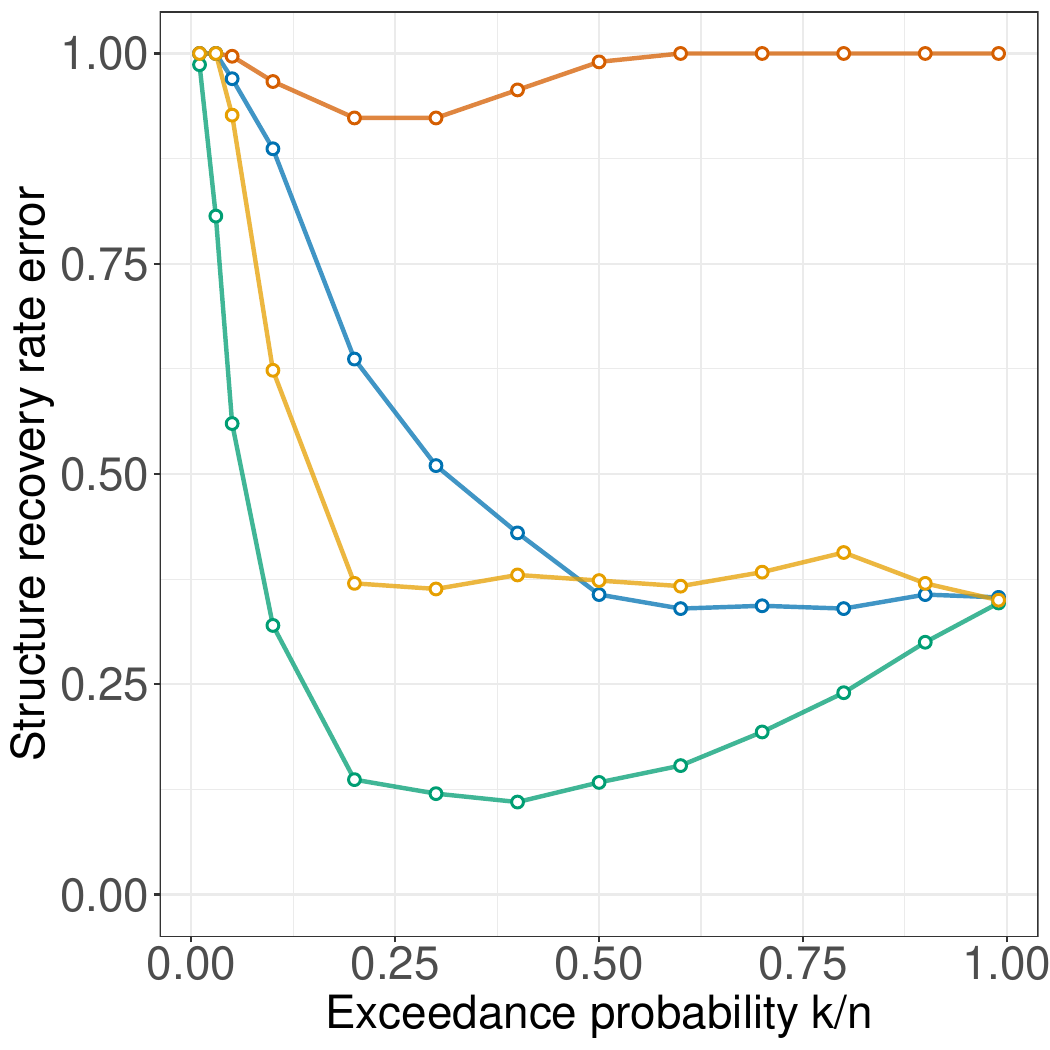}%
	\hspace{0em}\includegraphics[height=.33\textwidth]{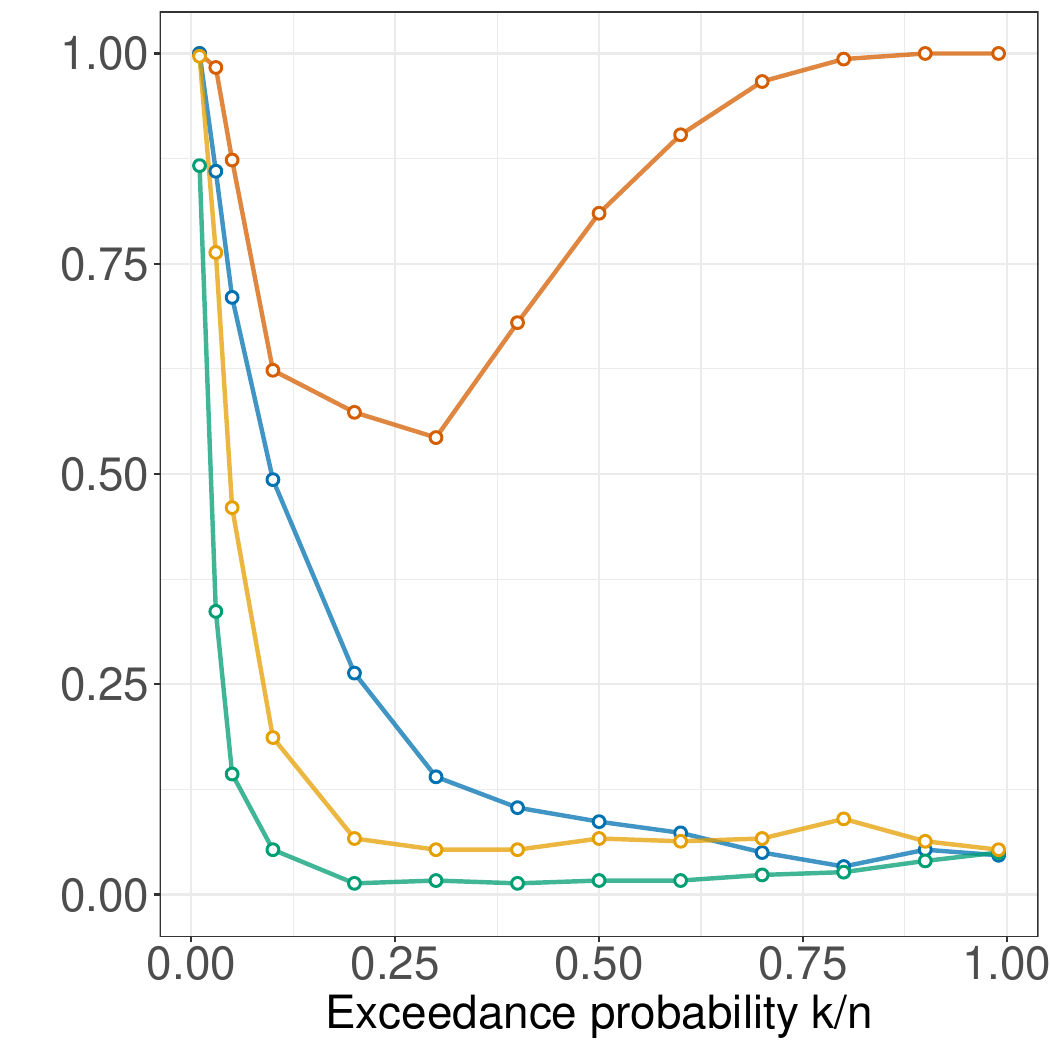}%
	\hspace{0em}\includegraphics[height=.33\textwidth]{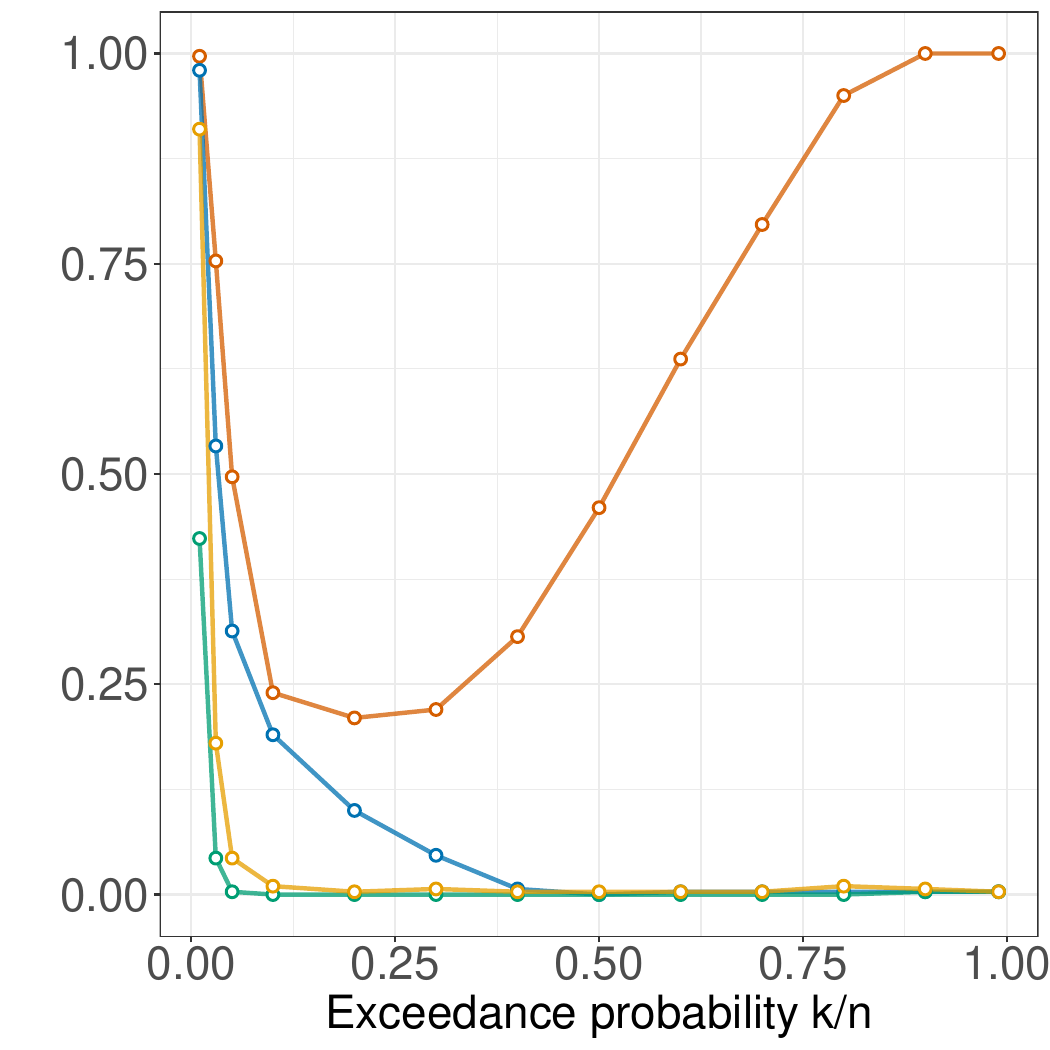}%
        
	\includegraphics[height=.33\textwidth]{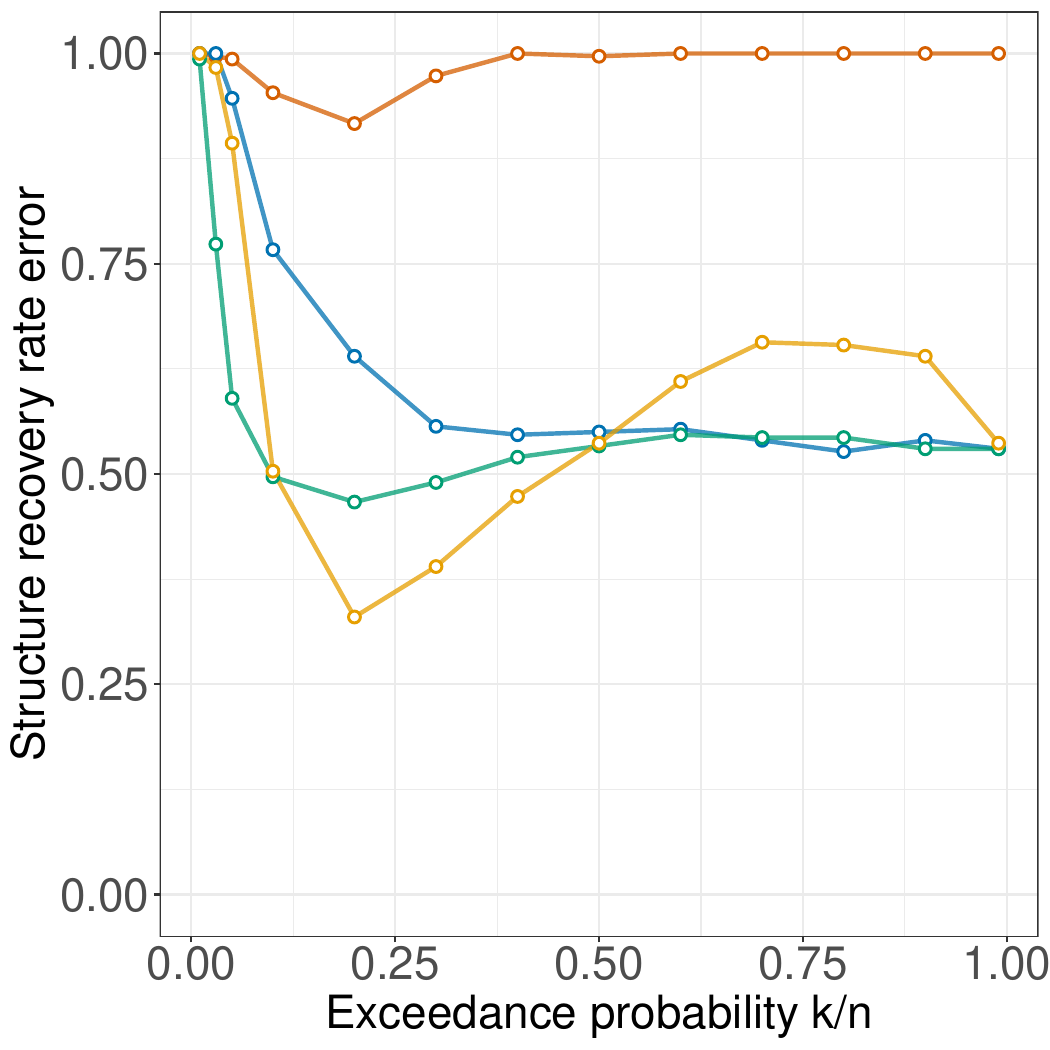}%
	\hspace{0em}\includegraphics[height=.33\textwidth]{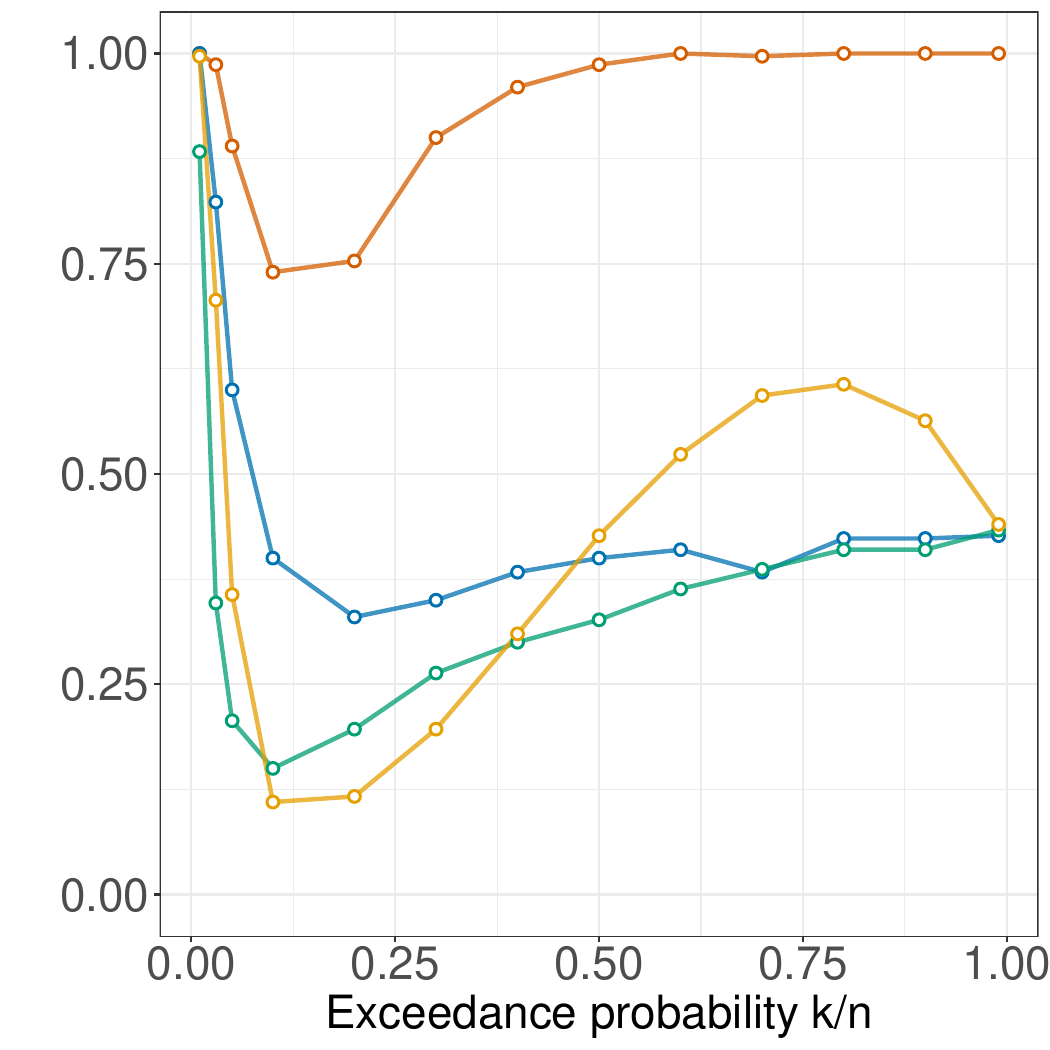}%
	\hspace{0em}\includegraphics[height=.33\textwidth]{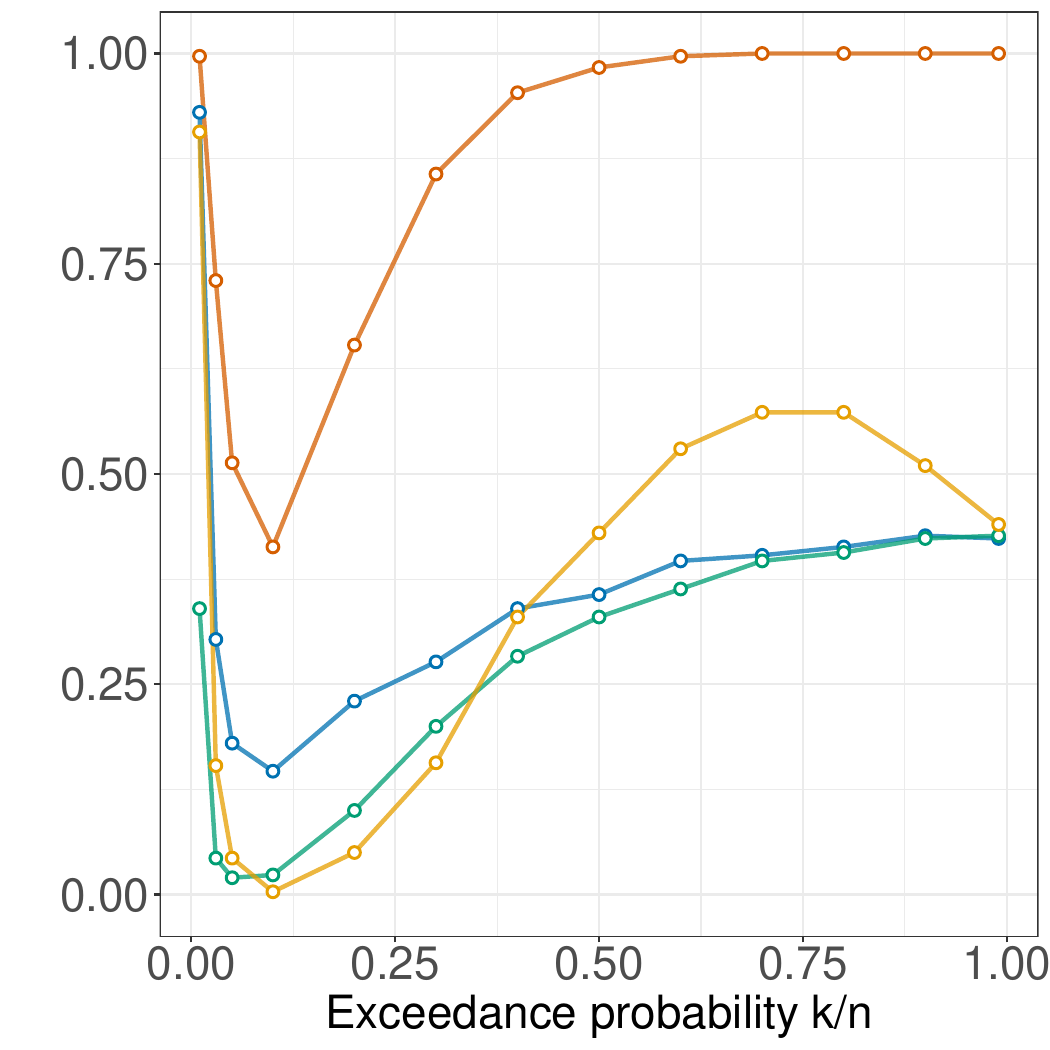}%

	\caption{
Structure recovery rate error of trees from the H\"usler--Reiss model (M1)  and independent noise (N1) (top) and tree noise (N2) (bottom) in dimension $d=20$ estimated based on empirical correlation (orange), extremal variogram with fixed $m\in V$ (blue), combined empirical variogram (green) and censored maximum likelihood (yellow) as a function of the exceedance probability $k/n$; sample sizes $n=500$ (left column), $n=1000$ (center column) and $n=2000$ (right column).
	}
	\label{fig_HR_kn}
      \end{figure}

Next we compare the performance of the different structure learning methods for varying sample size $n$. Since the value of $q^*$ which is required for consistent estimation is unknown in practice we choose $k = \lfloor n^{0.8}\rfloor$, which satisfies all assumptions of our theory.
The results for dimension $d=20$ are shown in the top row of Figure~\ref{fig_HR_srr} for the H\"usler--Reiss model (M1) and in the bottom row for the asymmetric Dirichlet model (M2). We observe that the two methods based on the extremal variogram perform consistently better that the extremal correlation based method. Intuitively this can be explained by the fact that the extremal variogram is a tree metric for conditional independence of multivariate Pareto distributions. The additivity on the tree results in a bigger loss in the minimum spanning tree algorithm when choosing a wrong edge, and therefore it is easier to identify the true structure. The extremal correlation only satisfies a weaker relation~\eqref{eq:orderchi2} on the tree, which might be a reason for the higher error rate. Additionally, the empirical variogram uses information from the entire multivariate Pareto distribution, while the extremal correlation evaluates its distribution at a single point only. A comparison with the censored maximum likelihood estimator (iv) yields several insights. First, this approach seems to lead to consistent estimation of the tree structure even in model (M2) where $\g Y$ is not a H\"usler--Reiss distribution and the likelihood is thus misspecified. This might be explained by the fact that the strength of dependence is still sufficiently well estimated and the minimum spanning tree does only require correct ordering of the edge weights, which is much weaker than consistency of the estimated weights. Second, the different types of noise distributions in (N1) and (N2) lead to opposing orderings of the best method: whereas $\hat T_{\text{CL}}$ has a slight advantage for noise (N2), $\hat T_\Gamma$ performs substantially better under (N1). Notably, this is even the case in model (M1) where the likelihood is well-specified. A possible explanation is that the likelihood is not exactly specified due to the added noise in the model and the use of ranks during estimation. This implies that classical results about asymptotic optimality of maximum likelihood methods do not apply here. Moreover, the added noise has different effects on the biases of the estimators, which changes the order of performance depending on the noise distribution.

  \begin{figure}
    \centering
  \includegraphics[height=.38\textwidth]{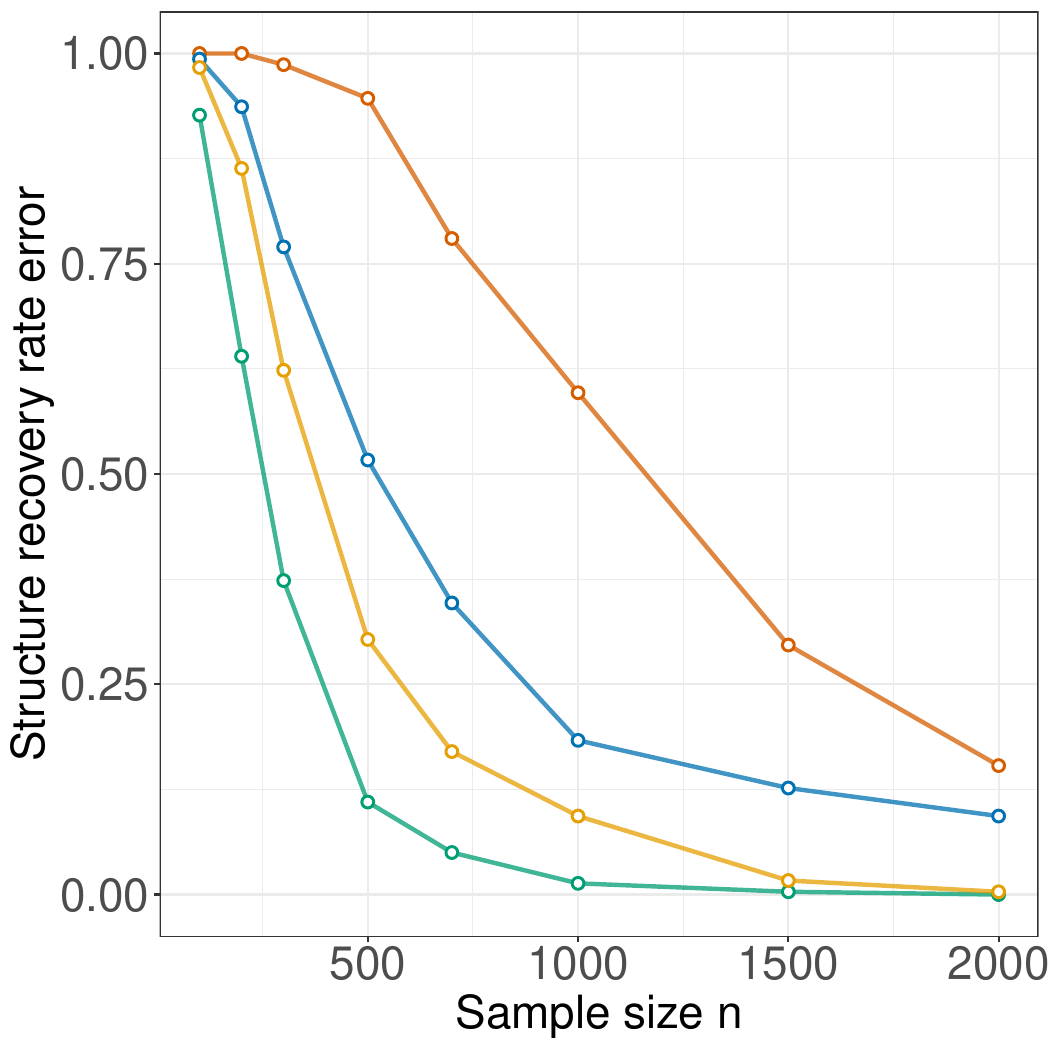}%
  \hspace{.6em}\includegraphics[height=.38\textwidth]{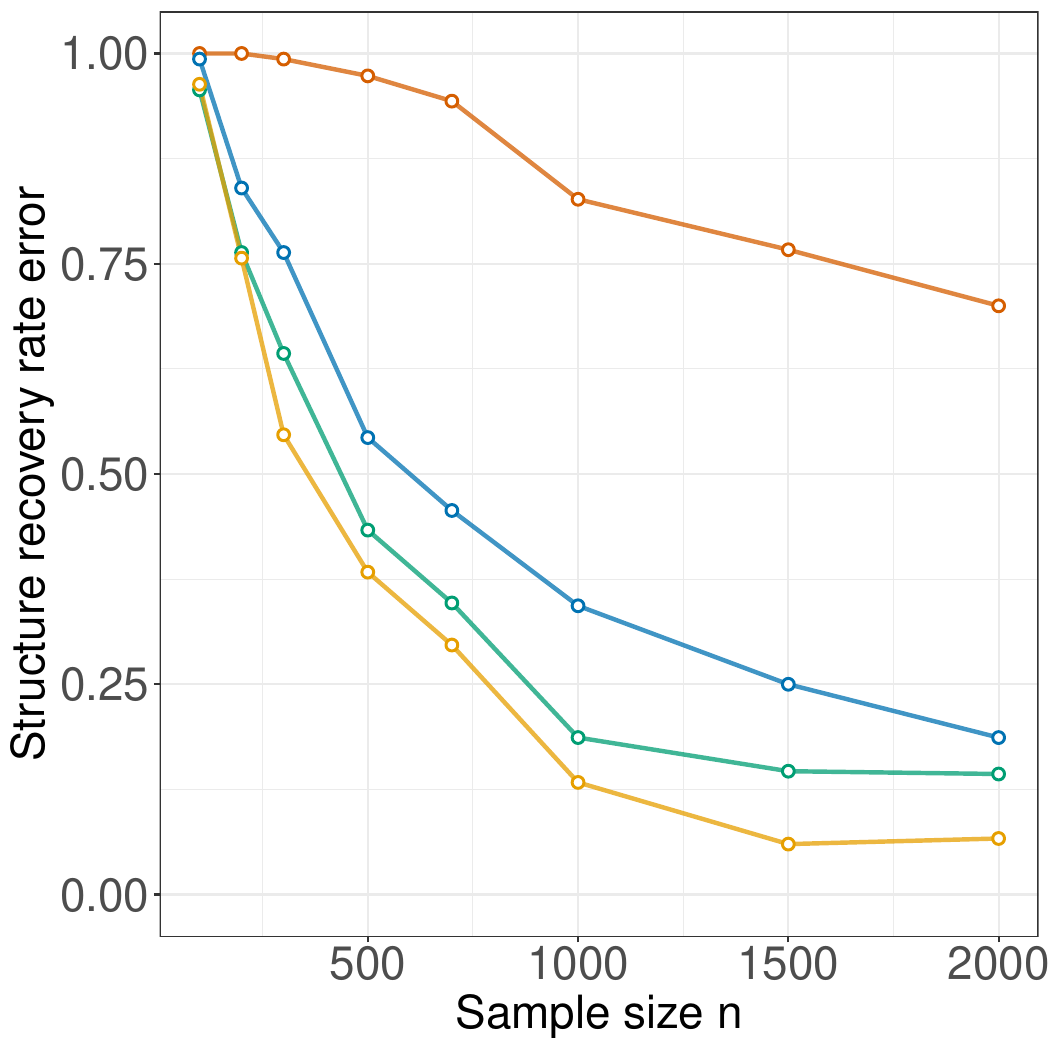}%
  
 \includegraphics[height=.38\textwidth]{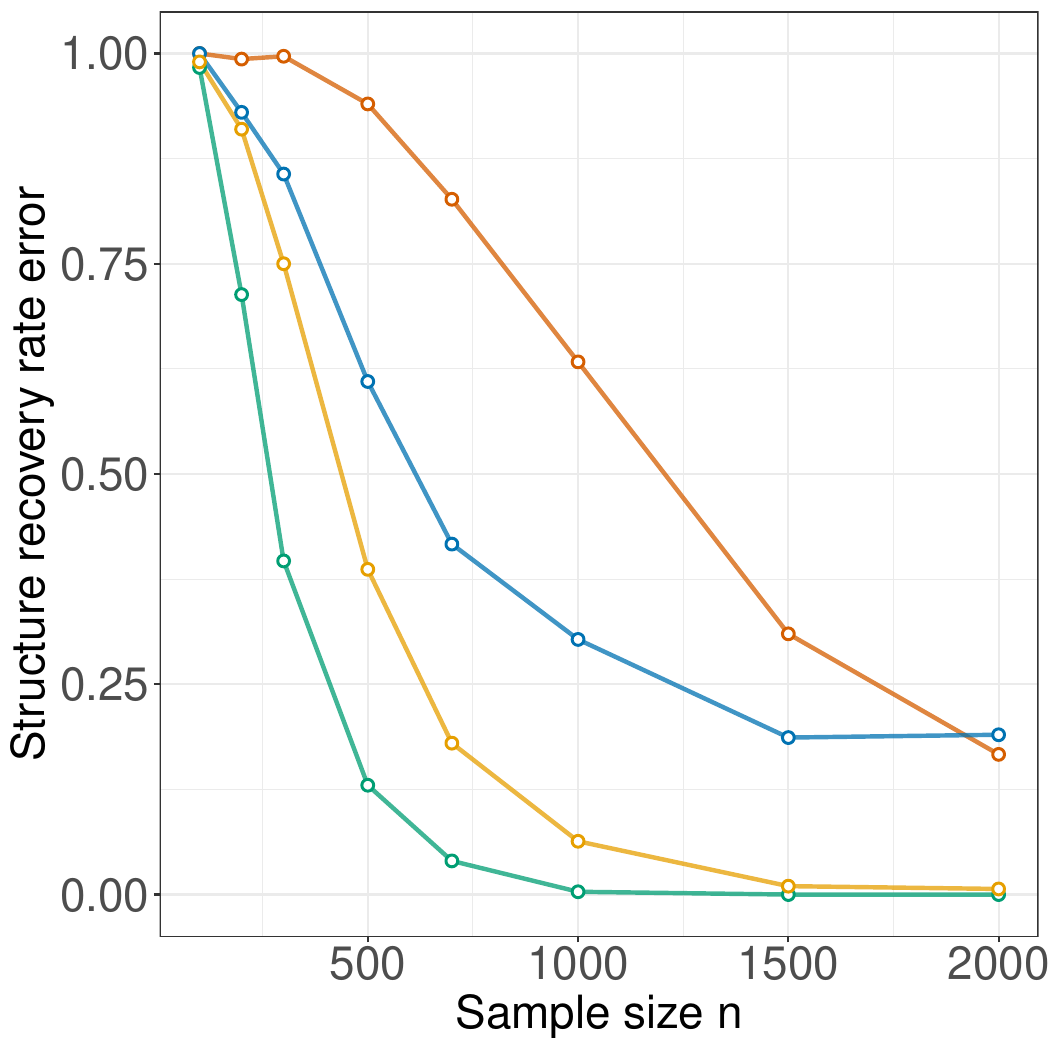}%
  \hspace{.6em}\includegraphics[height=.38\textwidth]{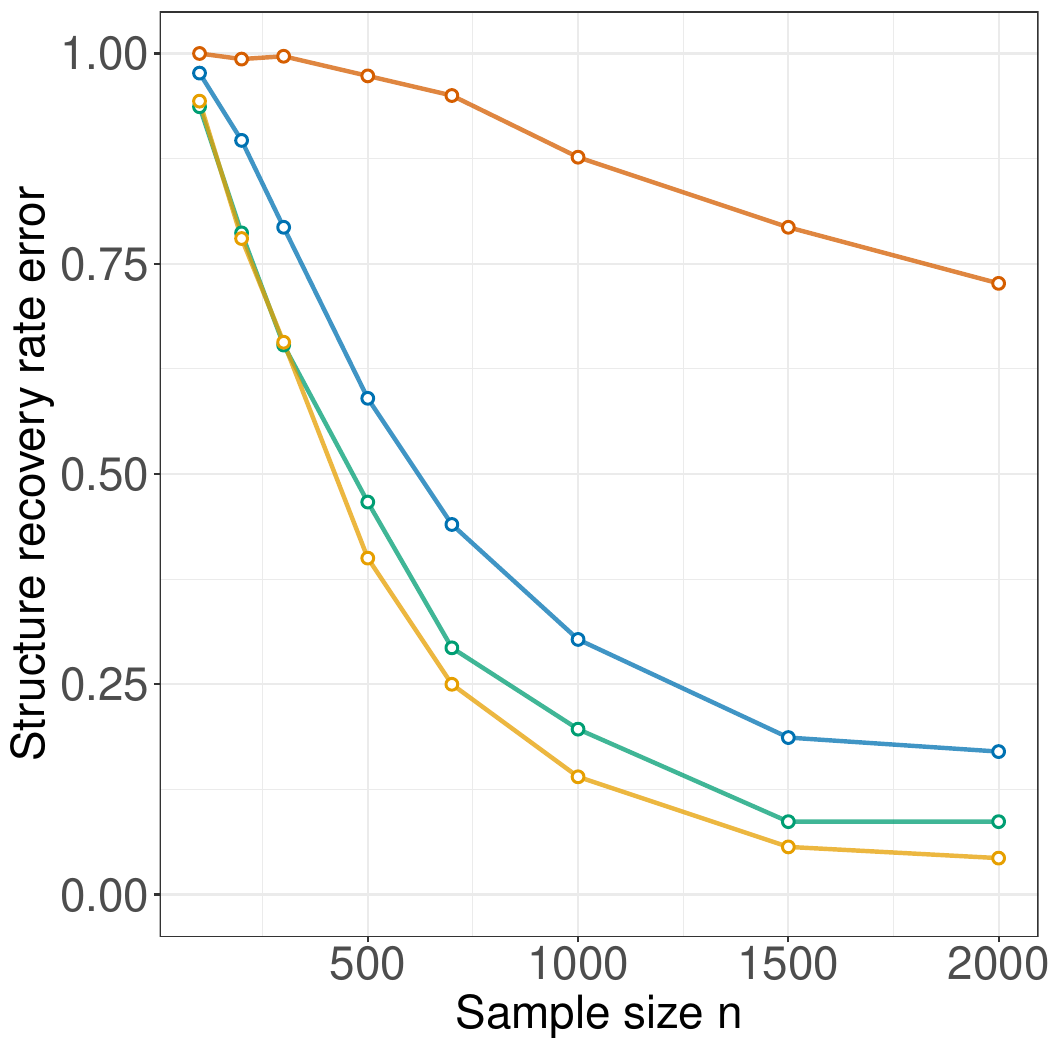}%
  \caption{Structure recovery rate error of trees from the H\"usler--Reiss model (M1) (top) and Dirichlet model (M2) (bottom) in dimension $d=20$ estimated based on empirical correlation (orange), extremal variogram with fixed $m\in V$ (blue), combined empirical variogram (green) and censored maximum likelihood (yellow); independent noise (N1) (left) and tree noise (N2) (right).}
  \label{fig_HR_srr}
  \end{figure}

For a given tree, the task of estimating the correct structure can largely differ according to the strength of dependence of the multivariate Pareto distribution. We therefore conduct a simulation study where we fix $n=500$ and $k = \lfloor n^{0.8}\rfloor$ and illustrate the performance of the structure estimation methods for a varying strength of tail dependence. For the H\"usler--Reiss model, we randomly generate a tree $T=(V,E)$ in dimension $d=20$ and for $(i,j) \in E$ we fix all $\Gamma_{ij} = \lambda$ to some constant $\lambda > 0$. Equivalently, that means that all neighboring nodes have extremal correlation $\chi_{ij} = 2 - 2 \Phi\big(\sqrt{\lambda}/2\big)$. The left panel of Figure~\ref{fig_HR_d10_dependence} shows the results for varying strength of extremal dependence between neighbors measured by the extremal correlation under noise model (N1). Unsurprisingly, the performance of all methods deteriorates at the boundaries, which correspond to the non-identifiable cases of independence and complete dependence. In general, it seems that the empirical variogram based estimators perform better under stronger dependence, which is probably due to the higher bias of the empirical extremal variogram under weak dependence. The same asymmetry can be observed for the censored maximum likelihood method, while the performance of the extremal correlation seems to be symmetric around $\chi = 1/2$. Comparing the performance of different methods, we observe that under noise (N1) the combined extremal variogram performs best uniformly in the values of $\chi$, and the advantage over all other methods can be substantial. The same analysis with noise (N2) is shown in the right panel of Figure~\ref{fig_HR_d10_dependence}. In line with the results in Figure~\ref{fig_HR_srr}, the performance of $\hat T_{\text{CL}}$ and $\hat T_\Gamma$ is fairly similar, with a slight advantage for $\hat T_{\text{CL}}$ at values of $\chi$ around $0.5$ and the converse for $\chi$ closer to $0.2$ and $0.8$.

  \begin{figure}
    \centering
  \includegraphics[height=.38\textwidth]{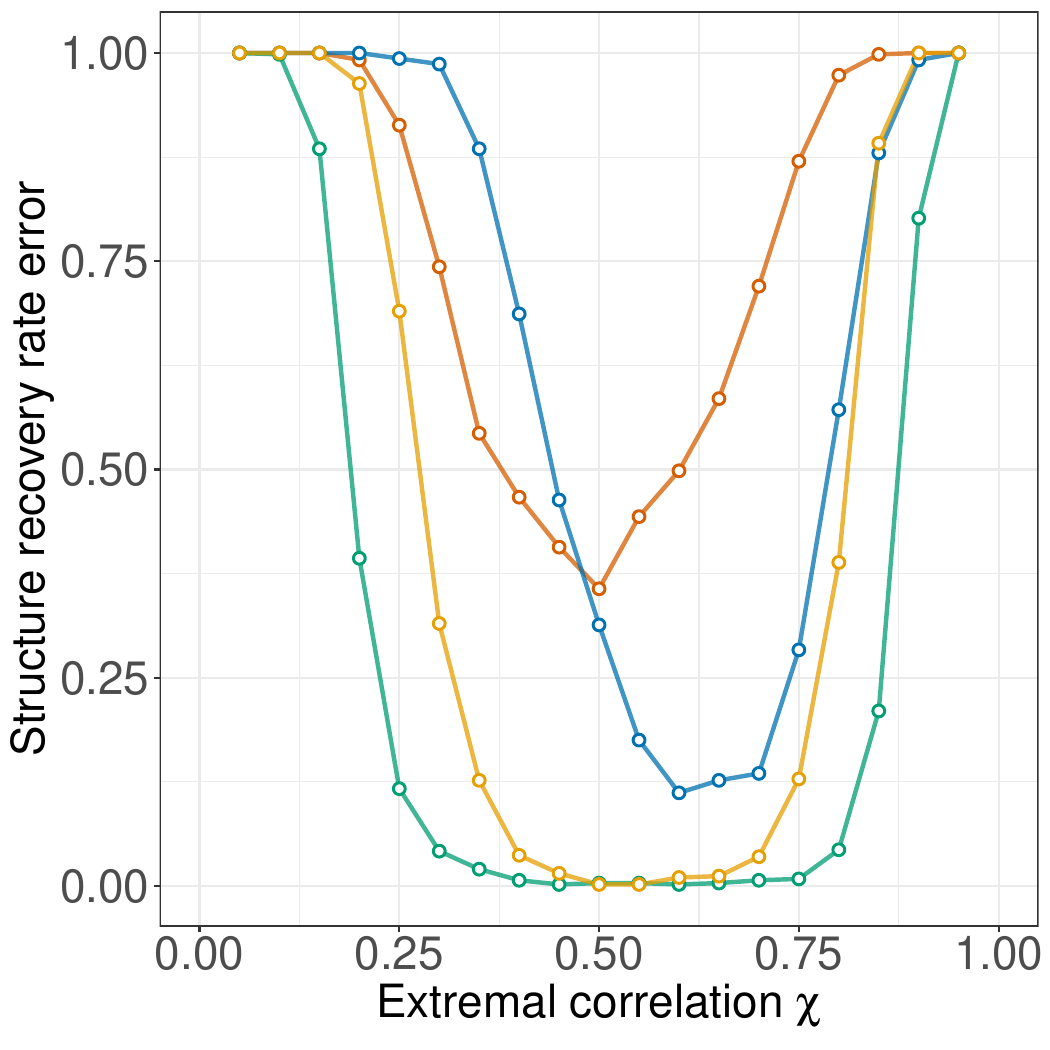}%
  \includegraphics[height=.38\textwidth]{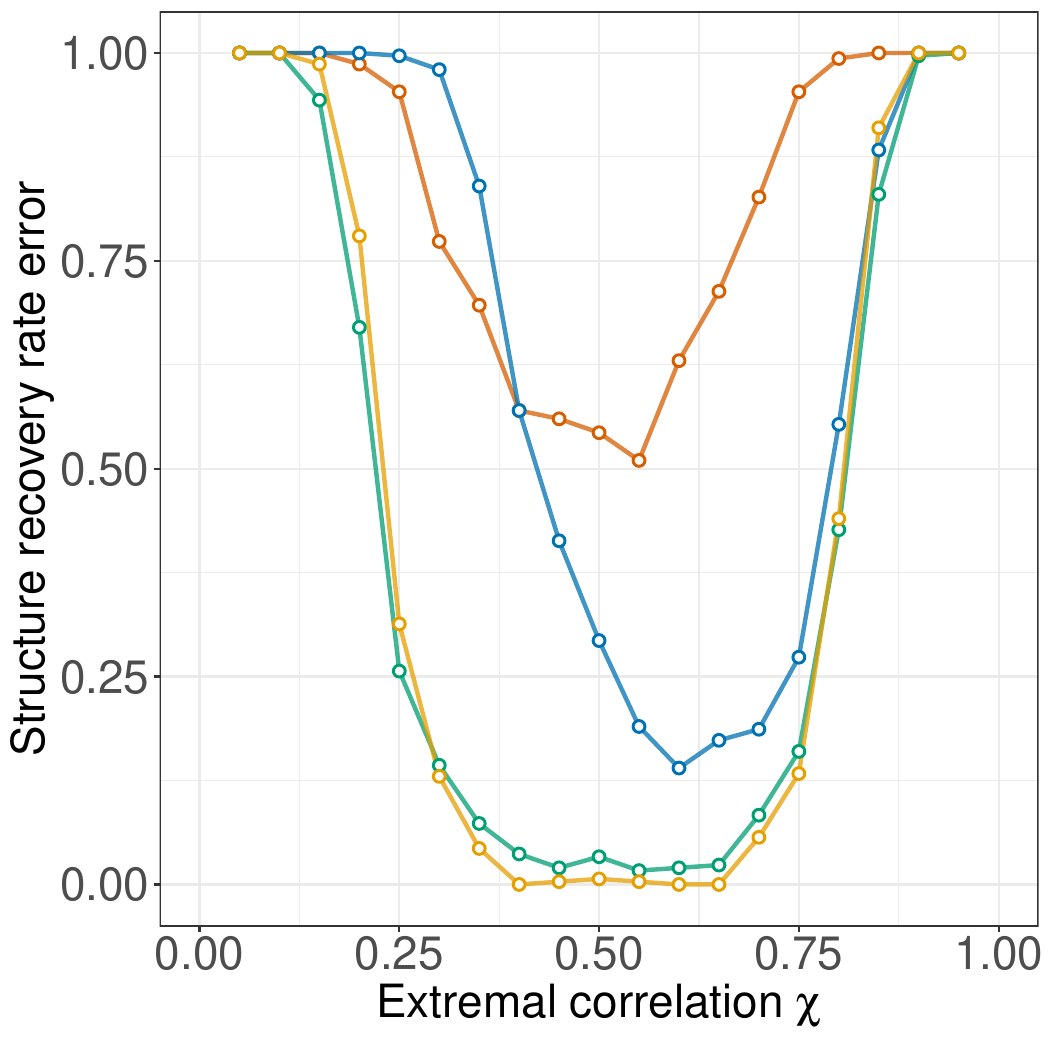}%
  \caption{Structure recovery rate error for the H\"usler--Reiss model (M1) with noise model (N1) (left) and (N2) (right) in dimension $d=20$ as a function of the extremal dependence between neighbors measured by the extremal correlation $\chi$; the different methods are based on empirical correlation (orange), extremal variogram with fixed $m\in V$ (blue), combined empirical variogram (green) and censored maximum likelihood (yellow).}
  \label{fig_HR_d10_dependence}
  \end{figure}

  For the final set of comparisons we study the performance of the methods for a growing dimension $d\in \{10,20,30,50,100,200,300\}$, where we fixed the sample size $n=1000$ and number of exceedances $k = \lfloor n^{0.8}\rfloor =251$. Figure~\ref{fig_HR_dimension} shows the structure recovery rate errors for the different methods. As expected, the errors increase for larger dimensions, but much slower for the combined extremal variogram than for the other methods. Theoretical pre-asymptotic error bounds for the error rates can be found in Section~\ref{sec:high_dim}.  We remark that for we were not able to run simulations in more than $d=100$ dimensions for the censored maximum likelihood estimator because of the prohibitive computational cost. For the same reason we have only included simulations in one model and one noise setting.

\begin{figure}
    \centering
  \includegraphics[height=.38\textwidth]{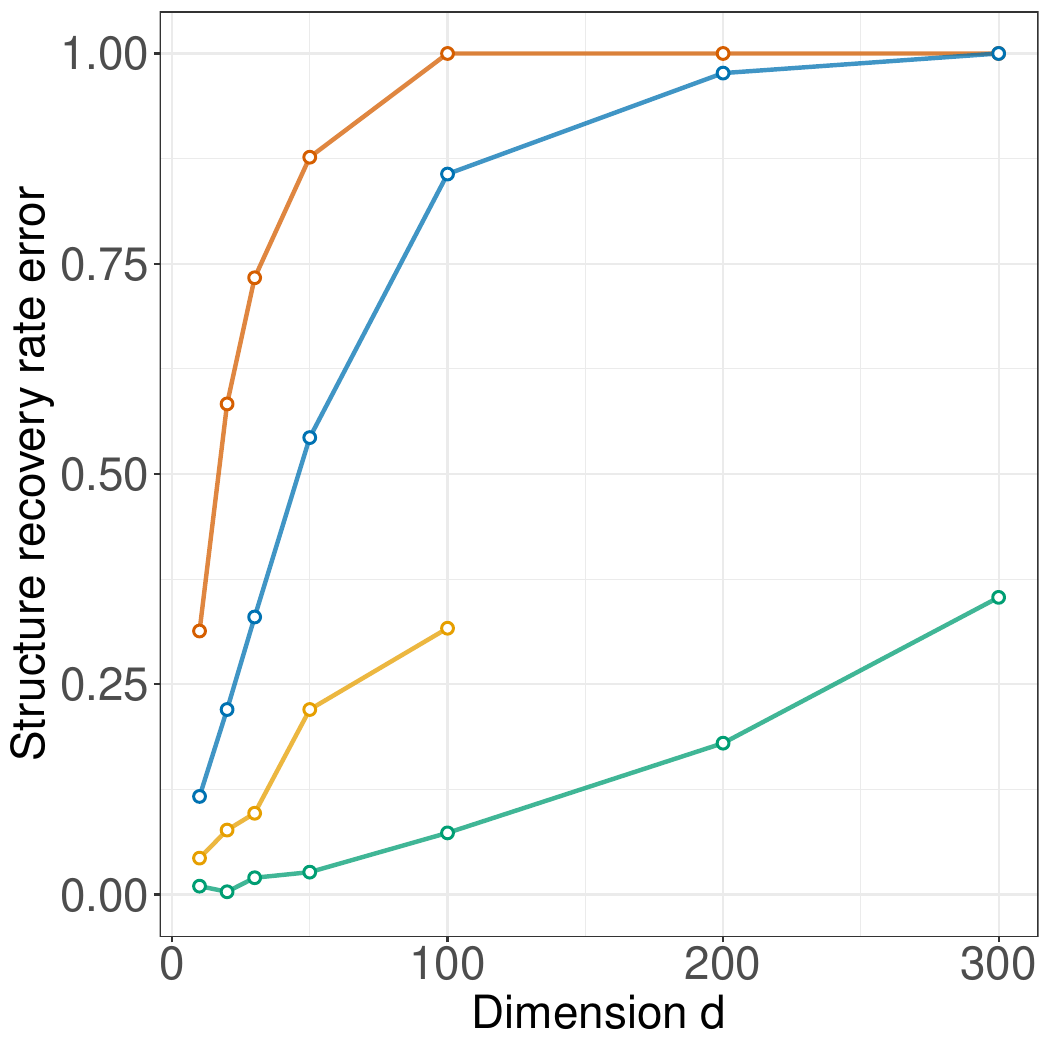}%
  \caption{Structure recovery rate error for the H\"usler--Reiss model (M1) with noise model (N1) (left) as a function of the number of nodes $|V| = d$ of the tree; the different methods are based on empirical correlation (orange), extremal variogram with fixed $m\in V$ (blue), combined empirical variogram (green) and censored maximum likelihood (yellow).}
  \label{fig_HR_dimension}
  \end{figure}

We close this section with some comments on computation times for the four estimators.
  The extremal correlation and variogram based trees rely on empirical estimators and are very efficient to compute. The censored likelihood estimator however requires numerical optimization for every weight $\rho_{ij}$, $i,j\in V$. Especially in higher dimensions this becomes prohibitively costly. Figure~\ref{comp_time} shows the average computation times for the four estimators in the simulations in Figures~\ref{fig_HR_srr} and~\ref{fig_HR_dimension}. It can be seen that the censored likelihood method is several orders of magnitude slower than the empirical methods. As seen in the right-hand panel of Figure~\ref{comp_time}, this quickly becomes prohibitive if the dimension grows.

\begin{figure}[h]
	\centering
	\includegraphics[height=.38\textwidth]{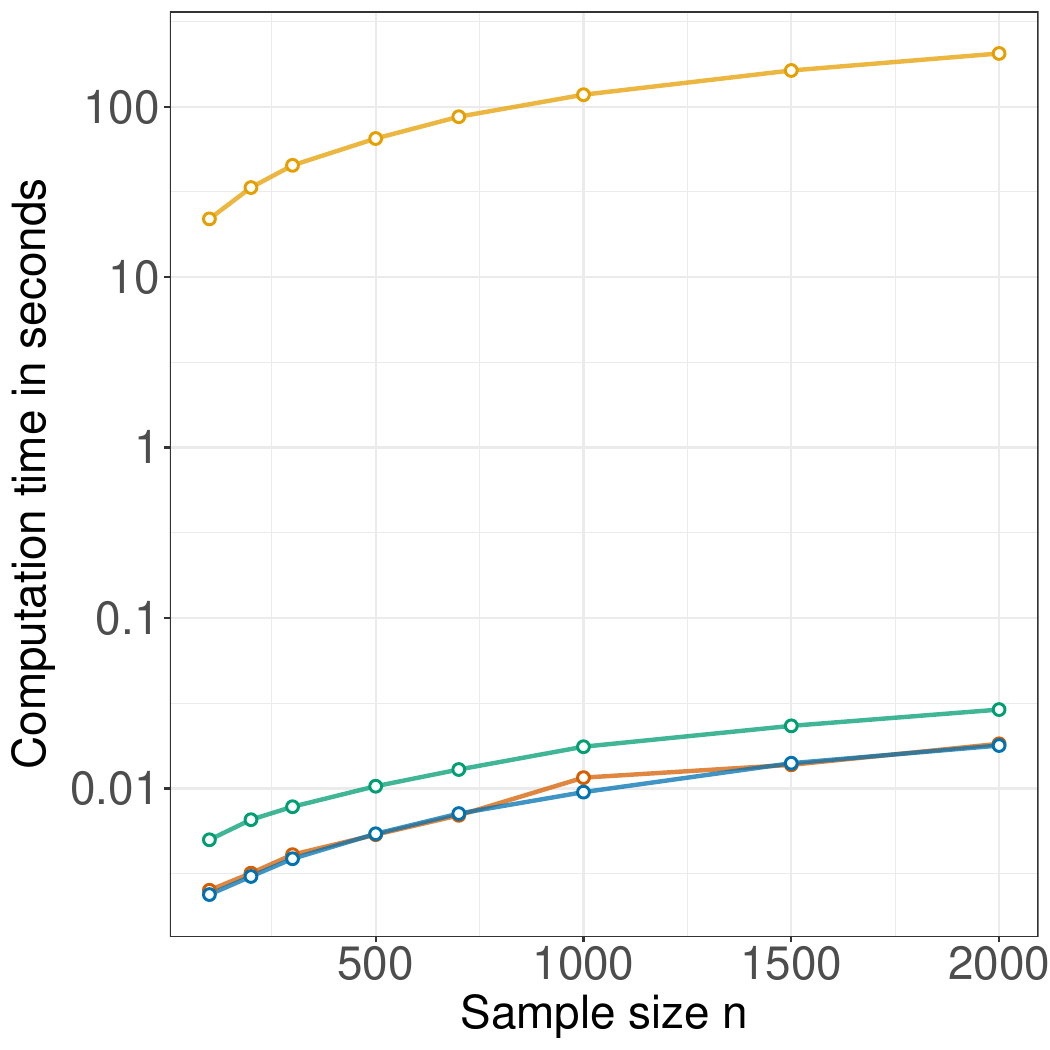}%
        \includegraphics[height=.38\textwidth]{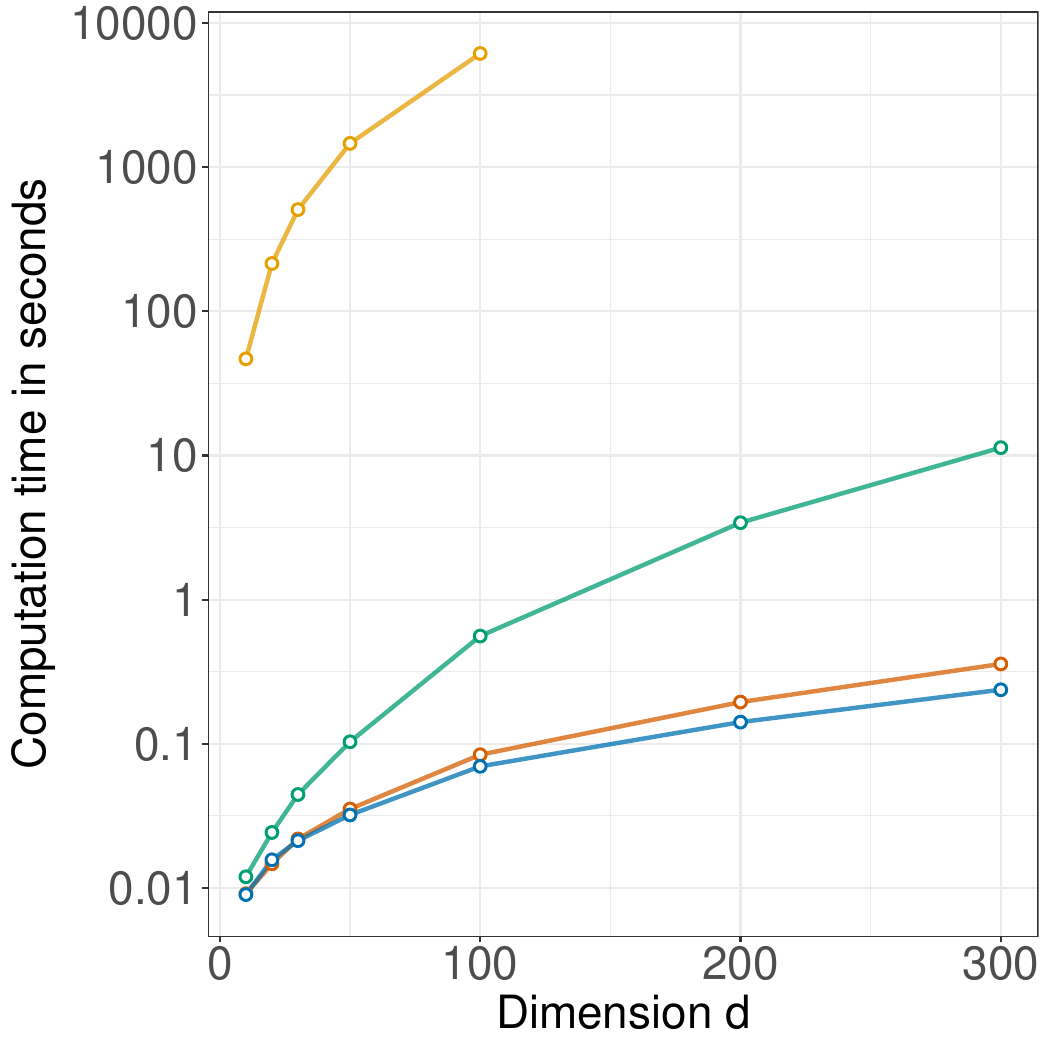}%
	\caption{Average computation times in seconds in the simulation studies in Section~\ref{sec:simu} of the four algorithms based on empirical correlation (orange), extremal variogram with fixed $m\in V$ (blue), combined empirical variogram (green) and censored maximum likelihood (yellow). {Left: for fixed dimension $d=20$; right: for fixed sample size $n=1000$.}}
	\label{comp_time}
      \end{figure}

  \section{Application}
\label{sec:application}
We illustrate the proposed methodology on foreign exchange rates of $d=26$ currencies expressed in terms of the British Pound sterling; see Table~\ref{tab:countries} in Appendix~\ref{app:codes} for the three-letter abbreviations of the respective countries. The data are available from the website of the Bank of England\footnote{\texttt{https://www.bankofengland.co.uk/}}. They consist of daily observations of spot foreign exchange rates in the period from 1 October 2005 to 30 September 2020, resulting in $n=3790$ observations.

In order to obtain time series without temporal dependence, we pre-process the data set.
  We first compute the daily log-returns $R_{ij}$, $i=1,\dots,n$, $j=1,\dots, d$, from the original time series. To remove the serial dependence, we then filter the univariate series by ARMA-GARCH processes; see \cite{hil2014} for a similar approach, and \cite{bol1992} and \cite{eng1982} for background on financial time series modeling. The AIC suggest that an $\text{ARMA}(0,2)$-$\text{GARCH}(1,1)$ model is the most appropriate for most of the univariate series.
  We derive the absolute values of the standardized filtered returns as
  $$ X_{ij} = \left |\frac{R_{ij} - \hat \mu_{ij}}{\hat \sigma_{ij}} \right|,$$
  where $\hat \mu_{ij}$ and $\hat \sigma_{ij}$ are the estimated mean and standard deviation of the ARMA-GARCH model. The absolute value means that we are interested in extremes in both directions.

  The data $\g X_i = (X_{i1}, \dots, X_{id})$ are approximately independent and identically distributed~for $i=1,\dots, n$, and we will model their tail dependence using an extremal tree model. We first check whether the assumption of asymptotic dependence is satisfied by inspecting the behavior of the function $q \mapsto \hat \chi_{ij}(q)$ for values $q=k/n$ close to 1. For most of the pairs this function seems to converge to a positive value and thus there is fairly strongly dependence in the tail between the filtered log-returns; see Figure~\ref{chiplots} in the Supplementary Material~\ref{sec:chiplots}
 for some examples.

 {Before estimating for this data set the extremal tree structure, we discuss the choice of the number of exceedances $k$, or equivalently the probability of exceedance $q=k/n$. This is an important practical issue and a long-standing problem in extreme value theory. In essence, it is a bias-variance trade-off as illustrated in the simulations in Figure~\ref{fig_HR_kn}.

  For tree structure estimation, we propose to leverage the specific structure of the tree learning problem. From~\eqref{gamma_additivity} it follows that {for an extremal graphical model on a tree $T$, the corresponding population $\Gamma = d^{-1} \sum_{m=1}^d \Gamma^{(m)}$ forms a tree metric on that tree}. In the sequel, for generic $\Gamma$ and tree $T$, denote the extremal variogram matrix completed on the tree $T$ by
    \begin{align*}
      \Gamma_{ij}^T = \sum_{(s,t) \in \ph(ij; T)} \Gamma_{st}, \quad i,j \in V.
    \end{align*}
{We propose to select $k$ so as to minimize the deviation of the empirical values of $\hat \Gamma$ from forming a tree metric on the estimated tree $\hat T$. More precisely, define
} 
\begin{align}\label{chi_err}
\hat \Delta(k/n) =  \sum_{i,j \in V} \left\{ g(\hat \Gamma_{ij}^{\hat T} (k/n))  - g(\hat \Gamma_{ij}(k/n)) \right\}^2,
\end{align}
where as function $g: \R^+ \to [0,1]$, we choose the transformation from $\Gamma$ to $\chi$ in H\"usler--Reiss models, that is, $ g(x) = 2-2\Phi(\sqrt{x}/2)$; see Example~\ref{ex_HR}. Here we indicate that these estimates depend on the exceedance probability $q = k/n$; note that also the estimated tree $\hat T$ depends on $k$. Additional motivation for the form of $\hat \Delta(k/n)$ and a literature review of classical approaches is given in Supplementary Material~\ref{sec:choicek}.

Motivated by the simulations in the previous section we estimate the extremal tree structure $\hat T_\Gamma = (V, \hat E_\Gamma)$ non-parameterically using the combined empirical extremal variogram $\hat \Gamma$. Figure~\ref{chi_error} shows the error $\hat \Delta(k/n)$ as a function of $q = k/n$ for the exchange rate data set. It can be seen that, indeed, the error seems to stabilize for values of $1-q$ above $0.97$. We therefore choose $q=0.03$ in this application, which corresponds to $k=114$.} The corresponding minimum spanning tree is shown in Figure~\ref{MST_exchange}; we note that the tree is very stable across different values of $q$ close to 0.

 The structure of the tree allows for a nice interpretation of extremal dependence. Extreme observations in the exchange rates with the Euro are strongly connected with extremes of other European currencies in Northern and Eastern countries. The graph suggests that extremes of exchange rates of these currencies are conditionally independent of exchange rates of other countries, given the value of Euro exchange rate. The Malaysian ringgit, the Chinese yuan, the Hong Kong dollar and the Taiwan dollar are strongly pegged to the US dollar and their closeness in the tree is therefore not surprising. Another branch of the tree contains several currencies of the Commonwealth. Finally, the connection between Japan and Switzerland is plausible because both currencies can be considered safe-haven currencies, which are both popular investments in times of crises.
  
  In order to address the stability of the tree structure we bootstrap our data $B=100$ times and fit each time the tree structure. {For generating each bootstrap sample, we draw with replacement $n$ data from the sample of filtered observations $\g X_1,\dots,\g X_n$. This is a heuristic approach to assess the overall stability of our empirical conclusions to small perturbations in the data and does not have a formal theoretical justification at this point.} Figure~\ref{MST_exchange_bootstrap} shows that graph where the width of each edge is proportional to the number of times it has been selected in an extremal tree. Overall, the tree seems to be fairly stable since there is only a small number of dominant edges. Moreover, we can identify clear clusters that are connected in most of the trees, such as the European currencies. On the other hand some currencies such as the Russian ruble that do not have a dominant connection to any of these clusters. In future research, it could therefore be interesting to study structure estimation for forests, which allow to have unconnected graphs whose connected components are trees \citep{LXGGLW2011}.

  So far we have not assumed any specific model for the extremal dependence on the edges since we are able to estimate the tree structure fully non-parametrically with the methods from this paper. If we were only interested in interpretation of the extremal graphical structure we could stop our analysis here. If we require a model for rare event simulation or risk assessment, in a second step we can choose arbitrary bivariate Pareto models for each edge. For simplicity, we choose here for all edges the H\"usler--Reiss model (see Example~\ref{ex_HR}) resulting in a H\"usler--Reiss tree. For this model, the bivariate parameter estimates $\hat \Gamma_{ij}$ can be chosen directly as the empirical extremal variogram estimates for all $\{i,j\} \in \hat E_\Gamma$. Alternatively, we could estimate them by censored maximum likelihood. In both cases, the remaining entries of the H\"usler--Reiss parameter matrix can be obtained from the additivity of the extremal variogram on the tree in~\eqref{gamma_additivity}. {We denote the corresponding parameter matrix completed on the tree $\hat T_\Gamma$ by $\hat \Gamma^{\hat T_\Gamma}$}. {Recall the relation between $\Gamma$ and $\chi$ for H\"usler--Reiss models from Example~\ref{ex_HR}.} Figure~\ref{MST_exchange_ECs} shows the extremal correlations implied by the fitted H\"usler--Reiss tree model, that is, $\hat \chi^{\hat T_\Gamma}_{ij} = 2 - 2 \Phi\big(\sqrt{\hat \Gamma_{ij}^{\hat T_\Gamma}}/2\big)$, against the empirical counterparts $\hat \chi_{ij}$, $i,j\in V$. Even though the tree structure is a very sparse graph with only $d-1$ edges, the extremal dependence between all variables is well-explained.

  \begin{figure}
    \centering
  \includegraphics[height=.65\textwidth, trim = 0 2cm 0 0, clip]{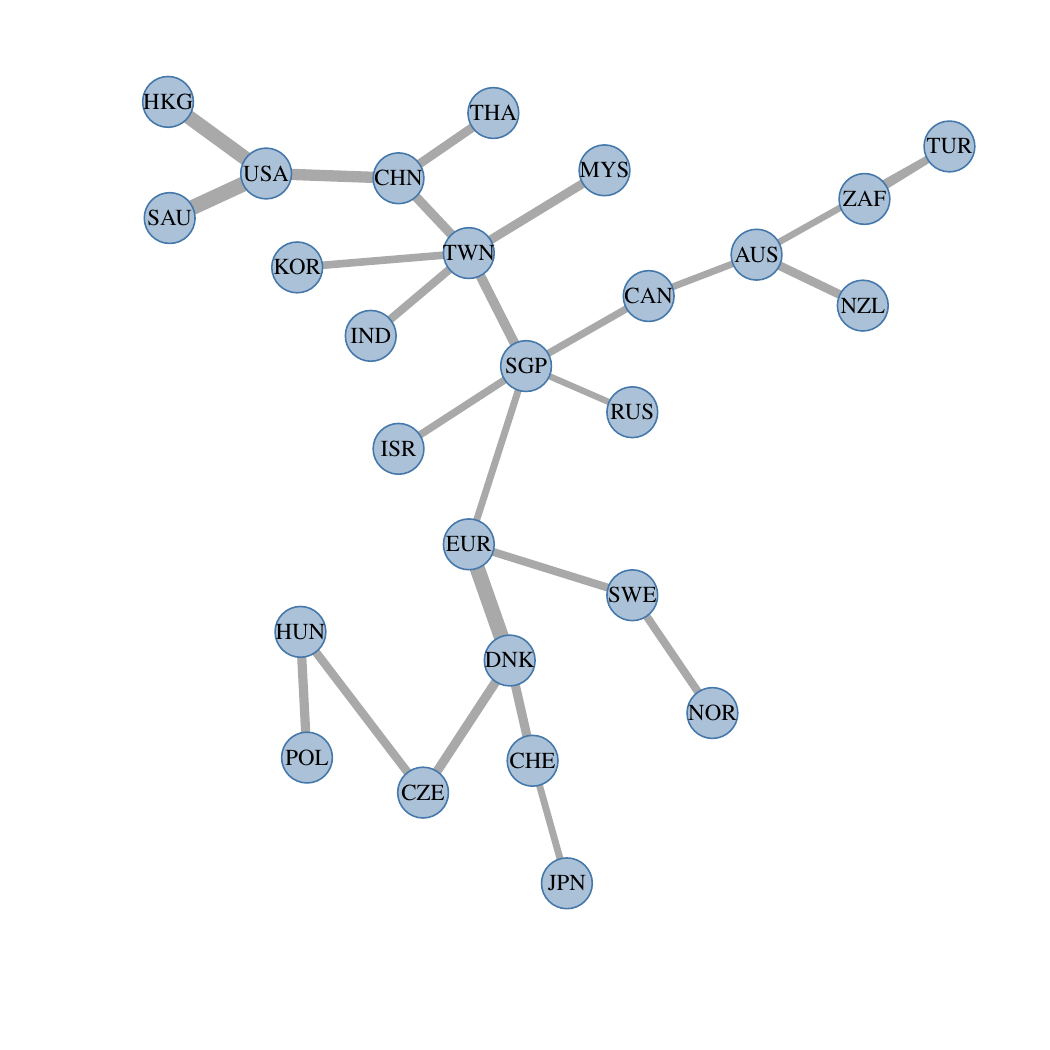}%
  \caption{Minimum spanning tree $\hat T_{\Gamma}$ of extremal dependence for the spot foreign exchange rate data based on the combined extremal variogram.
    The width of each edge $(i,j)\in \hat E_\Gamma$ is proportional to the extremal correlation $2 - 2 \Phi(\sqrt{\hat \Gamma_{ij}}/2)$, and therefore wider edges indicate stronger extremal dependence.}
  \label{MST_exchange}
  \end{figure}

  \begin{figure}
    \centering
  \includegraphics[height=.65\textwidth, trim = 0 2cm 0 0, clip]{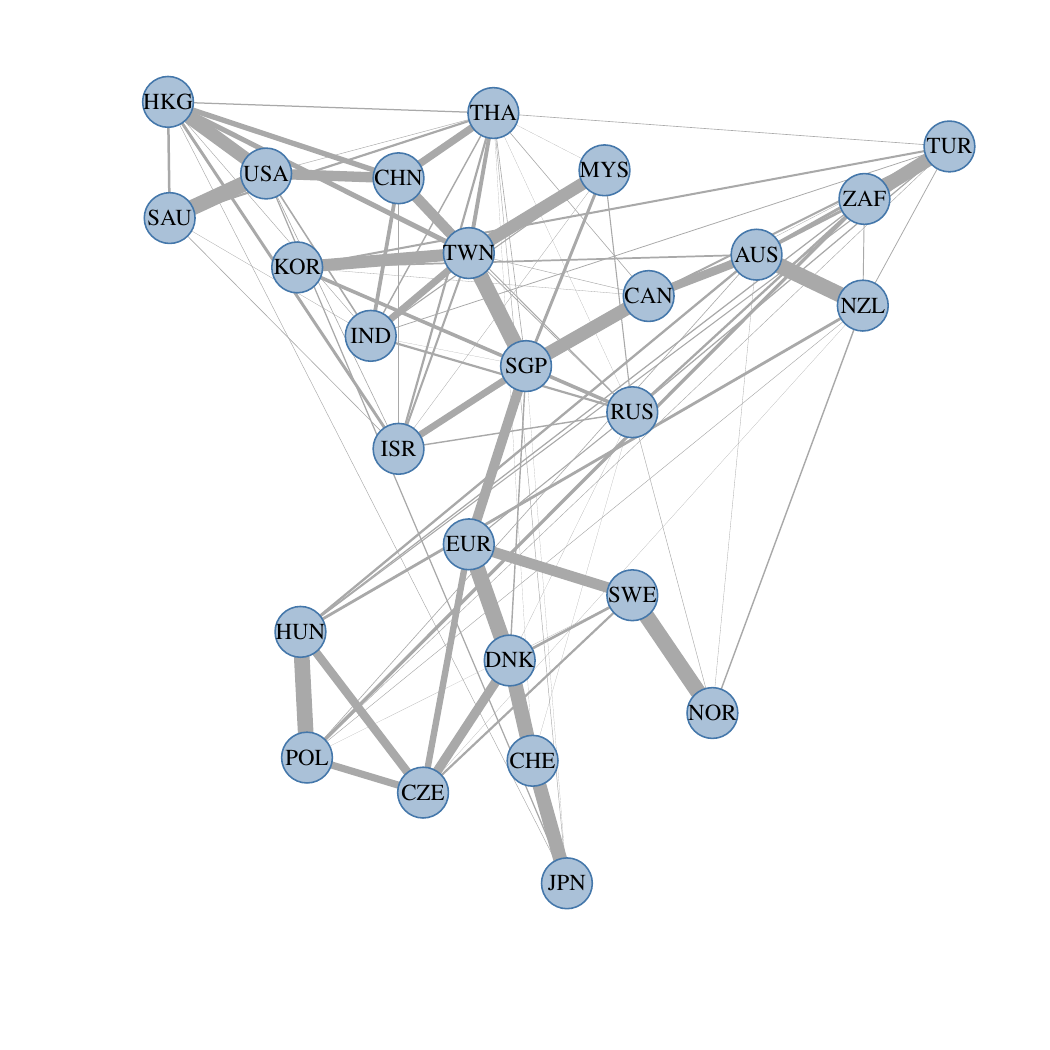}%
  \caption{Graph where the width of each edge is proportional to the number of times it has been selected in an extremal tree in the bootstrap procedure.}
  \label{MST_exchange_bootstrap}
  \end{figure}

  \begin{figure}
    \centering
  \includegraphics[height=.38\textwidth]{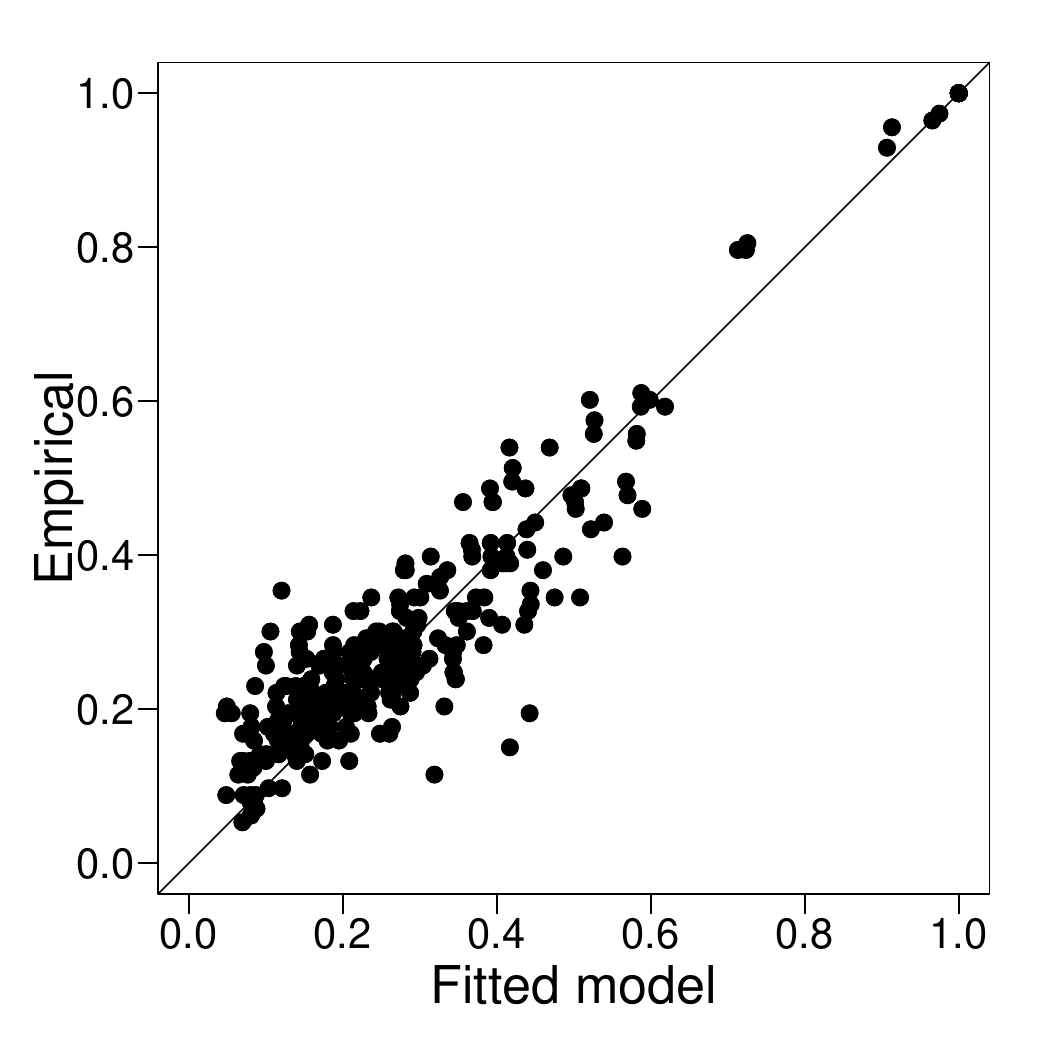}%
  \caption{Extremal correlations for the spot foreign exchange rate data implied by the fitted H\"usler--Reiss
tree model against the empirical counterparts.}
  \label{MST_exchange_ECs}
  \end{figure}

  \begin{figure}
    \centering
  \includegraphics[height=.38\textwidth]{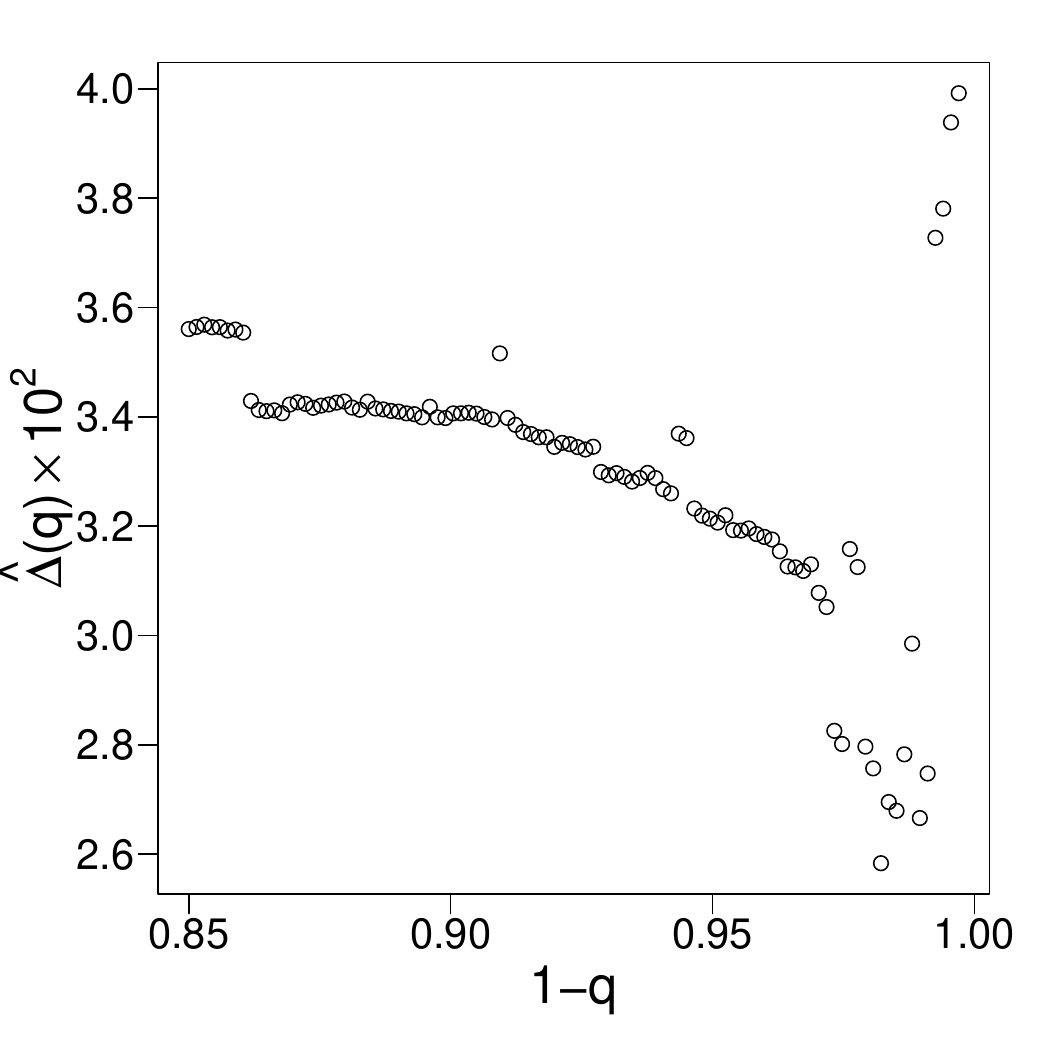}%
  \caption{The squared error $\Delta$ defined in~\eqref{chi_err} between empirical extremal correlations and those implied by the tree metric structure for different values of the exceedance probability $q=k/n$.}
  \label{chi_error}
  \end{figure}

\section*{Acknowledgments}
The authors would like to thank Jiaying Gu for pointing us to Hall's marriage theorem,{ and Johan Segers for pointing us to a mistake in an earlier version of Proposition \ref{prop:chimst}. We further thank} Nicola Gnecco, Adrien S.~Hitz, Micha\"el Lalancette and Chen Zhou for helpful comments. We also thank the Associate Editor and two anonymous Referees for comments which helped us to improve the presentation and for encouraging us to consider results in growing dimensions. Sebastian Engelke was supported by an Eccellenza grant of the Swiss National Science Foundation and Stanislav Volgushev was partially supported by a discovery grant from NSERC of Canada.

\appendix

\section*{Appendix}
 \setcounter{subsection}{0}
\setcounter{equation}{0}
\renewcommand{\thesubsection}{A.\arabic{subsection}}
\renewcommand{\theequation}{A.\arabic{equation}}

\renewcommand\theHsubsection{A.\thesubsection}
\renewcommand\theHequation{A.\theequation}

\subsection{Country codes used in the application in Section~\ref{sec:application}}
\label{app:codes}

Table~\ref{tab:countries} shows the three-letter country codes of the exchange rates into British Pound sterling.

\begin{table} 
	\caption{\label{tab:countries}Three-letter country codes.}
	\centering
	\begin{tabular}{l|l|l|l}
		Code & Foreign Exchange Rate (into GBP) & Code & Foreign Exchange Rate (into GBP)\\
		\hline 
		AUS  & Australian Dollar & NOR & Norwegian Krone \\
		CAN  & Canadian Dollar  & POL &Polish Zloty \\
		CHN  & Chinese Yuan & RUS & Russian Ruble\\
		CZE  & Czech Koruna & SAU &Saudi Riyal\\
		DNK &Danish Krone & SGP & Singapore Dollar\\
		EUR & Euro & ZAF & South African Rand\\
a		HKG & Hong Kong Dollar &  KOR &South Korean Won\\
		HUN  & Hungarian & SWE & Swedish Krona\\
		IND & Indian Rupee & CHE & Swiss Franc\\
		ISR & Israeli Shekel & TWN & Taiwan Dollar\\
		JPN &  Japanese Yen & THA & Thai Baht\\
		MYS & Malaysian ringgit & TUR & Turkish Lira\\
		NZL &New Zealand Dollar & USA &  US Dollar
	\end{tabular}
\end{table}

\subsection{Proof of~\eqref{biv_change} in Example~\ref{ex_biv}} \label{sec:proofchange}
Let $z \geq 1$ and define the set $A_z = \{\g x \in \mathcal L: x_2 > 1, x_1 > z x_2  \}$. It holds that
\[
\mathbb P( W_1^2 > z) = \mathbb P( \g Y^2 \in A_z) = \mathbb P( \g Y^1 \in A_z).
\]
The first equality holds since the representation $\g Y^2 = (P W_1^2, P)$ from~\eqref{extr_fct} and the fact that the Pareto variable $P \geq 1$ almost surely imply that $\{W_1^2 > z\} = \{ \g Y^2 \in A_z\}$ almost surely. The second equality holds since $A_z \in \mathcal L^1 \cap \mathcal L^2$ implies, by definition of $\g Y^i$,
\[
\mathbb P( \g Y^2 \in A_z) = \mathbb P( \g Y \in A_z) /  \mathbb P( Y_2 > 1) = \mathbb P( \g Y \in A_z) /  \mathbb P( Y_1 > 1) = \mathbb P( \g Y^1 \in A_z),
\] 
because $\mathbb P( Y_1 > 1) = \mathbb P( Y_2 > 1)$ from the discussion after~\eqref{mpd_limit}.

Note that we can rewrite $A_z = \{\g x \in \mathcal L: x_1 > z, 1 < x_2 <  x_1 / z \}$ and therefore 
\begin{align*}
\mathbb P( \g Y^2 \in A_z) &= \int_z^\infty u^{-2} \mathbb P(u W_2^1 \in (1, u/z)) \mathrm d u \\
&= \mathbb E \int_0^\infty u^{-2} \einsfun\{ u > z, u W_2^1 \in (1, u/z)\} \mathrm d u\\
& = \mathbb E \int_0^\infty v^{-2} W_2^1 \einsfun\{ v > zW_2^1, v \in (1, v/(zW_2^1))\} \mathrm d v     \\    
& = \mathbb E \int_0^\infty v^{-2} W_2^1 \einsfun\{ v > zW_2^1, v > 1, W_2^1 < 1/z\} \mathrm d v \\
& = \mathbb E \int_0^\infty v^{-2} W_2^1 \einsfun\{ v > 1, 1/W_2^1 > z\} \mathrm d v    \\
& = \mathbb E ( \einsfun\{1/W_2^1 > z\} W_2^1),   
\end{align*}
where for the third equality we used a change of variable $v = uW_2^1$.

If $z \in (0,1)$, using the independence of $P$ and $W_1^2$, we can write
$$ z \mathbb P(W_1^2 > z) = \mathbb P( P > 1/z,  W_1^2 > z) = \mathbb P( \g Y^2 \in \tilde A_z),$$
with $\tilde A_z = \{\g x \in \mathcal L: x_1 > 1, 1/z < x_2 <  x_1 / z \}$. Note that this is not possible if $z = 0$. Since $\tilde A_z \in \mathcal L^1 \cap \mathcal L^2$ we have
$ \mathbb P( \g Y^2 \in \tilde A_z) = \mathbb P( \g Y^1 \in \tilde A_z)$, and a similar computation as above yields 
$$ \mathbb P( \g Y^1 \in \tilde A_z) = z  \mathbb E ( \einsfun\{1/W_2^1 > z\} W_2^1).$$
It follows that for any $z > 0$ we have
$$ \mathbb P(W_1^2 > z) =  \mathbb E ( \einsfun\{1/W_2^1 > z\} W_2^1),$$
and consequently
$$ \mathbb P( W^2_1 = 0) = \lim_{z \to 0} \mathbb P(W_1^2 > z) =  \mathbb E ( W_2^1).$$
This yields the desired result.
\hfill $\Box$

\subsection{Proof of Proposition~\ref{prop_vario}}

The assertions of (i) and (ii) follow immediately from the definition of $\g W^m$ in \eqref{extr_fct} and the fact that $\Gamma^{(m)}$ is the variogram matrix of this random vector.

For (iii), the convergence $\chi_{n,im} \to 0$ {as $n \to \infty$} implies that the corresponding extremal functions $ W_{n,i}$ converge to $0$ almost surely. Indeed, we have for any $x \in (0,1)$
$$ 
\chi_{n,im} = \mathbb P(Y_i^m(n) > 1) = \mathbb P(P W_{n,i}^m > 1)  \geq x \mathbb P(P W_{n,i}^m > 1\mid P> 1/x) =  x  \mathbb P(W_{n,i}^m > x),
$$
which shows that $\mathbb P(W_{n,i}^m > x)\to 0$ as $n\to \infty$. 
This yields that  $\Gamma^{(m)}_{n,im} \to \infty$ as $n\to\infty$. \hfill $\Box$

\subsection{Proof of Proposition~\ref{prop_tree_gamma}}
In order to show that the extremal variogram $\Gamma^{(m)}$ defines a tree metric on $T$, we recall the stochastic representation of $\g Y^m$ in Proposition \ref{prop_tree}.
We compute
\begin{align*}
\Gamma_{ij}^{(m)}
  &=  \var \Big\{ \sum_{e\in \ph(mi; T^m)} \log W_e - \sum_{\tilde e\in \ph(mj; T^m)} \log W_{\tilde e} \Big\} \\
  &= \sum_{e\in \ph(mi; T^m) \Delta \ph(mj; T^m)} \var \left\{ \log W_e \right\}\\
  &= \sum_{(s,t) \in \ph(ij; T)} \Gamma^{(m)}_{st},
\end{align*}
where for two sets $A$ and $B$, $A \Delta B$ denotes the symmetric difference. The second to last equality follows from the independence of the $\{W_e:e\in E\}$.
Moreover, for the last equation we note that for two neighboring nodes $(s,t) \in E^m$ in the directed tree $T^m$, by applying the same argument as above, we have $\Gamma^{(s)}_{st} =  \var \left\{ \log W_t^s \right\} = \Gamma^{(m)}_{st}$. \hfill $\Box$

\subsection{Proof of Corollary~\ref{cor1}}
We have to show that for any tree $T' = (V, E')$ that differs from $T$ in at least one edge, it holds
\begin{align}\label{mst_diff0}
\sum_{(i,j)\in E'} \Gamma^{(m)}_{ij} - \sum_{(i,j)\in E} \Gamma^{(m)}_{ij}  > 0.
\end{align}
The terms for $(i,j)\in E \cap E'$ cancel directly between the two sums. For $(i,j)\in E \setminus E'$, the graph $(V, E\setminus \{(i,j)\})$ is disconnected with connected components, say, $V_1,V_2 \subset V$. Since $T'$ is connected, there must be a $h\in V_1$ and $l \in V_2$ such that $(h,l)\in E'$. Since the path $\ph(hl; T)$ must contain the edge $(i,j)$ and
\begin{align}\label{gamma_decomp0}
\Gamma^{(m)}_{hl} = \sum_{e \in \ph(hl; T)} \Gamma^{(m)}_{e},
\end{align}
this means that the first sum in \eqref{mst_diff0} contains $\Gamma^{(m)}_{ij}$ as part of $\Gamma^{(m)}_{hl}$, which cancels the corresponding term in the second sum.

There are the same number of edges in $E \setminus E'$ as in $E' \setminus E$ and every $\Gamma^{(m)}_{hl}$ for $(h,l) \in E \setminus E'$ is the sum of several terms in the decomposition \ref{gamma_decomp0}. Therefore, the difference on the left-hand side of \eqref{mst_diff0} is indeed strictly positive as long as none of the distances vanishes. \hfill $\Box$

\subsection{Proof of Corollary~\ref{cor2}}
For the true edge set $E$ we observe
\begin{multline*}
\sum_{(i,j)\in E} \dist_{ij} 
 =  \sum_{(i,j)\in E}  \sum_{m=1}^d w_m\Gamma^{(m)}_{ij}
<  \sum_{m=1}^d w_m \min_{T \neq T' = (V,E')} \sum_{(i,j)\in E'}  \Gamma^{(m)}_{ij} 
\\
 \leq \min_{T \neq T' = (V,E')}  \sum_{(i,j)\in E'} \sum_{m=1}^d w_m  \Gamma^{(m)}_{ij} 
= \min_{T \neq T' = (V,E')}  \sum_{(i,j)\in E'} \dist_{ij},
\end{multline*}
where the first inequality follows from the uniqueness of the minimum spanning tree with weights $\Gamma^{(m)}_{ij}$, $m\in V$.
It follows that $T = (V,E)$ must be the minimum spanning tree corresponding to the weights $\dist_{ij} =  \sum_{m=1}^d w_m\Gamma^{(m)}_{ij}$. \hfill $\Box$

\subsection{Proof of Proposition~\ref{prop:chimst}}

{We begin by proving~\eqref{eq:orderchi2}. To this end}, note that we can write the extremal correlation $\chi_{hl}$ in the extremal tree model $\g Y$ as
\begin{align*}
\chi_{hl} = \mathbb P(Y_l > 1 \mid Y_h > 1 ) = \mathbb P(Y^h_l > 1).
\end{align*}
From \eqref{tree_rep} we have that
$$Y^h_l =  P  \prod_{e\in \ph(hl; T^m)} W_e,$$
and therefore, by independence between $P$ and $\{W_e: e\in \ph(hl; T^m)\}$ and since $P$ follows a standard Pareto distribution,
\begin{align*}
\chi_{hl}
= \int_{1}^\infty u^{-2} \mathbb P\Big(u  > \prod_{e\in \ph(hl; T^m)} 1/W_e\Big) \mathrm d u = \E\Big[ \min\Big(  \prod_{e\in \ph(hl; T^m)} W_e, 1\Big) \Big],
\end{align*}
by changing the order of integration. Observe that for any two positive, independent random variables $A$ and $B$ with $\E A, \E B \leq 1$, we have from Jensen's inequality by concavity of $x  \mapsto \min(x,1)$
\begin{align}\label{AB_eq}
\E\left[ \min\left( AB ,1 \right) \right] &=  \E \left\{  \E\left[\min\left( AB ,1 \right) \mid A \right]\right\}
 \leq  \E \left\{ \min\left[ A \E(B\mid A) ,1 \right] \right\}
 =  \E \left[ \min\left(A ,1 \right) \right].
\end{align}
{Recall that we have $\E W_e \leq 1$ for all $e\in E$}. Since $(i,j) \in \ph(hl; T^m)$ we can apply the above successively to obtain
\begin{align*}
\chi_{hl} &= \E\Big[ \min\Big(  \prod_{e\in \ph(hl; T^m)} W_e, 1\Big) \Big] 
\leq
\E\left[ \min\left(W_{(i,j)}, 1\right) \right] = \chi_{ij}.
\end{align*}
Thus~\eqref{eq:orderchi2} follows.

We {show that the minimal spanning tree is unique provided that the inequality in~\eqref{eq:orderchi2} is strict.} We have to show that for any tree $T' = (V, E')$ that differs from $T$ in at least one edge, it holds
\begin{align}\label{mst_diff}
\sum_{(i,j)\in E'} \dist_{ij} - \sum_{(i,j)\in E} \dist_{ij}  > 0,
\end{align}
where we let $\dist_{ij} = -\log (\chi_{ij}) > 0$.

We will now compare the summands in the two sums in \eqref{mst_diff} in a pairwise fashion. To this end, we will construct a  bijective mapping $\tau: E \to E'$ such that for any $(i,j) \in E$, the corresponding edge $(h,l) = \tau\{ (i,j)\} \in E'$ satisfies $(i,j) \in \ph(hl; T)$.

Consider the undirected graph $G = (E+E',\mathcal{E})$ where $(i,j) \in E$ is connected to $(h,l)\in E'$ if and only if $(i,j) \in \ph(hl; T)$. In this formulation, our goal is to find an $E$-saturating matching, that is, a matching such that every element of $E$ is assigned one element in $E'$. A graphical illustration of this idea is provided in Figure~\ref{matching}.

By Hall's marriage theorem \citep{hal1935}, such a matching exists provided that for any subset $C \subset E$, the corresponding neighborhood $n(C)\subset E'$ of elements in $E'$ that are connected to at least one of the elements in $C$ satisfies
\begin{align}\label{marriage}
|C| \leq |n(C)|.
\end{align}
Let $e_1,\dots, e_p$ be the edges in $C$, where $p=|C|$. Removing these edges from the tree $T = (V,E)$ results in a graph $(V, E\setminus C)$ with $p+1$ connected components, which we denote by $V_1,\dots, V_{p+1}$. 

Starting with component $V_1$, we know from the connectedness of the tree $T'$ that there must be an edge in $E'$ between at least one of the elements of $V_1$, say $h_1$, to $l_1 \in V_{k_1}$ for some $k_1 \neq 1$. Since $h_1$ and $l_1$ are in different connected components in $(V, E\setminus C)$, the path $\ph(h_1l_1; T)$ must contain one of the edges in $C$, and therefore $e_1' = (h_1,l_1) \in n(C)$. 

Similarly, there must exist an edge $e_2' = (h_2,l_2)$ between an element $h_2 \in V_1 \cup V_{k_1}$ and some $l_2 \in V_{k_2}$, $k_2 \notin \{1,k_1\}$. This edge is necessarily different from $e_1'$ as it has a node in $V_{k_2}$, and the path $\ph(h_2l_2;T)$ must contain one of the edges in $C$ because $h_2,l_2$ are in different connected components of $(V, E\setminus C)$. Thus $e_2' \in n(C)$. 

Continuing this argument inductively we obtain $p$ different edges in $n(C)$ and therefore the condition \eqref{marriage} holds.

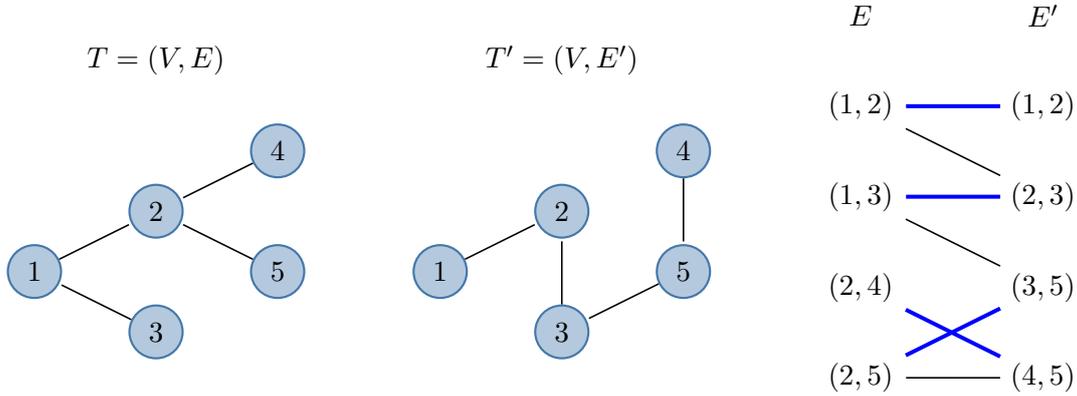
\begin{figure}[ht]
	\centering
	\begin{tikzpicture}
	[scale=.8,auto=left]
	\node (n0) at (2,3.5) {$T = (V,E)$};
	\node (n10) at (2,-2) {};
	\node[nodeObs] (n1) at (0,0) {$1$};
	\node[nodeObs] (n2) at (2,1)  {$2$};
	\node[nodeObs] (n5) at (4,0)  {$5$};
	\node[nodeObs] (n4) at (4,2)  {$4$};
	\node[nodeObs] (n3) at (2,-1)  {$3$};
	\draw[edgeObs] (n2) -- node[above] {} (n1);
	\draw[edgeObs] (n2) -- node[above] {} (n4);
	\draw[edgeObs] (n2) -- node[below] {} (n5);
	\draw[edgeObs] (n1) -- node[below] {} (n3);
	\end{tikzpicture}
	\hspace*{3em}
	\begin{tikzpicture}
	[scale=.8,auto=left]
	\node (n0) at (2,3.5) {$T' = (V,E')$};
	\node (n10) at (2,-2) {};
	\node[nodeObs] (n1) at (0,0) {$1$};
	\node[nodeObs] (n2) at (2,1)  {$2$};
	\node[nodeObs] (n5) at (4,0)  {$5$};
	\node[nodeObs] (n4) at (4,2)  {$4$};
	\node[nodeObs] (n3) at (2,-1)  {$3$};
	\draw[edgeObs] (n2) -- node[above] {} (n1);
	\draw[edgeObs] (n2) -- node[above] {} (n3);
	\draw[edgeObs] (n3) -- node[below] {} (n5);
	\draw[edgeObs] (n5) -- node[below] {} (n4);
	\end{tikzpicture}
	\hspace*{3em}
	\begin{tikzpicture}
	[scale=1.2,auto=left]
	\node (x1) at (0,1) {$E$};
	\node (x2) at (2,1) {$E'$};
	\node (x1) at (0,0) {$(1,2)$};
	\node (x2) at (0,-1) {$(1,3)$};
	\node (x3) at (0,-2) {$(2,4)$};
	\node (x4) at (0,-3) {$(2,5)$};
	\node (y1) at (2,0) {$(1,2)$};
	\node (y2) at (2,-1) {$(2,3)$};
	\node (y3) at (2,-2) {$(3,5)$};
	\node (y4) at (2,-3) {$(4,5)$};
	\draw[edgeObs, blue, line width=1.5pt] (x1) -- node[above] {} (y1);
	\draw[edgeObs] (x1) -- node[above] {} (y2);
	\draw[edgeObs, blue, line width=1.5pt] (x2) -- node[below] {} (y2);
	\draw[edgeObs] (x2) -- node[below] {} (y3);
	\draw[edgeObs, blue, line width=1.5pt] (x3) -- node[below] {} (y4);
	\draw[edgeObs, blue, line width=1.5pt] (x4) -- node[below] {} (y3);
	\draw[edgeObs] (x4) -- node[below] {} (y4);
	\end{tikzpicture}
	\caption{Left and center: two trees $T$ and $T'$. Right: bipartite graph between elements in $E$ and $E'$. A link from $(i,j)\in E$ to $(h,l)\in E'$ means that $(i,j) \in \ph(hl; T)$. The blue links indicate one possible matching $\tau: E \to E'$ in this case.}
	\label{matching}
\end{figure}

In order to show inequality~\eqref{mst_diff} we rewrite the left-hand side as
\begin{align}\label{mst_diff2}
\sum_{(i,j)\in E} \left(\dist_{\tau\{(i,j)\}} -  \dist_{ij} \right).
\end{align}
By construction of $\tau$, for $(h,l) = \tau\{(i,j)\}$, the path $\ph(hl; T)$ must contain the edge $(i,j)$ and thus by~\eqref{eq:orderchi2}
\begin{align}\label{gamma_decomp}
\dist_{hl} \geq \max_{e \in \ph(hl; T)} \dist_{e} \geq \dist_{ij}.
\end{align}
This means that all summands in \eqref{mst_diff2} are non-negative. {Recall that we assume in the second part of Proposition~\ref{prop:chimst} that the inequalities~\eqref{eq:orderchi2} are strict for $(i,j)\neq (h,l)$.}
Since there is at least one $(h,l) \in E'\setminus E$, the first inequality in~\eqref{gamma_decomp}
is strict for this edge and therefore, the difference on the left-hand side of \eqref{mst_diff} is indeed strictly positive. Thus the proof is complete. \hfill $\Box$

\subsection{A sufficient condition}\label{app:sufficient}

{
\begin{lemma}\label{lemma_sufficient}
  Let $\g Y$ be a multivariate Pareto distribution factorizing on the tree~$T = (V,E)$, such that
  that all extremal functions $W^i_j$ for $(i,j) \in E$ have support equal to the whole space $[0,\infty)$. Then for any $h,l \in V$ with $h\neq l$ we have
\begin{equation}\label{eq:orderchi3}
  \chi_{hl} < \chi_{ij},
\end{equation}
for all $(i,j) \in \ph(hl; T)$ such that $(i,j) \neq (h,l)$.
\end{lemma}
\textbf{Proof:}
  Let $h,l \in V$ with $h\neq l$ and $(i,j) \in \ph(hl; T)$ such that $(i,j) \neq (h,l)$. Following the proof of Proposition~\ref{prop:chimst}, it suffices to show that for any two non-negative random variables $A$ and $B$ with support equal to the whole space $[0,\infty)$ and $\E(A), \E(B) \leq 1$, we have that~\eqref{AB_eq} holds with strict inequality.

  Let $\mu_A$ and $\mu_B$ denote the probability measures corresponding to $A$ and $B$, respectively. By independence of $A$ and $B$ we can write
  \begin{align}\label{AB_eq2}
    \E\left[ \min\left( AB ,1 \right) \right] &=  \int_0^\infty  \E\left[\min\left( x B ,1 \right) \right] \mu_A(\mathrm dx).
  \end{align}
{Since $\E[B] \leq 1$ by assumption, we have for $x \in (0,1)$ that $\min(x\E[B],1) = x \E[B]$. We note further that for such $x\in (0,1)$
\[
\E[\min(xB,1)] = \int_{[0,1/x]} xb \mu_B(db) + \int_{(1/x,\infty)} 1 \mu_B(db) < \int_{[0,\infty)} xb \mu_B(db) = \E[xB] = \min(x\E[B],1).
\]
Here the inequality is strict because $\mu_B$ has full support and $xb > 1$ on $b > 1/x$. Since we always have $\E\left[\min\left( x B ,1 \right) \right] \leq \min(x\E[B],1)$ by Jensen's inequality and since $\mu_A$ has full support it follows that
\[
\E\left[ \min\left( AB ,1 \right) \right] = \int_0^\infty  \E\left[\min\left( x B ,1 \right) \right] \mu_A(\mathrm dx) < \int_0^\infty  \min\left( x \E\left[B\right] ,1 \right)  \mu_A(\mathrm dx) \leq \E[\min(A,1)]
\] 
as there is strict inequality between the integrands on $(0,1)$ and inequality otherwise. Following the lines after~\eqref{AB_eq} yields the result.}

\hfill $\Box$

}

\appendix

\section*{Supplementary Material}

\setcounter{subsection}{0}
\setcounter{equation}{0}
\renewcommand{\thesection}{S.\arabic{section}}
\renewcommand{\thesubsection}{S.\arabic{section}.\arabic{subsection}}
\renewcommand{\theequation}{S.\arabic{equation}}

\renewcommand\theHsubsection{S.\thesubsection}
\renewcommand\theHequation{S.\theequation}

\section{Illustration of graph notation}
\label{sm:graph}

{
  Figure~\ref{fig_graph_illus} shows an example of an undirected graph on the index set $V = \{1,\dots, 5\}$. In this example, there are several graph separations that can be stated. For instance, the set $B = \{2,3\}$ separates the set $A = \{1\}$ from $C = \{4,5\}$. Similarly, the set $B = \{2\}$ separates the set $A = \{1\}$ form the set $C = \{4\}$. Such separations are linked to probabilistic statements about a random vector $(Y_i)_{i\in V}$ through Markov properties, as for instance in \eqref{egm}.}
\begin{figure}[ht]
	\centering
	\begin{tikzpicture}
	[scale=.8,auto=left]
	\node (n10) at (2,-2) {};
	\node[nodeObs] (n1) at (0,0) {$1$};
	\node[nodeObs] (n2) at (2,1)  {$2$};
	\node[nodeObs] (n5) at (4,0)  {$5$};
	\node[nodeObs] (n4) at (4,2)  {$4$};
	\node[nodeObs] (n3) at (2,-1)  {$3$};
	\draw[edgeObs] (n2) -- node[above] {} (n1);
	\draw[edgeObs] (n2) -- node[above] {} (n4);
	\draw[edgeObs] (n2) -- node[below] {} (n5);
	\draw[edgeObs] (n1) -- node[below] {} (n3);
	\draw[edgeObs] (n2) -- node[below] {} (n3);
	\draw[edgeObs] (n5) -- node[below] {} (n3);
      \end{tikzpicture}
\caption{Undirected graph $G = (V,E)$ on the index set $V = \{1,\dots, 5\}$ with 6 undirected edges.}
	\label{fig_graph_illus}
    \end{figure}
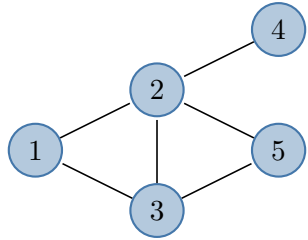

\section{Additional simulation results}
\label{sm:simu}
Figure~\ref{fig_HR_err} shows the wrong edge rate~\eqref{err} for the simulation study in Figure~\ref{fig_HR_srr}.

  \begin{figure}[h]
	\centering
	\includegraphics[height=.38\textwidth]{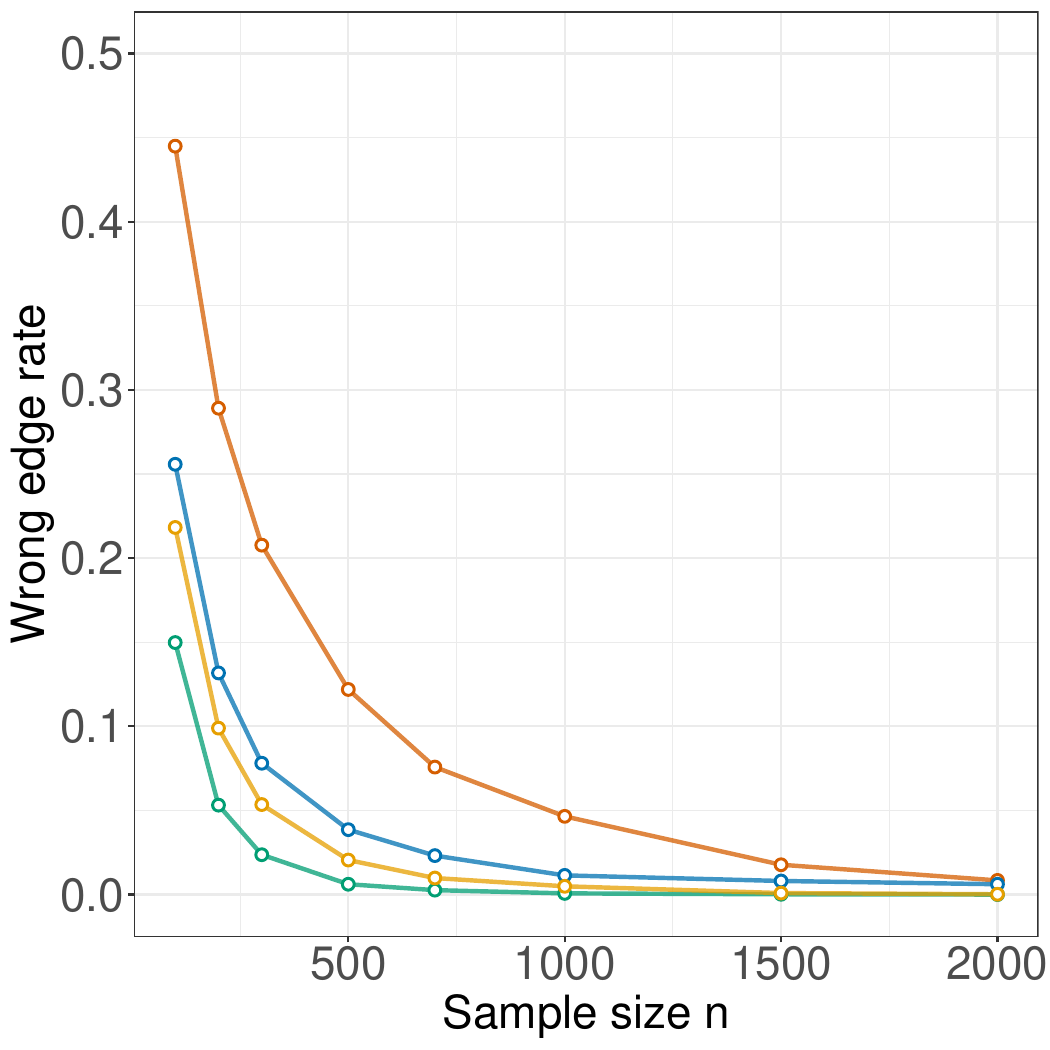}%
	\hspace{.6em}\includegraphics[height=.38\textwidth]{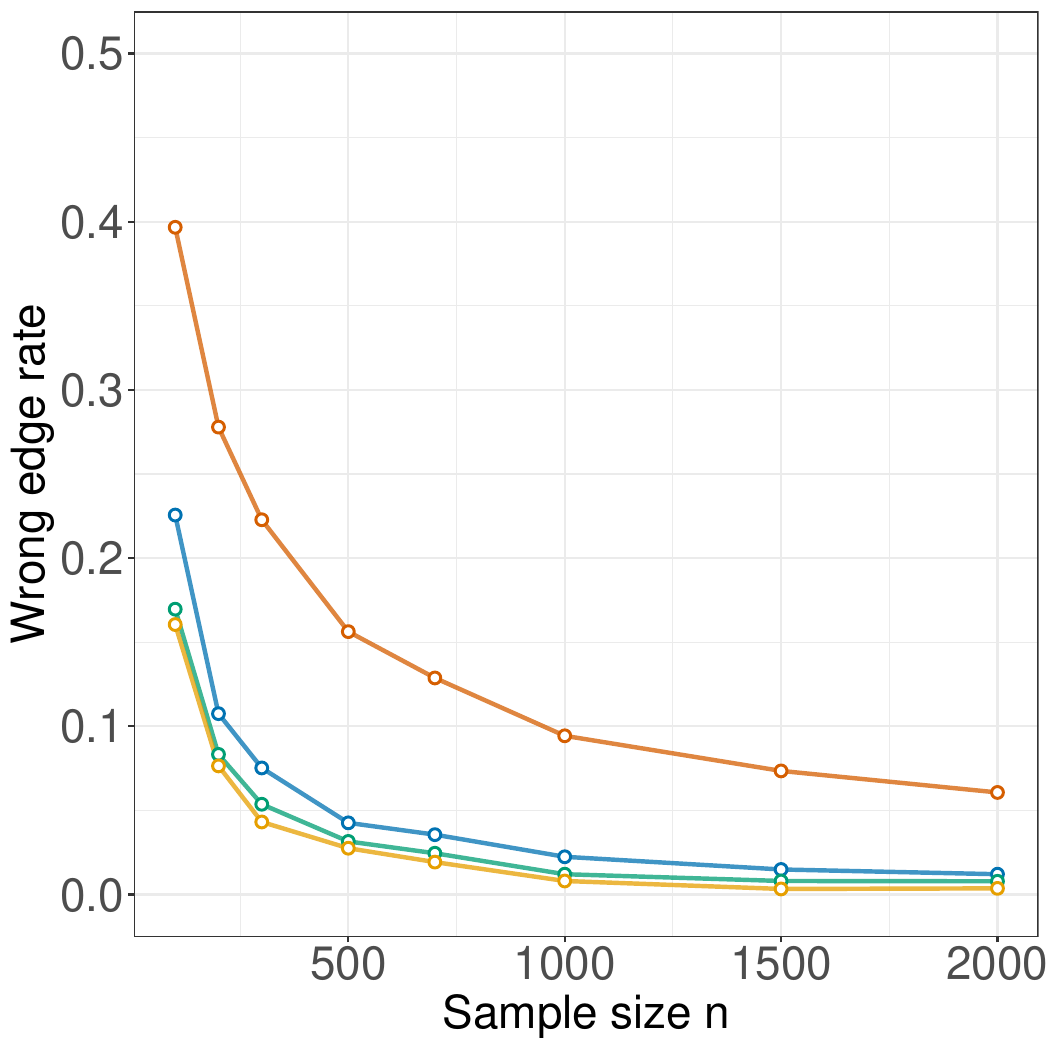}%

	\includegraphics[height=.38\textwidth]{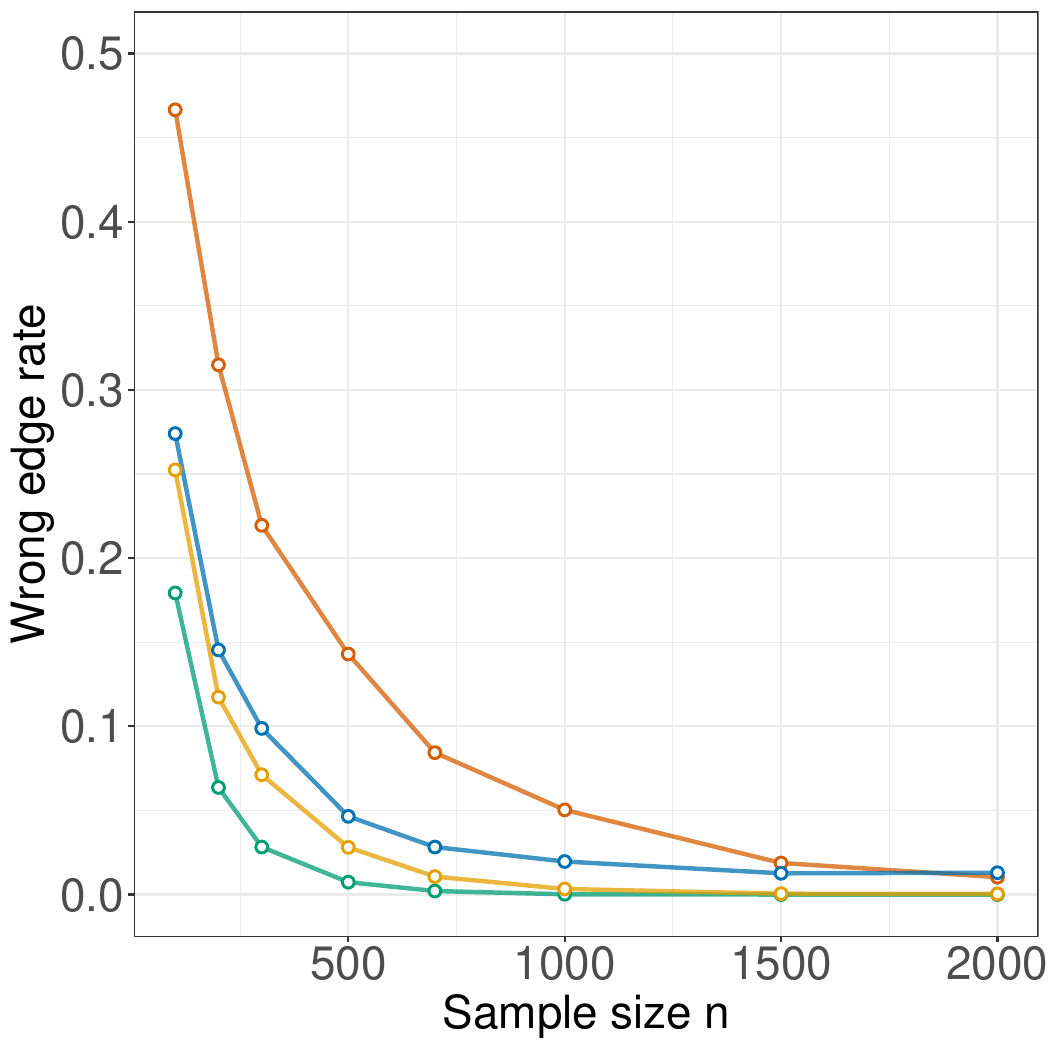}%
	\hspace{.6em}\includegraphics[height=.38\textwidth]{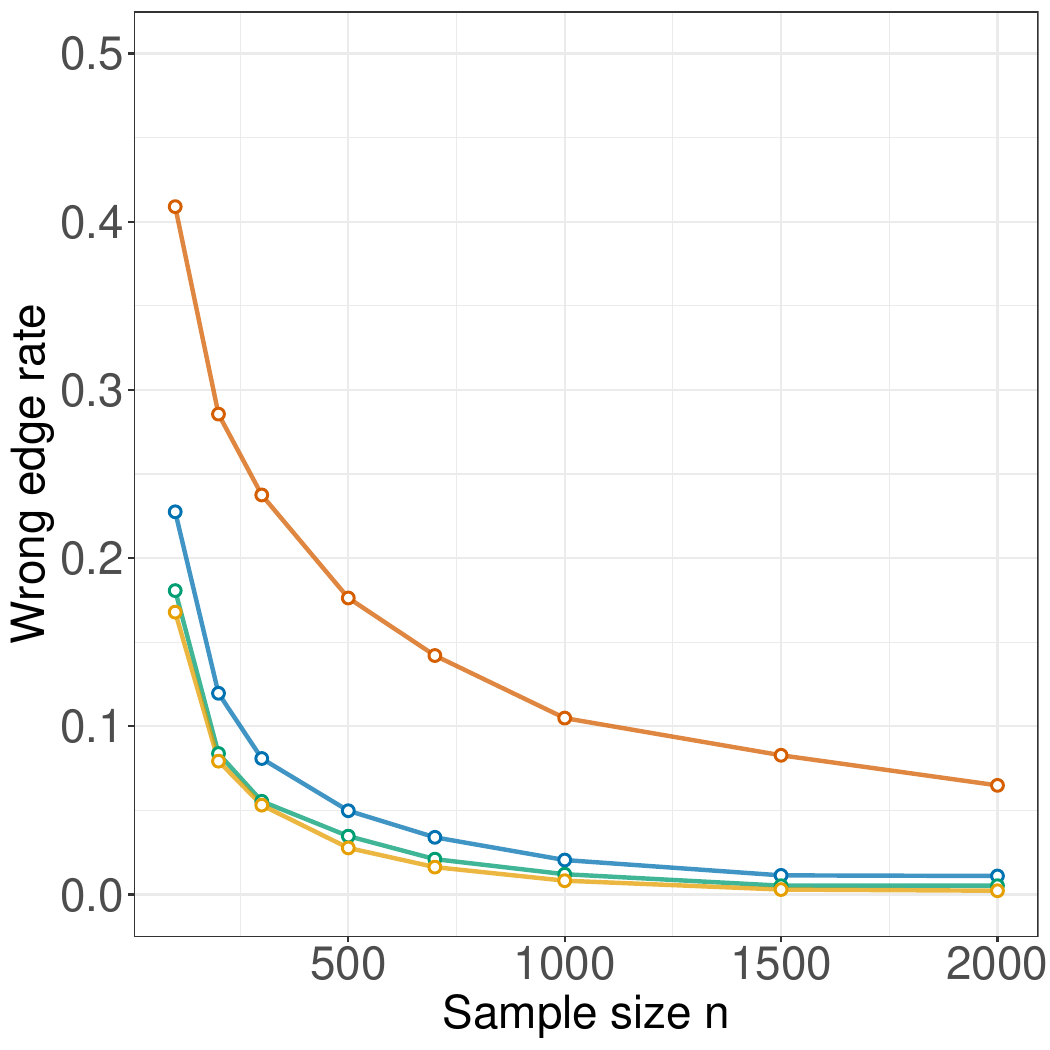}%
\caption{Wrong edge rate of trees from the H\"usler--Reiss model (M1) (top) and Dirichlet model (M2) (bottom) in dimension $d=20$ estimated by the different methods based on empirical correlation (orange), extremal variogram with fixed $m\in V$ (blue), combined empirical variogram (green) and censored maximum likelihood (yellow); independent noise model (N1) (left) and tree noise model (N2) (right).}
	\label{fig_HR_err}
\end{figure}

\newpage

\section{Additional discussion regarding the choice of $k$}
\label{sec:choicek}

We begin with a brief literature review on data-driven choices of $k$ for estimating parameters of extreme value distributions in the univariate case. Key approaches include minimizing asymptotic expansions of the mean squared error by plug-in procedures \citep{hall1985}, looking at stability of estimated parameters across $k$ \citep{drees2000}, and minimum distance procedures \citep{drees2020}; see also \cite{scarrott2012} for an overview of early results. While some of those procedures could be applied for obtaining a good bias-variance trade-off for estimating individual extremal coefficients or values of $\Gamma_{ij}^m$ for fixed $i,j,m\in V$, it is not clear how those choices could be aggregated to obtain one value for tree estimation. In addition, the discussion after Remark~\ref{rem:kn} and our simulations indicate that a good choice of $k$ for tree estimation can differ from $k$ that lead to a good bias-variance trade-off in estimating univariate parameters.

We next provide additional motivation for using the statistic defined in~\eqref{chi_err}.
Let $\Delta(k/n)$ be defined similarly by using the population $\Gamma$ and true underlying tree $T$ in~\eqref{chi_err}. 
The monotone transformation $g$ is applied in order to prevent large values of $\hat \Gamma_{ij}(k/n)$ from dominating the sum and leads to more stable selections in practice. We emphasize that, while the specific form of $g$ is motivated by the relation between $\Gamma$ and $\chi$ in H\"usler--Reiss models, no parametric assumptions are required for the motivation below. 	
Note that on population level, the completion $\Gamma^T_{ij}(0)$ on the true tree $T$ coincides with $\Gamma_{ij}(0)$ by the tree metric property so that $\Delta(0) = 0$. If $k$ is too large, we expect that there is pre-asymptotic bias in the $\hat \Gamma, \hat \Gamma^{\hat T}$ and $\hat T$ estimates, and thus the tree metric property should not hold, resulting in $\hat \Delta(k/n) > 0$ and even $\Delta(k/n) > 0$ on population level. On the other hand, if $k$ is too small, then there will be a high variance in the estimates, also increasing the squared error $\hat \Delta(k/n)$ in~\eqref{chi_err}.

\section{Plots of $\hat \chi$ the application in Section~\ref{sec:application}}
\label{sec:chiplots}

Figure~\ref{chiplots} shows plots of the estimated $\hat \chi_{ij}(q)$ coefficient for different pairs of exchange rates.

  \begin{figure}[h]
    \centering
  \includegraphics[height=.38\textwidth]{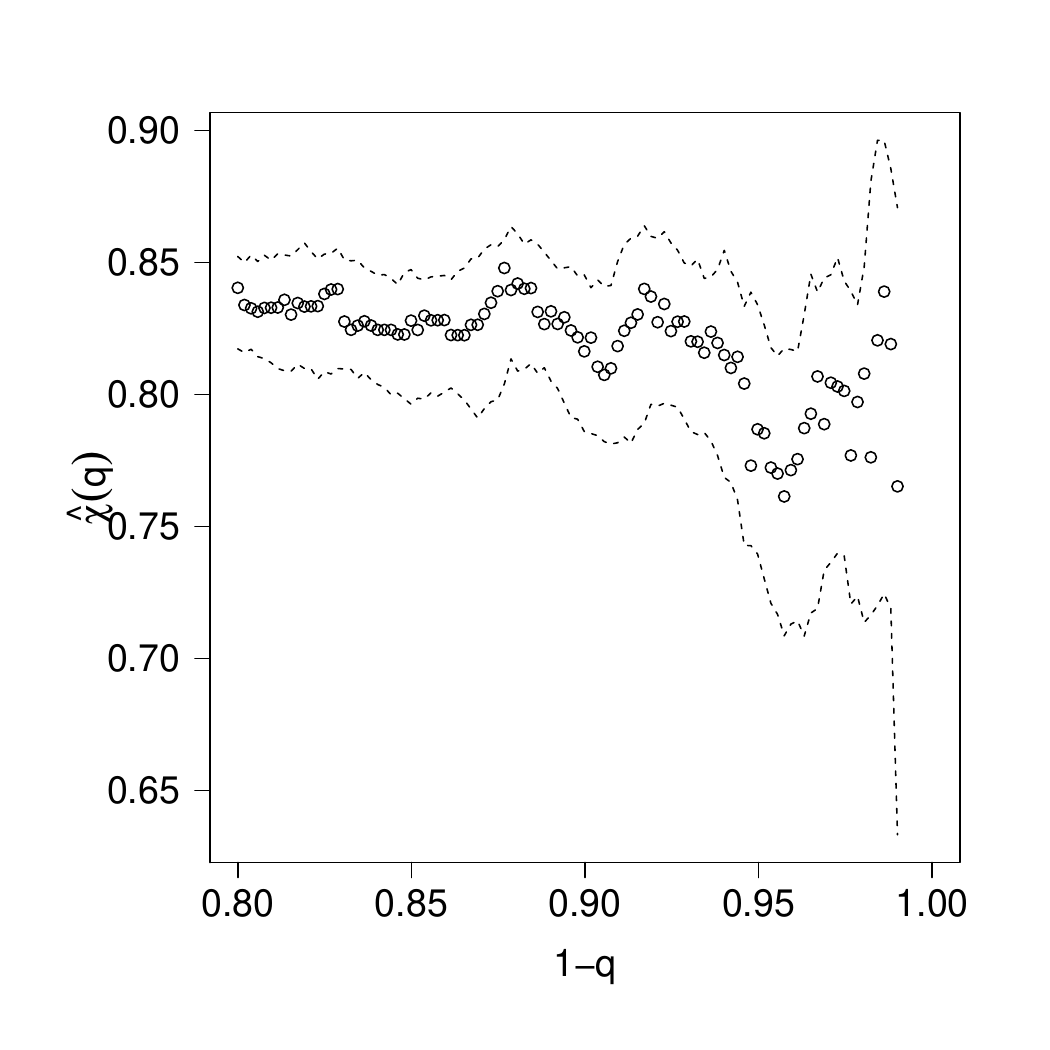}%
  \includegraphics[height=.38\textwidth]{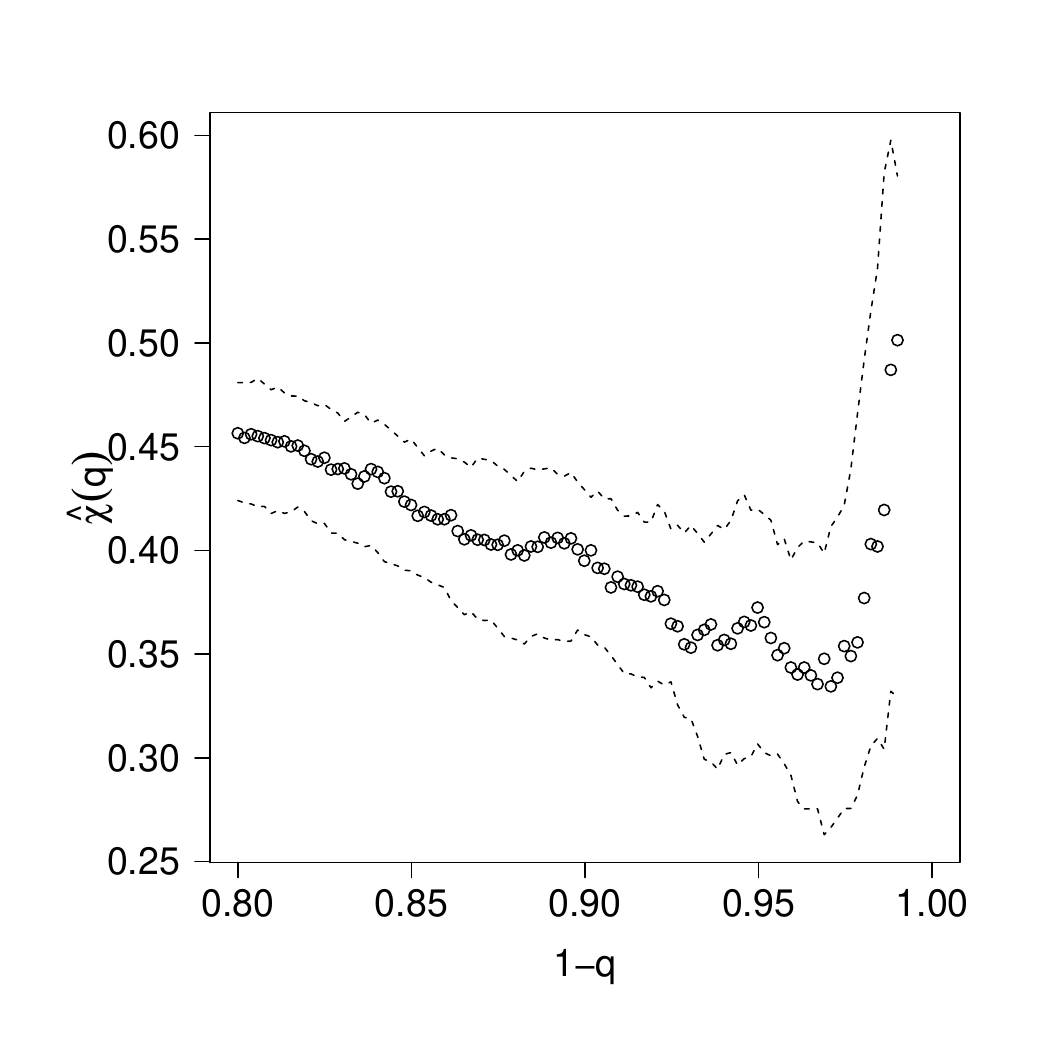}\\
  \includegraphics[height=.38\textwidth]{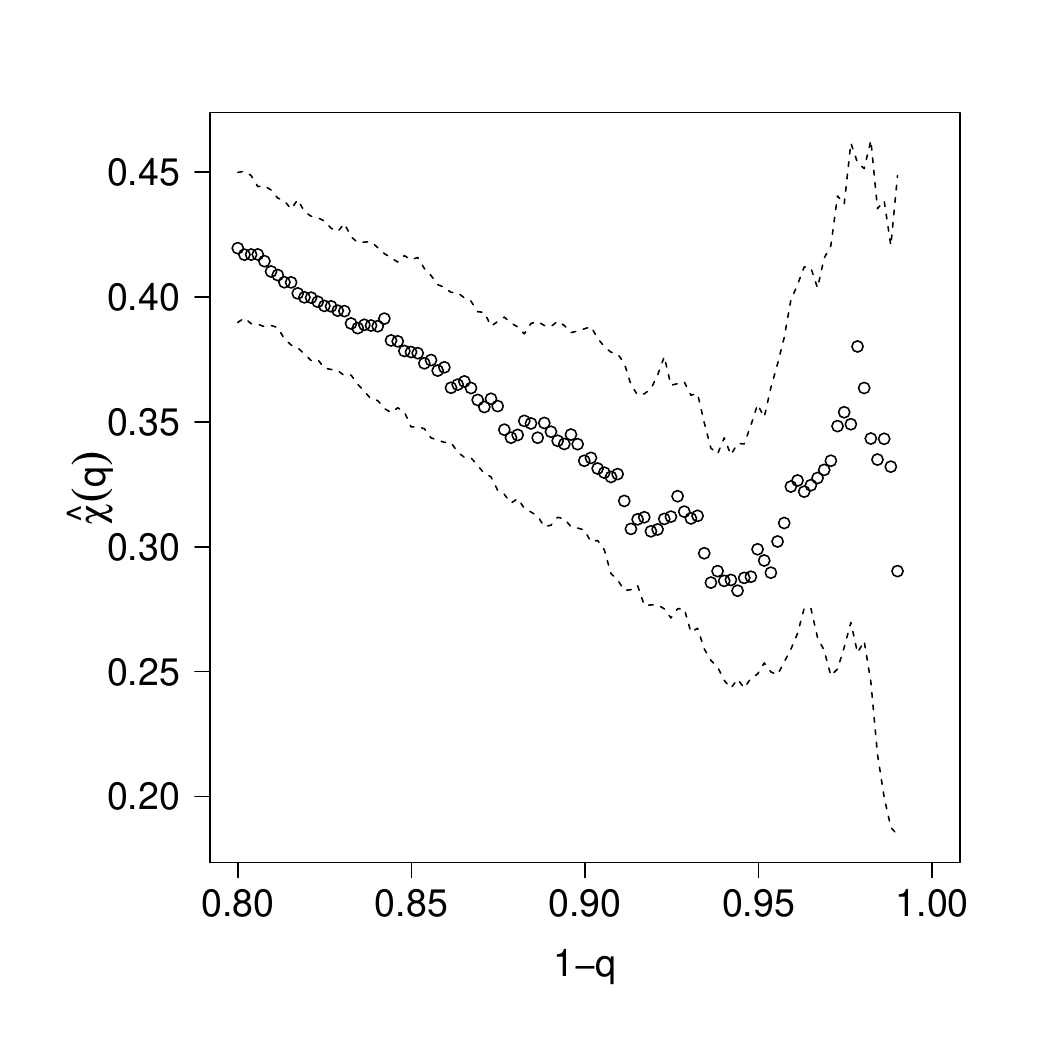}%
  \includegraphics[height=.38\textwidth]{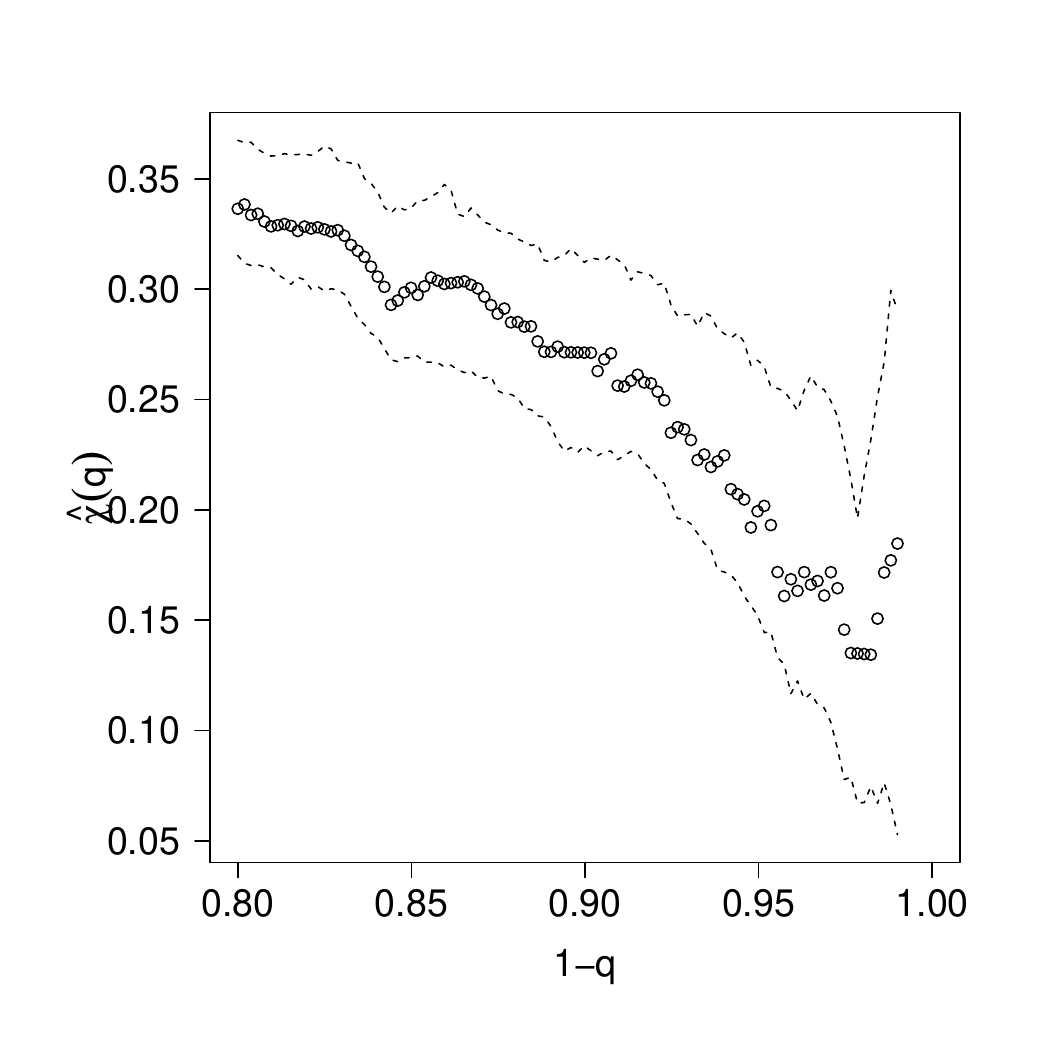}%
  \caption{Plots of the function $q \mapsto \hat \chi_{ij}(q)$ for values $q=k/n$ between $0.8$ and $1$ for four different pairs of exchange rates (all in terms of British Pound sterling); from left to right and top to bottom: CHN/USA, EUR/SGP, CAN/HKG, POL/RUS.}
  \label{chiplots} 
\end{figure}

\clearpage

\section{Technical details on multivariate Pareto distributions}

\label{sec:prop1}

\medskip

\begin{proposition}
	\label{prop_equiv}
	Let $\g V$ be a random vector on $[0,\infty)^d$ with eventually continuous marginal distributions $F_i$. Let $\g Z$ be a random vector supported on $\mathcal L$. The following are equivalent.
	\begin{itemize}
		\item[(i)] The random vectors $\g V$ and $\g Z$ satisfy 
		\begin{equation}\label{eq:lvz}
		\mathbb P(\g Z \le \g x) = \lim_{q \to 0} \mathbb P(F(\g V) \le 1 - q/\g x \mid F(\g V) \not \le 1 - q),
		\end{equation}   
		for all $\g x \in \mathcal L$ that are continuity points of the distribution function of $\g Z$, that is, $\g Z$ is a multivariate Pareto distribution and $\g V$ is in its domain of attraction.
		\item[(ii)] All equivalent conditions of Theorem 2 of~\cite{seg2019} hold for the random vector $\g X = 1/(1-F(\g V))$, index set $I = \{1,\dots, d\}$, regular varying function $b(t) = t$ with index $\alpha = 1$ and constants $c_i = 1$, $i\in I$. The limits in Theorem 2(c) of~\cite{seg2019} have the same laws as $\g Z \mid \g Z_i > 1$, and the limiting measure $\nu$ in Theorem 2(e) of~\cite{seg2019} satisfies $\mathbb P(\g Z \in A) = \nu(A)/\nu(\mathcal L)$ for all Borel sets $A \subset \mathcal L$. 
	\end{itemize}
	Moreover, any $\g Z$ appearing above is homogeneous as in~\eqref{MPD} and satisfies $\mathbb P(Z_1 > 1) = \dots = P(Z_d > 1)$. Conversely, if $\g V$ is homogeneous and satisfies $\mathbb P(V_1 > 1) = \dots = P(V_d > 1)$ then~\eqref{eq:lvz} holds with $\g Z = \g V$. In particular, any such $\g V$ is a multivariate Pareto distribution.  	
\end{proposition}

\textbf{Proof:}

We first show that (i) is equivalent to (ii) and later prove that (i) and (ii) imply the other statements at the end of the theorem.

We begin with preliminaries. Since by the assumption that $F_i$ are eventually continuous we have for sufficiently large $t$ that $1 - \mathbb P(1/(1-F_i(V_i))\leq t) = t$, we conclude that the assumptions of Theorem 2 in \cite{seg2019} are satisfied with $b(t) = t$, $\alpha=1$ and $c_i =1, i = 1\dots,d$. Hence it suffices to show that (i) implies Theorem 2(c) in~\cite{seg2019} and that Theorem 2(e) in~\cite{seg2019} implies (ii). 

The convergence in~\eqref{eq:lvz} is equivalent to
\begin{align}\label{mpd_limit-alt}
\mathbb P(\g Z \le \g x) = \lim_{q \to 0} \mathbb P\Big( q \frac{1}{1-F(\g V)} \le \g x \mid F(\g V) \not \le 1 - q\Big) 
\end{align}
for all $\g x \in \mathcal L$ that are continuity points of the distribution function of $\g Z$. Indeed, for $x_i > 0$ we have $q / (1-F_i(V_i)) \le x_i$ iff $F_i(V_i) \le 1-q/x_i$ by simple algebra while in the case $x_i = 0$ both conditions cannot hold. Note that~\eqref{mpd_limit-alt} is equivalent to stating that the probability measures
\[
\mu_q(\cdot) := \mathbb P\Big( q \frac{1}{1-F(\g V)} \in \cdot \mid F(\g V) \not \le 1 - q\Big) 
\]
defined on Borel subsets of $[0,\infty)^d$ converge in distribution as $q \to 0$ to the probability measure of $\g Z$, that is,
\begin{equation}\label{eq:weakmu}
\mu_q \weak \g Z,  \quad \text{as } q \to 0.
\end{equation} 
In what follows, we always interpret the boundary of sets $A \subset [0,\infty)^d$ in the sense of relative topology with respect to $[0,\infty)^d$.

\medskip

\noindent
\textbf{(i) implies Theorem 2(c) in \cite{seg2019}}

First we will prove that points $\g x$ with $\min_{i} x_i > 0$ are always continuity points of $\g Z$. It suffices to show that $\mathbb P(\exists i\in V: Z_i = x_i) = 0$.
Consider the sets $B_{i,y}^\eps := \{\g x \in [0,\infty)^d: x_i \in (y-\eps,y+\eps)\}$ and $B_{i,y} := \{\g x \in [0,\infty)^d: x_i = y\}$ with $y > 0, \eps < y/2$. By continuity of the marginal distribution function of $F_i$ in the tails, $1 / (1-F_i(V_i))$ has exact Pareto tails and thus for $q$ sufficiently small
\begin{align*}
\mu_q(B_{i,y}^\eps) &= \frac{\mathbb P\Big( \frac{1}{1-F(\g V)} \in q^{-1}B_{i,y}^\eps \cap \mathcal L\Big)}{\mathbb P(F(\g V)\not \le 1-q)} \leq \frac{\mathbb P\Big( \frac{1}{1-F(\g V)} \in q^{-1}B_{i,y}^\eps \Big)}{q} = \frac{\mathbb P\Big( \frac{1}{1-F_i(V_i)} \in q^{-1}(y-\eps,y+\eps)\Big)}{q}
\\
& = \frac{1}{q}  \Big(\frac{q}{y-\eps} - \frac{q}{y+\eps}\Big) = \Big(\frac{1}{y-\eps} - \frac{1}{y+\eps}\Big). 
\end{align*} 
The right-hand side does not depend on $q$ and converges to zero for any fixed $y>0$ as $\eps$ tends to zero.
Since the sets $B_{i,y}^\eps$ are open in $[0,\infty)^d$ it follows that for any fixed and sufficiently small $\eps >0$
\[
\mathbb P(\g Z \in B_{i,y}) \leq \mathbb P(\g Z \in B_{i,y}^\eps) \leq \liminf_{q\to 0} \mu_q(B_{i,y}^\eps).
\] 
Thus, taking $\eps \to 0$ on the right-hand side, we obtain 
\begin{align}\label{eq:nomass}
\mathbb P(\g Z \in B_{i,y}) = 0 \quad \forall i, y > 0.
\end{align} 
Finally, observe that
\[
\mathbb P(\exists i\in V: Z_i = x_i) \leq \mathbb P\Big(\g Z \in \cup_{i=1}^d B_{i,x_i}\Big) = 0,	
\]
for $\g x$ with $\min_i x_i > 0$. This completes the proof that such $\g x$ are continuity points of the distribution of $\g Z$.

\medskip

Now we prove that~\eqref{eq:weakmu} implies that the statement in Theorem 2(c) of \cite{seg2019} holds for the vectors $(1/(1-F(\g V)))$ and limits $\g Z \mid Z_i > 1$. To this end we will show that for $i=1,\dots, d$ and all $\g x \in \R^d$ that are continuity points of the distribution of $\g Z \mid Z_i > 1$
\begin{equation} \label{eq:segers}
\lim_{t \to \infty} \mathbb P\Big( \frac{1}{t}\frac{1}{1-F(\g V)} \le \g x \mid \frac{1}{1-F_i(V_i)} > t\Big) = \mathbb{P}(\g Z \le \g x\mid Z_i > 1). 
\end{equation}
Note that this statement is trivial if $x_i \le 1$ or $\min_j x_j < 0$ since in that case both sides are identically zero. Hence it suffices to consider $\g x \in \mathcal L^i := \{ \g x \in \mathcal L: x_i > 1\}$. Recall that all $\g x \in \mathcal{L}$ with $\min_i x_i > 0$ are continuity points of the distribution of $\g Z$. Combined with the fact that convergence on the complement of $\mathcal L^i$ holds trivially, it suffices to prove~\eqref{eq:segers} for $\g x \in \mathcal L^i: \min_i x_i > 0$ which are always continuity points of the distribution of $\g Z$. In what follows, fix such an $\g x$.

For all sufficiently small $q$ we have 
\begin{align*}
\mathbb P\Big( q \frac{1}{1-F(\g V)} \in \mathcal L^i \mid F(\g V) \not \le 1 - q\Big) &= \mathbb P\Big( F_i(V_i) > 1-q \mid F(\g V) \not \le 1 - q\Big) = \frac{\mathbb P( F_i(V_i) > 1-q)}{\mathbb P( F(\g V) \not \le 1 - q)}
\\
&= \frac{q}{\mathbb P( F(\g V) \not \le 1 - q)}.
\end{align*}
By~\eqref{eq:nomass} we have $\mathbb P(\g Z \in \partial \mathcal L^i) = \mathbb P(\g Z \in B_{i,1}) = 0$ and the limit of the first expression in the chain of equalities above equals $\mathbb P(\g Z \in \mathcal L^i) = P(Z_i > 1)$. This shows that
\begin{align}\label{eq:zi1}
\lim_{q\to 0} \frac{q}{\mathbb P( F(\g V) \not \le 1 - q)} = P(Z_i > 1).
\end{align}
Next set $t = 1/q$ and note that for $\g x$ as discussed earlier
\begin{align*}
\mathbb P\Big( \frac{1}{t}\frac{1}{1-F(\g V)} \le \g x \mid \frac{1}{1-F_i(V_i)} > t\Big) 
& =~ \frac{\mathbb P\Big(q\frac{1}{1-F(\g V)} \in [\g 0, \g x]\cap \mathcal L^i \Big)}{\mathbb P( F(\g V) \not \le 1 - q)} \frac{\mathbb P( F(\g V) \not \le 1 - q)}{q}
\\
& {\stackrel{q \to 0}{\longrightarrow}}~ \frac{\mathbb P(\g Z \in [\g 0, \g x]\cap \mathcal L^i)}{\mathbb P( Z_i > 1)} = \mathbb P(\g Z \leq \g x \mid Z_i > 1).
\end{align*}
Here we used~\eqref{eq:zi1} and the weak convergence of $\mu_q$ to the law of $\g Z$ together with the fact that $\g Z$ puts no mass on the boundary of sets of the form $[\g 0, \g x]\cap \mathcal L^i$ provided that $\g x$ is a continuity point of the law of $\g Z$ which we assumed. This proves~\eqref{eq:segers}. 

\medskip 

\noindent
\textbf{(ii) implies (i)}

Condition (e) of Theorem 2 in \cite{seg2019} and the discussion preceding this result imply that the measures
\[
\nu_t(\cdot) := t \mathbb P\Big(\frac{1}{t} \frac{1}{1-F(\g V)} \in \cdot \Big)
\]
converge to the measure $\nu$, in the sense of $\mathcal{M}_\mathbb{O}$ convergence in~\cite{lindskog2014} with, in the notation of the latter paper, $\mathbb O = [0,\infty)^d\backslash \g 0$, $\mathbb C = \g 0$. Since sets $A \subset \mathcal L$ are bounded away from $\g 0$, Theorem 2.1(iv) in~\cite{lindskog2014} implies that for any Borel set $A \subset \mathcal L$ with $\nu(\partial A) = 0$ we have
\begin{equation}\label{eq:seg2e}
\lim_{t \to \infty } t \mathbb P\Big(\frac{1}{t} \frac{1}{1-F(\g V)} \in A\Big) = \nu(A).
\end{equation} 
Repeating the arguments leading to~\eqref{eq:nomass} we deduce that $\nu(\partial \mathcal L^i) = 0$; this in turn implies that $\nu$ puts no mass on $\partial \mathcal L$. Thus for $A \subset \mathcal L$ with $\nu(\partial A) = 0$
\begin{align*}
\lim_{q \to 0} &\mathbb P\Big( q \frac{1}{1-F(\g V)} \in A \mid F(\g V) \not \le 1 - q\Big) 
\\
= & \lim_{q \to 0} \mathbb P\Big( q \frac{1}{1-F(\g V)} \in A \mid q \frac{1}{1-F(\g V)} \in \mathcal{L} \Big)
\\
= & \lim_{q \to 0} \frac{\mathbb P\Big( q \frac{1}{1-F(\g V)} \in A \Big)}{q} \frac{q}{ \mathbb P \Big( q \frac{1}{1-F(\g V)} \in \mathcal{L} \Big) }
\\
=& \frac{\nu(A)}{\nu(\mathcal L)} = \mathbb P(\g Z \in A).
\end{align*}
In other words the law of 
\[
q \frac{1}{1-F(\g V)} \in A \mid F(\g V) \not \le 1 - q
\] 
converges to the law of $\g Z$ and~\eqref{eq:weakmu} follows.

\medskip

\noindent
\textbf{Now suppose that (i) and (ii) hold}.

From condition (e) in Theorem 2 of~\cite{seg2019} it follows that the
limiting measure $\nu$ is homogeneous. By the discussion following Theorem 2 in~\cite{seg2019} the limiting measure $\nu$ in Theorem 2(e) of the latter reference is homogeneous.
Since $\mathbb P(\g Z \in A) = \nu(A)/\nu(\mathcal L)$, this implies that
$\g Z$ is homogeneous as in~\eqref{MPD}. 
On the other hand, since (i) holds, $\mathbb P(Z_1 > 1) = \dots = P(Z_d > 1)$ follows directly from~\eqref{eq:zi1} in the proof above.

Conversely, suppose that $\g V$ is homogeneous and satisfies $\mathbb P(V_1 > 1) = \dots = P(V_d > 1)$. We show that $\g V$ is its own domain of attraction as in~\eqref{eq:lvz} and is therefore a multivariate Pareto distribution. From homogeneity it follows that for any $i=1,\dots, d$ we have $\mathbb P(V_i \leq  x \mid V_i > 1) = 1- 1 /x$ for $x\geq 1$, that is, the distribution function of $V_i$ is eventually continuous and satisfies $F_i(x) = 1 - c/x$, where $c = \mathbb P(V_i >1)$ does not depend on $i$. For $z\geq 1 -c$ we thus have $F_i^{\leftarrow}(z) = c / (1-z)$.
For $t$ large enough we have $\mathbb P(F_i (V_i) \leq 1 - 1/t) = 1 - 1/t$ and thus for any $\g x \in \mathcal L$
\begin{align*}
\mathbb P(F(\g V) \le 1 - 1/(t\g x) \mid F (\g V) \not \le 1 - 1/t) & = \mathbb P\left( \g V \in A_{\g x,tc} \cap  C_{tc} \right) /  \mathbb P\left( \g V \in C_{tc}\right), 
\end{align*}
where the sets 
$$ A_{\g x,s} = \{\g y \in [0,\infty)^d:  y_j \leq s \g x \text{ for all } j=1,\dots, d\}, \quad C_{s} = \{\g y \in [0,\infty)^d:  \max y_j > s \},$$
satisfy $A_{\g x,tc} \cap  C_{tc} = tc (A_{\g x,1} \cap  C_{1})$ and $C_{tc} = tc C_{1} = tc \mathcal L$.
By homogeneity of $\g V$, we thus have 
\begin{align*}
\mathbb P(F(\g V) \le 1 - 1/(t\g x) \mid F (\g V) \not \le 1 - 1/t)  &= \mathbb P\left( \g V \in A_{\g x,i,1} \cap  C_{1}\right) /  \mathbb P\left( \g V \in \mathcal L\right)\\
&= \mathbb P\left( \g V \leq \g x\right),
\end{align*}
and thus (i) follows. \hfill $\Box$

\section{Proof of Proposition~\ref{prop_tree}}

Without loss of generality, let $m=1$ and suppose that $d$ is a terminal node of the directed tree $T^m$ and node $d-1$ is its only parent; this can always be achieved by renaming nodes. It follows from the global Markov property that
\[  
Y^{d-1}_{d}\ci \g Y^{d-1}_{\setminus\{d-1,d\}} \mid \g Y^{d-1}_{d-1}.
\]  
Recalling the representation $\g Y^{d-1} = P \g W^{d-1}$ from \eqref{extr_fct}, we can rewrite this to
\[ 
P  W^{d-1}_{d}\ci P \g W^{d-1}_{\setminus d} \mid P,
\]
where $P$ is a  standard Pareto random variable, which is independent of $W^{d-1}$ and since $W^{d-1}_{d-1} = 1$ almost surely. This implies that $W^{d-1}_{d}\ci \g W^{d-1}_{\setminus\{d-1,d\}}$ are unconditionally independent. In the sequel, we use an identity that relates the distributions of extremal functions with respect to the different components $m = 1$ and $d-1$. For any continuous, bounded function $h:[1,\infty)\times [0,\infty)^{d-1}\to [0,\infty)$ we have
\begin{align}\label{switch_EF}  
\E\left[ \einsfun\{W^{m}_{d-1} > 0\} h(\g W^m)  \right] =  \E\left[h(\g W^{d-1}/ W^{d-1}_m)  W^{d-1}_m  \right];
\end{align}
see for instance \citet[Cor.~3]{seg2019}, or similar representations in \citet[Prop.~4.2]{dom2013a} and \citet[Prop.~1]{dom2016}.

We first consider the distribution of $\g Y^m$ on the set $\{ Y^m_{d-1} > 0\}$. Observe that for any continuous, bounded function $f:[1,\infty)\times [0,\infty)^{d-1}\to [0,\infty)$ and any $u >0$ we have
\begin{align*}
\E[f(u \g W^{d-1}/W_m^{d-1}) W_m^{d-1}] 
&= \E[\E\{f(u \g W^{d-1}/W_m^{d-1}) W_m^{d-1} \mid W_{d}^{d-1}\}]
= \E[g(u,W_{d}^{d-1})]
\end{align*}
where, noting that $W_{d-1}^{d-1} = 1$ almost surely 
\begin{align*}
g(u,v) &:= \E[f(u \g W^{d-1}_{\setminus d} / W_m^{d-1}, uvW_{d-1}^{d-1}/W_m^{d-1}) W_m^{d-1}]
\\
&= \E[f(u \g W^{m}_{\setminus d}, uvW^{m}_{d-1})\einsfun\{W^{m}_{d-1} > 0\}]
\end{align*}
Here, the second equality in the representation for $g$ follows by~\eqref{switch_EF} applied with 
\[
h(w_1,\dots,w_d) := f(uw_1,\dots,uw_{d-1},uvw_{d-1}).
\] 
Thus we obtain for bounded, continuous functions $f:[1,\infty)\times [0,\infty)^{d-1}\to [0,\infty)$ 
\begin{align}
\notag& \E\left[ \einsfun\{ Y^m_{d-1} > 0\} f(\g Y^m)\right] =   \E\left[ \einsfun\{ W^m_{d-1} > 0\} f(P \g W^m) \right] 
\\
\notag=~~& \int_1^\infty u^{-2}   \E\left[ \einsfun\{ W^m_{d-1} > 0\} f(u \g W^m) \right] \mathrm du  
\\     
\notag \stackrel{by~\eqref{switch_EF}}{=}& \int_1^\infty u^{-2}  \E\left[ f(u \g W^{d-1}/ W^{d-1}_m) W^{d-1}_m \right] \mathrm du 
\\
\notag \stackrel{(a)}{=}~& \int_1^\infty u^{-2} \int_{[0,\infty)}  \E\left[ \einsfun\{ W^m_{d-1} > 0\} f(u \g W^{m}_{\setminus d}, u W^{m}_{d-1} w_d ) \right] \mathbb P \left( W^{d-1}_d = \mathrm d w_d \right) \mathrm du
\\
\label{markov_decomp} =~&  \E\left[ \einsfun\{ Y^m_{d-1} > 0\} f(\g Y^m_{\setminus d}, Y^m_{d-1} \tilde W^{d-1}_{d})\right]      
\end{align}
where $\tilde W^{d-1}_{d}$ is an independent copy of $W^{d-1}_d$, also independent of all the other random variables in the above equation, and equation (a) uses the representation for $\E[f(u \g W^{d-1}/W_m^{d-1}) W_m^{d-1}]$ derived earlier.

If $ \mathbb P(Y^m_{d-1} = 0) > 0$, it remains to consider the term $\E\left[ \einsfun\{ Y^m_{d-1} = 0\} f(\g Y^m)\right]$. From $ Y^m_d \ci \g Y^m_{\setminus \{{d-1},d\}} \mid Y^m_{d-1}$ it follows for any $s \geq 1$ that
\begin{align}\label{eqYim}
\mathbb P\left( Y^m_m > s, Y^m_d > s \mid Y^m_{d-1} = 0 \right) =
\mathbb P\left( Y^m_m > s \mid Y^m_{d-1} = 0 \right)
\mathbb P\left(Y^m_d > s \mid Y^m_{d-1} = 0 \right).
\end{align}
Since $ \mathbb P(Y^m_{d-1} = 0) > 0$, the first factor on the right-hand side is positive as
\[
\mathbb P(Y_m > 1)\mathbb P\left( Y^m_m > s,  Y^m_{d-1} = 0 \right) = \mathbb P\left( Y_m > s, Y_{d-1} = 0 \right) = s^{-1}\mathbb P\left( Y_m > 1, Y_{d-1} = 0 \right) > 0,
\]
where the last equation follows from the homogeneity of $\g Y$. Thus, equation~\eqref{eqYim} can be written as
\begin{align*}
\mathbb P\left(Y^m_d > s \mid Y^m_{d-1} = 0 \right) = 
\frac{ \mathbb P\left( Y_m > s, Y_d > s, Y_{d-1} = 0 \right)}{\mathbb P\left( Y_m > s, Y_{d-1} = 0 \right)}= 
\frac{ \mathbb P\left( Y_m > 1, Y_d > 1, Y_{d-1} = 0 \right)}{\mathbb P\left( Y_m > 1, Y_{d-1} = 0 \right)},
\end{align*}
where we used again the homogeneity of $\g Y$. The right-hand side does therefore not depend on $s$ and thus $\mathbb P\left(Y^m_d > s \mid Y^m_{d-1} = 0 \right) = 0$ for all $s\geq 1$ by taking the limit $s \to \infty$. Using this, we also obtain
\begin{multline*}
\mathbb P\left(Y_d > s^{-1}, Y_{d-1} = 0, Y_m > 1 \right) = s \mathbb P\left(Y_d > 1, Y_{d-1} = 0, Y_m > s \right)
\\ 
\leq {s} \mathbb P\left(Y^m_d > 1,  Y^m_{d-1} = 0 \right) {\mathbb P(Y_m > 1)} = 0,
\end{multline*}
 which implies $\mathbb P\left(Y^m_d = 0 \mid Y^m_{d-1} = 0 \right) = 1$ and thus 
\begin{align*}
\E\left[ \einsfun\{ Y^m_{d-1} = 0\} f(\g Y^m)\right] &= \E\left[ \einsfun \{ Y^m_{d-1} = 0\} f(\g Y^m_{\setminus d}, 0)\right].
\end{align*}   
Combining this with \eqref{markov_decomp} yields
\begin{align*}
\E\left[ f(\g Y^m)\right] &= \E\left[ f(\g Y^m_{\setminus d}, Y^m_{d-1} W^{d-1}_d)\right],
\end{align*}
and by induction we can use the representation~\eqref{tree_rep} for $ \g Y^m_{\setminus d}$
to conclude the first part of the proof.

For the converse statement, let $\g Y^1, \dots, \g Y^d$ be random vectors defined as in~\eqref{tree_rep} for independent random variables $\{W_i^j, W_j^i; \{i,j\} \in E\}$, where $W_i^j$ and $W_j^i$ satisfy the duality~\eqref{biv_change}. We first show that the extremal functions are mutually consistent on the intersections of their domains. For $m,m'\in V$, let $A \subset \mathcal L^m \cap \mathcal L^{m'}$ be a Borel subset, then
\begin{align}\label{Ym_def}
  \mathbb P( \g Y^m \in A) =  \int_1^\infty u^{-2}  \mathbb P \Big\{ u  \Big( \prod_{e\in \ph(mi; T^m)} W_e\Big)_{i\in V} \in A \Big\} \mathrm{d} u,
\end{align}
where the empty product is defined as one, and we explicitly specify with respect to which directed tree the path is taken.
Note that $\g Y \in \mathcal L^m \cap \mathcal L^{m'}$ implies that all $W_e>0$ for $e \in \ph(m m'; T^m)$ and $W_{e'}>0$, where for $e = (i,j)$ the edge $e' = (j,i)$ has reversed orientation. 
From the duality in~\eqref{biv_change} we get for any bounded, measurable function $h:[0,\infty) \to [0,\infty)$
\begin{equation} \label{biv_change2}
\E[h(W_{i}^j) \einsfun\{W_i^j > 0\})] = \E[ h(1/W_j^i) W_j^i].
\end{equation} 
By swapping the order of integration in~\eqref{Ym_def} we get
\begin{align*}
&  \mathbb E \Big[ \int_1^\infty u^{-2}   \einsfun\Big \{ u  \Big( \prod_{e \in E_i} W_e\Big)_{i\in V} \in A \Big\} \mathrm{d} u\Big]
\\
=~&  \mathbb E \Big[ \int_1^\infty u^{-2} \einsfun\Big\{ u \Big(\prod_{e \in E_i \backslash S} W_e \prod_{e \in E_i \cap S} W_e\Big)_{i \in V} \in A\} \mathrm{d} u \Big]
\\
=~& \mathbb E \Big[ \int_1^\infty u^{-2}  \einsfun\Big\{ u \Big(\prod_{e \in E_i \backslash S} W_e \prod_{e \in E_i \cap S} 1/W_{e'} \Big)_{i\in V} \in A \Big\} \prod_{e \in S} W_{e'}   \mathrm{d} u \Big]\\
=~&  \mathbb E \Big[ \int_{1/\prod_{e \in S} W_{e'}}^\infty v^{-2}\einsfun\Big\{ v \prod_{e \in S} W_{e'} \Big(\prod_{e \in E_i \backslash S} W_e \prod_{e \in E_i \cap S} 1/W_{e'} \Big)_{i\in V} \in A \Big\}  \mathrm{d} v \Big]
\\
=~&  \mathbb E \Big[ \int_{1/\prod_{e \in S} W_{e'}}^\infty v^{-2}\einsfun\Big\{ v \Big(\prod_{e \in \ph(m'i;T^{m'})} W_{e}  \Big)_{i\in V} \in A \Big\}  \mathrm{d} v \Big]
\\
=~&   \mathbb E \Big[  \int_{1}^\infty v^{-2}\einsfun\Big\{ v \Big(\prod_{e \in \ph(m'i;T^{m'})} W_{e} \Big)_{i\in V} \in A \Big\}  \mathrm{d} v  \Big],
\end{align*}
where we used the abbreviated notation $E_i := \ph(mi;T^m)$ and $S := \ph(mm';T^m)$. Here we used independence of the $W_e$ in the first equality, the identity~\eqref{biv_change2} in the second (noting that by the assumptions we made on $A$ the $\einsfun\{W_i^j > 0\}$ in that identity can be dropped), the substitution $u = v \prod_{e \in S} W_{e'} $ in the third equality. The fourth equality follows from elementary considerations upon observing that the edges $e \in T^m$ have the same orientation in $T^{m'}$ whenever $e \notin \ph(mm'; T^m)$ and reversed orientation otherwise. 
For the last equality recall that by the assumption $A \subset \mathcal L^m \cap \mathcal L^{m'}$ and by the representation $Y_m^{m'} = P \prod_{e \in \ph(m'm;T^{m'})} W_e$ we have that $v > 1$ and $v > 1/\prod_{e \in S} W_e$ whenever the indicator function is non-zero. This shows that $\mathbb P( \g Y^{m'} \in A) = \mathbb P( \g Y^{m} \in A)$. 

We can now define the random vector $\g Y$ on $\mathcal L$ by
\begin{align}\label{def_Y}
  \mathbb P(\g Y \in A) := c \sum_{i=1}^d \mathbb P(\g Y^{i} \in A \cap B_i), \quad A \subset \mathcal L,
\end{align}
where $c>0$ is an appropriate normalizing constant to make this a probability measure and $B_1,\dots,B_d$ define a disjoint partition of the set $\mathcal L$ and have the additional property $B_i \subset \mathcal L^i, i=1,\dots,d$. Further note that by this definition, we have
$$ \mathbb P(Y_m > 1) =  c \sum_{i=1}^d \mathbb P(\g Y^{i} \in \mathcal L^m \cap B_i) =  c \sum_{i=1}^d \mathbb P(\g Y^{m} \in \mathcal L^m \cap B_i) = c \mathbb P(\g Y^{m} \in \mathcal L^m) = c,$$
where we used for the second equality that $\mathbb P(\g Y^{i} \in \mathcal L^m \cap B_i) = \mathbb P(\g Y^{m} \in \mathcal L^m \cap B_i)$ for all $i\in V$, since $\mathcal L^m \cap B_i\subset \mathcal L^m \cap \mathcal L^{i}$. This shows that $c = \mathbb P(Y_1 > 1) = \dots = \mathbb P(Y_d > 1)$. Moreover, all $\g Y^m$ are homogeneous as in~\eqref{MPD} since for any Borel set $A \subset \mathcal L^m$ and $t\geq 1$
\begin{align*}
  \mathbb P(\g Y^m \in t A) = \mathbb P(P \g W^m \in t A) &= \int_1^\infty  u^{-2} \mathbb P(u \g W^m \in t A) \mathrm{d} u \\
                                                          &= \int_{t}^\infty  u^{-2} \mathbb P(u \g W^m \in t A) \mathrm{d} u \\
                                                          &= t^{-1} \int_{1}^\infty  v^{-2} \mathbb P(v \g W^m \in A) \mathrm{d} v\\
   &= t^{-1}\mathbb P(\g Y^m \in A).
\end{align*}
The third equality follows from the fact that $A \subset \mathcal L^m$ implies that $\mathbb P(u \g W^m \in t A) = 0$ for all $u \in [1,t]$, since $W^m_m = 1$ and $x_m > t$ for all $\g x \in tA$.
Thus, it follows that $\g Y$ is also homogeneous from its definition in~\eqref{def_Y}. By Proposition~\ref{prop_equiv} it is therefore a $d$-dimensional Pareto distribution.

The conditioned random vector $\g Y \mid Y_m > 1$ has the same distribution as $\g Y^m$ since for $A\subset \mathcal L^m$
\begin{align*}
\mathbb P(\g Y \in A \mid Y_m > 1)  
= \frac{\sum_{i=1}^d \mathbb P(\g Y^{i} \in A \cap \mathcal{L}^m \cap B_i)}{\sum_{i=1}^d \mathbb P(\g Y^{i} \in \mathcal L^m \cap B_i)}
= \frac{\sum_{i=1}^d \mathbb P(\g Y^{m} \in A \cap \mathcal{L}^m \cap  B_i)}{\sum_{i=1}^d \mathbb P(\g Y^{m} \in \mathcal L^m \cap B_i)} = \mathbb P(\g Y^{m} \in A ),
\end{align*}
because of the consistency between $\g Y^m$ and $\g Y^i$ and since for any $i\in V$ the sets $A \cap \mathcal{L}^m \cap B_i$ and $\mathcal L^m \cap B_i$ are subsets of $\mathcal L^m \cap \mathcal L^i$; note further that $\g Y^m \in \mathcal{L}^m$ with probability one. Finally, it is readily seen that $\g Y^m$ satisfies the global Markov property on $T$, and thus $\g Y$ is an extremal graphical model on $T$. \hfill $\Box$

\section{Proof of the expression of $\Gamma^{(m)}_{ij}$ in  Example~\ref{ex_logistic}}
\label{proof_log}

Recall the representation of extremal function $\g W^m$ for the logistic distribution in  Example~\ref{ex_logistic}. For $i,j\neq m$, we have
  $$\Gamma_{ij}^{(m)} = \var\left( \log U_i - \log U_j\right) = \var\left( \log U_i\right) +\var \left( \log U_j\right).$$ 
  Since the logarithm of a Fr\'echet distribution is a Gumbel distribution, the result follows after some algebra.

  If $i\neq j=m$, we need to compute
  $$\Gamma_{ij}^{(m)} = \var\left( \log U_i\right) +\var \left( \log U_m\right).$$ 
  The density of $\log U_m$ is
  $$ f_{\log U_m}(z) = \frac{e^{-z/\theta + z}}{\theta G(1-\theta)^{1/\theta}} \exp\left\{- G(1-\theta)^{-1/\theta} e^{-z/\theta}\right\}.$$
  We can write it as an exponential tilting
  $$ f(z) = f_{\log U_m}(z) = e^{z} g(z),$$
  where $g$ is the density of $\text{Gumbel}(\text{location}=- \log G(1-\theta), \text{scale}=\theta)$ distribution. We need to find the moments $\E(X^k e^X)$, where $X$ is the above Gumbel distribution, $k=1,2$.

Recall that the moment generating function of a random variable $X$ is defined as $m(t) = \E[e^{tX}]$. Since derivatives and expectation in this example can be interchanged  we obtain for the $k$th derivative $m^{(k)}(t) = \E[X^k\exp(tX)]$, and thus $\E[X^ke^X] = m^{(k)}(1)$. The moment generating function of a Gumbel(scale = $\mu$, shape = $\beta$) is $m(t) = G(1-\beta t) e^{\mu t}$. Hence we obtain after some simple calculations
\begin{align*}
E[Xe^X]  = - \theta\psi^{(0)}(1-\theta) - \log G(1-\theta)
\end{align*}
where $\psi^{(0)} = G'/G$ is the digamma function. For the second moment note that for $m(t) = G(1-\beta t) e^{\mu t}$ we have
\[
m''(1) = \Big(\beta^2 G''(1-\beta) - 2 \mu \beta G'(1-\beta t) + \mu^2 G(1 - \beta) \Big)e^\mu.
\]
Hence, plugging in $\beta = \theta, \mu = - \log G(1-\theta)$,
\begin{align*}
E[X^2e^X] & =\theta^2 \frac{G''(1-\theta)}{G(1-\theta)} + 2 \theta \log G(1-\theta) \frac{G'(1-\theta)}{G(1-\theta)} + \{\log G(1-\theta)\}^2
\\
& =\theta^2\Big(\psi^{(1)}(1-\theta) + \psi^{(0)}(1-\theta)^2 \Big) + 2\theta \log G(1-\theta) \psi^{(0)}(1-\theta) + \{\log G(1-\theta)\}^2.  
\end{align*} 
where the last equation uses $G''(t)/G(t) = \psi^{(1)}(t) + \psi^{(0)}(t)^2$ for the trigamma function $\psi^{(1)}$. Combining the above expressions some simple algebra yields 
\begin{align*}
E[X^2e^X] - (E[Xe^X])^2 = \theta^2 \psi^{(1)}(1-\theta). 
\end{align*}
\hfill $\Box$

\section{Proof of Theorem~\ref{th:consMST}}

We give a detailed proof for $\hat T_\Gamma^{(m)}$, all other proofs are similar. For an arbitrary tree $T'=(V,E')$ define $\dist_{\Gamma^{(m)}(q)}(T') := \sum_{(i,j)\in E'} \Gamma_{ij}^{(m)}(q)$ and $\hat \dist_{\Gamma^{(m)}}(T') := \sum_{(i,j)\in E'} \hat \Gamma_{ij}^{(m)}$. By Corollary~\ref{cor1} we know that 
\[
\min_{T' \neq T} \dist_{\Gamma^{(m)}(0)}(T') - \dist_{\Gamma^{(m)}(0)}(T) > 0.
\]
By Proposition~\ref{prop:consvar} we have as $q \to 0$
\[
\min_{T' \neq T} \dist_{\Gamma^{(m)}(q)}(T') - \dist_{\Gamma^{(m)}(q)}(T) \to \min_{T' \neq T} \dist_{\Gamma^{(m)}(0)}(T') - \dist_{\Gamma^{(m)}(0)}(T) > 0,
\]
hence there exists a $1 \geq q^* > 0$ such that
\[
\min_{T' \neq T} \dist_{\Gamma^{(m)}(q)}(T') - \dist_{\Gamma^{(m)}(q)}(T) > 0 \quad \forall q \in [0,q^*].
\] 
Now for $k/n \to q \in [0,q^*]$ {as $n \to \infty$} Theorem~\ref{th:Gammacons} implies that
\[
\min_{T' \neq T} \hat \dist_{\Gamma^{(m)}}(T') - \hat \dist_{\Gamma^{(m)}}(T) = \min_{T' \neq T} \dist_{\Gamma^{(m)}(q)}(T') - \dist_{\Gamma^{(m)}(q)}(T) + o_\bP(1), \quad \text{{as $n \to \infty$,}}
\]
which yields 
\[
P\Big( \min_{T' \neq T} \hat \dist_{\Gamma^{(m)}}(T') - \hat \dist_{\Gamma^{(m)}}(T) > 0 \Big) \to 1 \quad \text{{as $n \to \infty$.}}
\]
The claim for $\hat T_\Gamma^{(m)}$ follows. For the corresponding result on $\hat T_\Gamma^w$ we apply Corollary~\ref{cor2} instead of Corollary~\ref{cor1}. To prove the consistency for $\hat T_\chi$ we use~\eqref{eq:chicons} instead of Theorem~\ref{th:Gammacons}, Proposition~\ref{prop:chimst} instead of Corollary~\ref{cor1}, and note that $\chi_{ij}(q) \to \chi_{ij}$ as $q \to 0$ follows from~\eqref{mpd_limit} \hfill $\Box$

\section{Proofs of Proposition~\ref{prop:consvar} and Theorem~\ref{th:Gammacons}}
\label{sm_prop3}

\subsection{Technical preliminaries and equivalences of various conditions} \label{seq:equivalence}

We first connect condition (B) with a second order condition on the stable tail dependence function $\ell$ that is standard in the literature. Throughout this section we
will assume that the marginal distribution functions of $\g x$ are continuous everywhere on the support of $\g X$. Assume that~\eqref{mpd_limit} holds and define 
\[
\ell(\g x) := \frac{\mathbb P(\g Y \not \le 1/\g x)}{\mathbb P(Y_1 > 1)}, \quad \g x \in \mathcal L.
\]
The stable tail dependence function $\ell$ is a popular object for describing multivariate extremes. 
An assumption that is routinely imposed in the literature is that for all $q \in (0,1)$
\begin{equation}\label{eq:secorell}
\sup_{\g x \in [0,1]^d} \Big|q^{-1}\mathbb P(F(\g X) \not \le 1-q\g x) - \ell(\g x)\Big| \leq K_\ell q^\xi
\end{equation}  
see  for instance Assumption (C2) in Theorem 4.2 of \cite{einmahl2012} or (6) in \cite{fougeres2015} for assumptions that are similar in spirit. We will show that this is equivalent to the existence of $K_Y$ such that for $q \in (0,1)$
\begin{equation}\label{eq:secory}
\sup_{\g x \in [1,\infty]^d} \Big| \mathbb P\left( F(\g X) \leq  1 - q/\g x \mid F(\g X) \leq  1 - q\right) - \mathbb P(\g Y \nleq \g x)\Big| \leq K_Y q^\xi
\end{equation} 
for the same $\xi$. The constants $K_\ell, K_Y$ depend only on each other and on $d$. Thus condition (B) is equivalent to the following assumption which we will use in most of the proofs that follow.
\begin{enumerate}
	\item[(B')] There exist $\xi >0, K_B' < \infty$ such that for any $I \subset V$ with $|I| \in \{2,3\}$ and all $q \in (0,1)$
	\begin{equation}
	\sup_{\g x \in [0,1]^{|I|}} \Big| q^{-1}\mathbb P(F_I(\g X_I) \not \le 1-q\g x) - \ell_I(\g x)\Big| \leq K_B' q^\xi.
	\end{equation}
\end{enumerate} 
For notational convenience we define the random variables $ U_i := 1 - F_i(X_i)$. Denote the joint distribution of $\g U := (U_1,\dots,U_d)$ by $C$ and for $I \subset V$ with $|I| \in\{2,3\}$ let 
\begin{equation}\label{eq:defR}
R_I(\g x) := \lim_{q \to 0} q^{-1} \bP(F_I(\g X_I) > 1 - q\g x), \quad \g x \in [0,1]^{|I|}.
\end{equation}
Those limits exist by condition (B') and simple manipulations involving the inclusion-exclusion formula. Note that $R_I$ can be represented as a linear combination of the functions $\ell_I$ for various $I \subset \{1,\dots,d\}$ and is thus homogeneous of order $1$, i.e., $R_I(c\g x) = c R_I(\g x)$ for $\g x, c\g x \in [0,1]^{|I|}, c > 0$. This homogeneity property can be used to extend the domain of $R_I$ to $[0,\infty)^{|I|}$. 

Observe that condition (B') implies the existence of a constant $K_R$ such that for all $q \in (0,1)$ and $|I| = 2,3$
\begin{equation} \label{eq:bounRnkR'}
\sup_{\g x \in [0,1]^{|I|}} \Big|\frac{1}{q} \mathbb P(F_I(\g X_I) \geq 1 - q\g x) - R_{I}(\g x) \Big| = 
\sup_{\g x \in [0,1]^{|I|}} \Big|\frac{1}{q} C_I(q \g x) - R_{I}(\g x) \Big| \leq K_R q^\xi.
\end{equation}
Moreover, for any  $I=\{i,m\}$ (T) implies for all $M > 1$
\begin{align}
\notag|R_I(M,1) - 1| &= \mathbb P(Y_i^m \leq 1/M)
\\
\notag  &= \int_1^{\infty} x^{-2} \mathbb P\left( xW^{m}_i \leq  1/M  \right) dx
\\
\notag  &= \int_1^{\infty} x^{-2} \mathbb P\left( 1/W^{m}_i \geq  xM  \right) dx
\\
&\leq \mathbb E \left(W^{m}_i\right)^{-\gamma} M^{-\gamma} \int_1^{\infty} x^{-2-\gamma} dx \leq M^{-\gamma} \frac{K_W}{1+\gamma}, \label{eq:tailRij'}
\end{align}
by Markov's inequality.

\bigskip

\textbf{Proof of equivalence between~\eqref{eq:secorell} and~\eqref{eq:secory}}.
Observe that by the definition of $\ell$ and the fact that $\mathbb P(\g Y \not \le \g 1) = 1$ we have $\ell(\g 1) = 1/\mathbb P(Y_1 > 1)$. Further 
\[
1 \leq \mathbb P(Y_1 > 1) + \dots + \mathbb P(Y_d > 1) = d \mathbb P(Y_1 > 1)
\]
so that $\mathbb P(Y_1 > 1) \geq 1/d$. We begin by proving that~\eqref{eq:secorell} implies~\eqref{eq:secory}. Observe that for all $\g x \in [1,\infty)^d$
\begin{align*}
& \mathbb P\left( F(\g X) \nleq  1 - q/\g x \mid F(\g X) \nleq  1 - q\right) - \mathbb P(\g Y \nleq \g x)\\
=~ & \frac{\mathbb P\big( F(\g X) \nleq  1 - q/\g x\big) }{\mathbb P\big( F(\g X) \nleq  1 - q\big)} - \ell(1/\g x) \mathbb P(Y_1>1)
\\
=~ & \frac{q^{-1}\mathbb P\big( F(\g X) \nleq  1 - q/\g x\big) }{q^{-1}\mathbb P\big( F(\g X) \nleq  1 - q\big)} - \frac{\ell(1/\g x)}{\ell(\g 1)}.
\end{align*} 
A simple Taylor expansion taking into account that $\g x \in [1,\infty]^d$ iff $1/\g x \in [0,1]^d, \ell(\g 1) \geq 1$ and $\ell(1/\g x) \leq d$ now shows that that~\eqref{eq:secorell} implies~\eqref{eq:secory}. 

Next we will prove the converse implication. Let $\g x = (1,\infty,\dots,\infty)$. Then
$\mathbb P(\g Y \not \le \g x) = \mathbb P(Y_1 > 1)$ and $\mathbb P(F(\g X) \not \le 1-q/\g x) = \mathbb P(F_1(X_1) > 1-q) = q$. Then~\eqref{eq:secory} implies that for sufficiently small $q$
\[
\Big|\frac{q}{\mathbb P(F(\g X) \not \le 1-q)} - \mathbb P(Y_1 > 1)\Big| \leq K_Y q^\xi.
\]
By a Taylor expansion this yields for $q \in (0,1)$ and a constant $K$ that depends on $K_Y, d$ only
\[
\Big|q^{-1}\mathbb P(F(\g X) \not \le 1-q) - \frac{1}{\mathbb P(Y_1 > 1)}\Big| \leq K q^\xi.
\]
Combining this with the representation 
\begin{align*}
&q^{-1}\mathbb P(F(\g X) \not \le 1-q\g x) - \ell(\g x)
\\
=~ & \mathbb P\left( F(\g X) \nleq  1 - q/\g x \mid F(\g X) \nleq  1 - q\right) q^{-1}\mathbb P(F(\g X) \not \le 1-q) - \mathbb P(\g Y \not\le1/\g x) \ell(\g 1)
\end{align*}
we find that~\eqref{eq:secory} indeed implies~\eqref{eq:secorell}. This completes the proof of the equivalence of those two conditions. \hfill $\Box$

\subsection{Alternative representations for estimators and limiting objects}

Next we discuss several representations that will be useful for both proofs. Since 
\[
\Gamma_{i,j}^{(m)} = \Var(\log(1/Y_{i}^m) - \log(1/Y_j^m))
\]
we have
\begin{equation}\label{eq:repgamma}
\Gamma_{i,j}^{(m)} = e_{i}^{(m),2} + e_{j}^{(m),2} - 2 e_{i,j}^{(m)} - \Big(e_{i}^{(m),1} - e_{j}^{(m),1}\Big)^2
\end{equation}
where
\begin{align*}
e_{i,j}^{(m)} &= \E[\log(1/Y_i^m)\log(1/Y_j^m)]  
\\
e_{i}^{(m),\ell} &= \E[(\log (1/Y_i^m))^\ell] 
\end{align*}
For the pre-asymptotic versions, define the random vectors $\g U^m(q)$ with distribution on 
\[
\mathcal D^m(q) := [0,1]^{m-1}\times[0,q]\times[0,1]^{d-m}
\] 
given by 
\[
\mathbb P(\g U^m(q) \in A) = q^{-1}\mathbb P(\g U \in A \cap \mathcal D^m(q))
\]
With this notation the pre-asymptotic variogram can be represented as
\begin{align}
\Gamma_{ij}^{(m)}(q) &= \var \Big(-\log\{U_i^m(q)/q\} + \log\{U_j^m(q)/q\} \Big) \nonumber
\\
& =  e_{i}^{(m),2}(q) + e_{j}^{(m),2}(q) - 2 e_{i,j}^{(m)}(q) - \Big(e_{i}^{(m),1}(q) - e_{j}^{(m),1}(q)\Big)^2 \label{eq:repgammaq}
\end{align}
with
\begin{align*}
e_{i,j}^{(m)}(q) &= \E[\log(U_i^m(q)/q)\log(U_j^m(q)/q)] 
\\
e_{i}^{(m),\ell}(q) &= \E[(- \log(U_i^m(q)/q))^\ell].
\end{align*}
The quantities above have alternative representations which we will use in the following proof. First note that for $i \neq m$ we have
\begin{align}
\label{eq:epowalt}
e_{i}^{(m),\ell} &= \int_{0}^{1} \frac{R_{i,m}(x,1)\ell(-\log x)^{\ell-1}}{x} dx + \int_{1}^{\infty} \frac{(R_{i,m}(x,1) - 1)\ell(- \log x)^{\ell-1}}{x} dx,
\\ \label{eq:epowalt-q} 
e_{i}^{(m),\ell}(q) &= \int_{0}^{1} \frac{q^{-1} C_{i,m}(qx,q)\ell(-\log x)^{\ell-1}}{x} dx + \int_{1}^{q^{-1}} \frac{(q^{-1} C_{i,m}(qx,q) - 1)\ell( - \log x)^{\ell-1}}{x} dx.
\end{align}
For the next representations assume that $i,j,m$ are all different. Then
\begin{align} \label{eq:emixalt-jj}
e_{i,j}^{(j)} &= \int_{0}^{1} \int_{0}^{1} \frac{R_{i,j}(x,y)}{xy} dx dy + \int_{0}^{1} \int_{1}^{\infty} \frac{R_{i,j}(x,y) - y}{xy} dx dy,
\\
e_{i,j}^{(j)}(q) &= \int_{0}^{1} \int_{0}^{1} \frac{C_{i,j}(qx,qy)}{qxy} dx dy + \int_{0}^{1} \int_{1}^{q^{-1}} \frac{C_{i,j}(qx,qy) - C_{i,j}(1,qy)}{qxy} dx dy, \label{eq:emixalt-jj-q}
\end{align}
and
\begin{align}
e_{i,j}^{(m)} =~ & \int_0^1\int_0^1 \frac{R_{i,j,m}(x,y,1)}{xy} dxdy  \nonumber
\\
&\quad+  \int_0^1\int_1^{\infty} \frac{R_{i,j,m}(x,y,1) - R_{i,j,m}(\infty,y,1)}{xy} dxdy  \nonumber
\\
&\quad+ \int_1^{\infty}\int_0^1 \frac{R_{i,j,m}(x,y,1) - R_{i,j,m}(x,\infty,1)}{xy} dxdy  \nonumber
\\
&\quad +  \int_1^{\infty}\int_1^{\infty} \frac{R_{i,j,m}(x,y,1) - R_{i,j,m}(\infty,y,1) - R_{i,j,m}(x,\infty,1) + R_{i,j,m}(\infty,\infty,1)}{xy} dxdy \label{eq:emixalt-jm}
\end{align}
as well as
\begin{align}
e_{i,j}^{(m)}(q) =~ & \int_0^1\int_0^1 \frac{C_{i,j,m}(qx,qy,q)}{qxy} dxdy  \nonumber
\\
&\quad+  \int_0^1\int_1^{1/q} \frac{C_{i,j,m}(qx,qy,q) - C_{i,j,m}(1,qy,q)}{qxy} dxdy  \nonumber
\\
&\quad+ \int_1^{1/q}\int_0^1 \frac{C_{i,j,m}(qx,qy,q) - C_{i,j,m}(qx,1,q)}{qxy} dxdy  \nonumber
\\
&\quad +  \int_1^{1/q}\int_1^{1/q} \frac{C_{i,j,m}(qx,qy,q) - C_{i,j,m}(1,qy,q) - C_{i,j,m}(qx,1,q) + C_{i,j,m}(1,1,q)}{qxy} dxdy. \label{eq:emixalt-jm-q}
\end{align}

Next we discuss similar representations for the empirical version of the extremal variogram. Define the random variables $U_{ti} := 1 - F_i(X_{ti})$ (here $X_{ti}$ denotes the $i$'th entry of the vector $\g X_t$) and the vectors $\g U_t := (U_{t1},\dots,U_{td})^\top$. Let $R(\g x) := \Lambda([\g 1/ \g x, \g \infty))$ and denote by $\hat F_i$ the empirical distribution function $U_{1i},\dots,U_{ni}$. Define the vector $ \widehat{\g F}^-(\g x) := (\hat F_1^-(x_1),\dots, \hat F_d^-(x_d))$, the function
\begin{equation} \label{eq:Ccirc}
\hat C^\circ(\g x) := \frac{1}{n} \sum_{i=1}^n I\{U_{t1} \leq x_1,\dots,U_{td} \leq x_d\}
\end{equation}
and $\hat C(k \g x/n) := \hat C^\circ(\widehat{\g F}^-(k\g x/n)).$ Introduce the notation
\[
\widehat R_I(\g x) := \frac{n}{k} \hat C_I(k\g x/n).
\]
Note that the estimator $\hat \Gamma^{(m)}$ depends only on the marginal ranks of $X_{ti}$; thus we have almost surely
\begin{align*}
\hat \Gamma_{i,j}^{(m)} &= \widehat{\Var}\Big(\log (\hat F_i(U_{ti})) - \log(\hat F_j(U_{tj})) : \hat F_m(U_{tm}) \leq \frac{k}{n}  \Big)
\\
&= \widehat{\Var}\Big(- \log (n\hat F_i(U_{ti})/k) + \log(n\hat F_j(U_{tj})/k) : \hat F_m(U_{tm}) \leq \frac{k}{n}  \Big). 
\end{align*}
Now observe the following representation
\begin{equation}\label{eq:repgammahat}
\hat \Gamma_{i,j}^{(m)} = \hat e_{i}^{(m),2} + \hat e_{j}^{(m),2} - 2 \hat e_{i,j}^{(m)} - \Big(\hat e_{i}^{(m),1} - \hat e_{j}^{(m),1}\Big)^2
\end{equation}
where
\begin{align*}
\hat e_i^{(m),\ell} &:= \frac{1}{k} \sum_{t=1}^n \Big\{- \log \Big(\frac{n \hat F_{i}(U_{ti})}{k}\Big)\Big\}^\ell I\{\hat F_m(U_{tm}) \leq k/n \}, \quad \ell = 1,2
\\
\hat e_{ij}^{(m)} &:= \frac{1}{k} \sum_{t=1}^n \log \Big(\frac{n \hat F_{j}(U_{tj})}{k}\Big) \log \Big(\frac{n \hat F_{i}(U_{ti})}{k}\Big) I\{\hat  F_m(U_{tm}) \leq k/n \}.
\end{align*}
The quantities above have alternative representations which we will use frequently. The exact form of those representations depends on whether $m \in \{i,j\}$ or $m \notin\{i,j\}$ and those two cases will be considered separately. 

We start with the case $m \in \{i,j\}$. Assume without loss of generality that $j = m$. Then, {as $n \to \infty$}, 
\begin{align} \label{eq:hatemixalt}
\hat e_{ij}^{(j)} = & \int_{1/k}^{1} \int_{1/k}^{1} \frac{\widehat R_{ij}(x,y)}{xy} dx dy + \int_{1/k}^{1} \int_{1}^{n/k} \frac{\widehat R_{ij}(x,y) - \widehat R_{ij}(n/k,y) }{xy} dx dy + O((\log n)^2/k),
\\ \label{eq:hatepowalt}
\hat e_{i}^{(j),\ell} = & \int_{1/k}^{1} \frac{\widehat R_{ij}(x,1) \ell( -\log x)^{\ell-1}}{x} dx + \int_{1}^{n/k} \frac{(\widehat R_{ij}(x,1) - 1)\ell (-\log x)^{\ell-1}}{x} dx + O((\log n)^2/k).
\end{align}
We also note that, {as $n \to \infty$},
\begin{align*}
\hat e_i^{(i),\ell} &= \frac{1}{k} \sum_{t=1}^n \Big\{- \log \Big(\frac{n \hat F_{i}(U_{ti})}{k}\Big)\Big\}^\ell I\{\hat F_i(U_{ti}) \leq k/n \} = \frac{1}{k} \sum_{t=1}^k \{- \log(t/k)\}^\ell
\\
& = \int_0^1 \{\log(1/x)\}^\ell dx  + o(1) = \E[(\log Y_i^{(i)})^\ell] + o(1)
\end{align*}
where we used that $Y_i^{(i)}$ is unit Pareto and the difference between the integral and the sum is $o(1)$ by a standard Riemann approximation.
When $m \notin\{i,j\}$ we have, {as $n \to \infty$},
\begin{align}
\hat e_{i,j}^{(m)} &=  \int_{1/k}^1 \int_{1/k}^1 \frac{\widehat R_{i,j,m}(x,y,1)}{xy} dxdy  \nonumber
\\
&\quad+  \int_{1/k}^1\int_1^{n/k} \frac{\widehat R_{i,j,m}(x,y,1) - \widehat R_{i,j,m}(\infty,y,1)}{xy} dxdy  \nonumber
\\
&\quad+ \int_1^{n/k}\int_{1/k}^1 \frac{\widehat R_{i,j,m}(x,y,1) - \widehat R_{i,j,m}(x,\infty,1)}{xy}dxdy  \nonumber
\\
&\quad +  \int_1^{n/k}\int_1^{n/k} \frac{\widehat R_{i,j,m}(x,y,1) - \widehat R_{i,j,m}(\infty,y,1) -\hat R_{i,j,m}(x,\infty,1) + 1}{xy} dxdy + O((\log n)^2/k)
\label{eq:hatemixalt2}
\end{align}
while the representation for $\hat e_{i}^{(m),\ell}$ does not change. 

All representations defined above will be established in section~\ref{sec:altrepr}. After this preparation, we proceed to proving the main asymptotic results.

\subsection{Proof of Proposition~\ref{prop:consvar}} \label{sec:proofconsvar}

{Throughout this subsection, all $o(\cdot), O(\cdot)$ terms and convergences are understood as $q \to 0$ unless otherwise stated.} We begin by proving some useful technical results: under~\eqref{eq:bounRnkR'} and~\eqref{eq:tailRij'} we have for any $0 < \delta < 1$ such that $(1-\delta)\xi-\delta > 0$, 
\begin{align}
\sup_{2 \leq |I| \leq 3}{\sup_{\g x \in [0,q^{-\delta}]^{|I|-1}\times [0,1]}} \Big|\frac{1}{q}C_I(q \g x) - R_I(\g x) \Big| &= O(q^{(1-\delta)\xi-\delta}) \label{eq:cbound1}
\\
\sup_{2 \leq |I| \leq 3} {\sup_{\g x \in [q^{-\delta},q^{-1}]^{|I|-1}\times [0,1]}} \Big|\frac{1}{q}C_I(q \g x) - R_I(\g x) \Big| &= O(q^{(1-\delta)\xi-\delta} + q^{\gamma\delta}) \label{eq:cbound2}
\\
\sup_{|I|=3} \sup_{\g x \in [q^{-\delta},q^{-1}]\times [0,q^{-\delta}] \times [0,1]} \Big|\frac{1}{q}C_I(q \g x) - R_I(\g x) \Big| & = O(q^{(1-\delta)\xi-\delta} + q^{\gamma\delta}
) \label{eq:cbound3}
\end{align}
Note that combining the above bounds and setting $\delta = \xi/(\xi+\gamma+1)$ implies
\begin{equation} \label{eq:bounRnkR'-unif}
\sup_{2 \leq |I| \leq 3} \sup_{\g x \in [0,q^{-1}]^{|I|-1}\times [0,1]} \Big|\frac{1}{q} C_I(q \g x) - R_{I}(\g x) \Big|= O(q^{\xi\gamma/(\xi+\gamma+1)}), \qquad \text{as } q\to 0
\end{equation}
The key difference to~\eqref{eq:bounRnkR'} is that some components of $\g x$ are now allowed to vary over a growing set as $q \to 0$. The price for this generalization is a strictly smaller power of $q$ in the corresponding upper bound. 

Next we derive a general bound on $R$. Let $I \subset \{1,\dots,d\}$ with $|I|=3$ be arbitrary. Note that $R_I$ can be seen as the distribution function of a measure on $[0,\infty)^3$ and that $R_I(\infty,\infty,x) = x$, and thus slightly abusing notation we have
\begin{multline} 
\sup_{x \geq 1, y \in [0,\infty], z \in [0,1]} x^\gamma \Big|R_I(x,y,z) - R_I(\infty,y,z)\Big| = \sup_{x \geq 1, y \in [0,\infty], z \in [0,1]} x^\gamma R_I((x,\infty)\times[0,y]\times[0,z]) 
\\ \label{eq:xgammaR3}
= \sup_{x \geq 1} x^\gamma R_I((x,\infty)\times[0,\infty)\times[0,1])   < \infty, 
\end{multline}
where the finiteness of the last display follows by~\eqref{eq:tailRij'}. In particular this implies 
\begin{equation} \label{eq:tailR3}
\sup_{x \geq q^{-\delta}, y \in [0,\infty], z \in [0,1]} \Big|R_I(x,y,z) - R_I(\infty,y,z)\Big| = O(q^{\gamma\delta}).  
\end{equation}

For a proof of~\eqref{eq:cbound1}, note that by~\eqref{eq:bounRnkR'} and homogeneity of $R$ we have for sequences $\g x = \g x_q$ with $\|\g x\|_\infty \leq q^{-\delta}$, 
\begin{align*}
\frac{1}{q}C_I(q \g x) &= \frac{1}{q}C_I(q^{1-\delta}q^\delta \g x) = \frac{q^{1-\delta}}{q}\Big\{R_I(q^{\delta} \g x) + O(q^{(1-\delta)\xi})\Big\}
= R_I(\g x) + O(q^{(1-\delta)\xi - \delta}). 
\end{align*}

For a proof of~\eqref{eq:cbound3} observe that have for $q^{-1} \geq x \geq q^{-\delta}, y \leq q^{-\delta}, I= (i_1,i_2,i_3)$
\begin{align*}
\frac{1}{q}C_I(1,qy,qz) &\geq \frac{1}{q}C_I(qx,qy,qz) 
\geq \frac{1}{q}C(q^{1-\delta},qy,qz) 
= \frac{1}{q}C_I(qq^{-\delta},qy,qz)
\\
& = R_I(q^{-\delta},y,z) + O(q^{(1-\delta)\xi - \delta})
= R_I(\infty,y,z) + O(q^{(1-\delta)\xi - \delta}+q^{\gamma\delta}) 
\\
& = \frac{1}{q}C_I(1,qy,qz) + O(q^{(1-\delta)\xi - \delta}+q^{\gamma\delta}) 
\end{align*}
where the last equality follows by~\eqref{eq:cbound1} applied with $I = (i_2,i_3)$, the second equality follows by~\eqref{eq:cbound1}, the third equality follows by~\eqref{eq:tailR3} and all $O$ terms are uniform in $q^{-1} \geq x \geq q^{-\delta}, y \leq q^{-\delta}, I= (i_1,i_2,i_3)$. This implies, 
\begin{equation}\label{eq:cbound-tail}
\sup_{|I|=3} \sup_{\g x \in [q^{-\delta},q^{-1}]\times [0,q^{-\delta}] \times [0,1]} \Big| \frac{1}{q}C_I(qx,qy,qz) - \frac{1}{q} C_I(1,qy,qz)\Big| = O(q^{(1-\delta)\xi - \delta}+q^{\gamma\delta})
\end{equation}
and also, 
\[
\sup_{|I|=3} \sup_{\g x \in [q^{-\delta},q^{-1}]\times [0,q^{-\delta}] \times [0,1]} \Big| \frac{1}{q}C_I(qx,qy,qz) - R_I(\infty,y,z)\Big| = O(q^{(1-\delta)\xi - \delta}+q^{\gamma\delta}).
\]
Combined with~\eqref{eq:tailR3} this completes the proof of~\eqref{eq:cbound3}. 

The proof of~\eqref{eq:cbound2} is similar. Indeed we have for $\g x \in [q^{-\delta},\infty)^{|I|-1}\times [0,1]$ 
\[
x_3 \geq R_I(\g x) \geq R_I(q^{-\delta},q^{-\delta},x_3) = R_I(\infty,q^{-\delta},x_3) + O(q^{\gamma\delta}) = x_3 + O(q^{\gamma\delta})
\]
where~\eqref{eq:tailR3} is applied twice: first with $y = q^{-\delta}$ and second with $I = (i_1,i_2)$. By~\eqref{eq:cbound1} and the bound above 
\begin{align*}
x_3 \geq \frac{1}{q}C_I(q \g x) &\geq \frac{1}{q}C_I(q q^{-\delta},q q^{-\delta},x_3) = R_I(q^{-\delta},q^{-\delta},x_3) + O(q^{(1-\delta)\xi - \delta}) = x_3 + O(q^{(1-\delta)\xi - \delta}+q^{\gamma\delta}).
\end{align*}
Combining the two chains of inequalities above we find that $q^{-1}C_I(q\g x) = x_3 + O(q^{(1-\delta)\xi - \delta}+q^{\gamma\delta})$ and $R_I(\g x) = x_3 + O(q^{\gamma\delta})$ implies \eqref{eq:cbound2}.

\medskip

We will now show that $e_i^{(m),\ell}(q) \to e_i^{(m),\ell}$, $e_{i,j}^{(m)}(q) \to e_{i,j}^{(m)}$ and $e_{i,m}^{(m)}(q) \to e_{i,m}^{(m)}$. Combined with the representations in~\eqref{eq:repgamma},~\eqref{eq:repgammaq} this will complete the proof of Proposition~\ref{prop:consvar}. 

\bigskip 

\textbf{Proof of $e_i^{(m),\ell}(q) \to e_i^{(m),\ell}$ {as $q \to 0$}}. To keep the notation simple we only consider the case $\ell =1$, the case $\ell = 2$ follows by exactly the same arguments. We have  
\begin{align*}
0 \leq \int_{0}^{q^{\xi}} \frac{1}{x} \frac{1}{q} C_{i,m}(qx,q) dx \leq \int_{0}^{q^{\xi}} \frac{1}{xq} xq dx = q^{\xi}
\end{align*}
and
\begin{align*}
\int_{q^{\xi}}^1 \frac{1}{x} \frac{1}{q} C_{i,m}(qx,q) dx &= \int_{q^{\xi}}^1 \frac{1}{x} \Big\{R_{i,m}(x,1) + O(q^\xi)\Big\} dx = \int_{q^{\xi}}^1 \frac{1}{x} R_{i,m}(x,1) dx + O(q^\xi |\log q|)
\\
&= \int_0^1 \frac{1}{x} R_{i,m}(x,1) dx + O(q^\xi |\log q|) 
\end{align*}
where the last line follows since $0 \leq \frac{1}{x} R_{i,m}(x,1) = R_{i,m}(1,1/x) \leq R_{i,m}(1,\infty) = 1$. Next observe that by~\eqref{eq:bounRnkR'-unif} 
\begin{align*}
\int_{1}^{q^{-1}} \frac{1}{x} \Big\{\frac{1}{q}C_{i,m}(qx,q) - 1\Big\}dx - \int_{1}^{q^{-1}} \frac{1}{x} \Big\{R_{i,m}(x,1) - 1\Big\}dx = O(q^{\gamma\xi/(\gamma+\xi+1)} |\log q|). 
\end{align*}
The claim follows by combining this with~\eqref{eq:epowalt},~\eqref{eq:epowalt-q} and the fact that by~\eqref{eq:tailRij'}
\[
\int_{q^{-1}}^\infty \frac{1}{x} \Big\{R_{i,m}(x,1) - 1\Big\}dx = o(1).
\]

\bigskip

\textbf{Proof of $e_{i,j}^{(m)}(q) \to e_{i,j}^{(m)}$ and $e_{i,m}^{(m)}(q) \to e_{i,m}^{(m)}$ {as $q \to 0$}} 

Since the proof of $e_{i,m}^{(m)}(q) \to e_{i,m}^{(m)}$ is similar but simpler we will only provide details for $e_{i,j}^{(m)}(q) \to e_{i,j}^{(m)}$. For the sake of a lighter notation we will drop the index $i,j,m$ from $C,R$ in all calculations that follow. 

Fix $\alpha > \xi$. Then
\begin{align*}
\int_0^1\int_0^1 \frac{1}{xy} \frac{1}{q}C(qx,qy,q) dxdy = \int_{[q^\alpha,1]^2} \frac{1}{xy} \frac{1}{q}C(qx,qy,q) dxdy + \int_{([q^\alpha,1]^2)^C} \frac{1}{xy} \frac{1}{q}C(qx,qy,q) dxdy
\end{align*}
First, by~\eqref{eq:bounRnkR'},
\begin{align*}
\int_{[q^\alpha,1]^2} \frac{1}{xy} \frac{1}{q}C(qx,qy,q) dxdy &= \int_{[q^\alpha,1]^2} \frac{1}{xy} R(x,y,1) dxdy + O((\log q)^2 q^\xi).
\end{align*} 
Second, by the upper Fr\'echet--Hoeffding bound
\begin{align*}
0 &\leq \int_{([q^\alpha,1]^2)^C} \frac{1}{xy} \frac{1}{q}C(qx,qy,q) dxdy \leq \Big(\int_0^{q^\alpha}\int_{0}^1 + \int_{0}^1\int_0^{q^\alpha}\Big) \frac{1}{xy} \frac{1}{q}C(qx,qy,q) dxdy
\\
&\leq 2 \int_0^{q^\alpha}\int_{0}^1 \frac{x \wedge y}{xy}  dxdy = o(q^\xi)
\end{align*}
where the last line follows since
\[
\int_0^{q^\alpha}\int_{0}^1 \frac{x \wedge y}{xy}  dxdy = \int_0^{q^\alpha} \frac{1}{y} \Big(\int_0^{y} \frac{x}{x} dx + \int_y^{1} \frac{y}{x} dx\Big)dy = \int_0^{q^\alpha} 1 - \log(y) dy.
\]
Lastly, since $R(x,y,1) \leq x \wedge y$, we have proved
\begin{equation} \label{eq:eijm-h1}
\int_0^1\int_0^1 \frac{1}{xy} \frac{1}{q}C(qx,qy,q) dxdy = \int_0^1\int_0^1 \frac{1}{xy} R(x,y,1) dxdy + O(q^\xi (\log q)^2). 
\end{equation} 
Next consider the decomposition
\begin{align*}
&\int_0^1\int_1^{1/q} \frac{1}{xy} \frac{1}{q}\{C(qx,qy,q) - C(1,qy,q)\}dxdy
\\
=&~ \Big(\int_0^{q^\alpha}\int_1^{1/q} + \int_{q^\alpha}^1\int_1^{1/q}\Big) \frac{1}{xy} \frac{1}{q}\{C(qx,qy,q) - C(1,qy,q)\}dxdy.
\end{align*}
Noting that 
\[
\Big| C(qx,qy,q) - C(1,qy,q) \Big| \leq qy
\] 
the first term can be bounded as follows  
\begin{align*}
0 \leq \int_0^{q^\alpha}\int_1^{1/q} \frac{1}{xy} \frac{1}{q}\{C(1,qy,q) - C(qx,qy,q)\}dxdy \leq \int_0^{q^\alpha}\int_1^{1/q} \frac{1}{x} dxdy = q^\alpha |\log(q)|.
\end{align*}
Similarly we have
\[
0 \leq \int_0^{q^\alpha}\int_1^{1/q} \frac{1}{xy} \{R(\infty,y,1) - R(x,y,1)\}dxdy \leq \int_0^{q^\alpha}\int_1^{1/q} \frac{1}{x} dxdy = q^\alpha |\log(q)|.
\]
Next, by~\eqref{eq:bounRnkR'-unif}
\begin{align*}
\int_{q^\alpha}^1\int_1^{q^{-1}} \frac{1}{xy} \frac{1}{q}\{C(qx,qy,q) - C(1,qy,q)\}dxdy =& \int_{q^\alpha}^1\int_1^{q^{-1}} \frac{1}{xy} \{R(x,y,1) - R(\infty,y,1)\}dxdy
\\
& + O(q^{\gamma\xi/(\gamma+\xi+1)}(\log q)^2).
\end{align*}
By~\eqref{eq:xgammaR3} we have  
\begin{align*}
\Big|\int_{q^\alpha}^1\int_{q^{-1}}^{\infty} \frac{1}{xy} \{R(x,y,1) - R(\infty,y,1)\} dxdy\Big| &\leq |\log q| O(1) \int_{q^{-1}}^\infty x^{-1-\gamma} dx = O(|\log q| q^{\gamma}).
\end{align*}
Finally, since by~\eqref{eq:xgammaR3} 
\[
\Big| R(x,y,1)-R(\infty,y,1)) \Big| \leq K (y \wedge x^{-\gamma})
\]
we have
\[
\int_0^{q^\alpha} \int_{q^{-1}}^\infty \frac{1}{xy} \{R(x,y,1) - R(\infty,y,1)\} dx dy= O(q^\alpha |\log q|).
\]
In summary, we have proved 
\begin{multline} \label{eq:eijm-h2}
\int_0^1\int_1^{1/q} \frac{1}{xy} \frac{1}{q}\{C(qx,qy,q) - C(1,qy,q)\}dxdy
\\
= \int_0^1\int_1^{\infty} \frac{1}{xy} \{R(x,y,1) - R(\infty,y,1)\}dxdy + o(1) 
\end{multline}
Next observe that by~\eqref{eq:bounRnkR'-unif}
\begin{multline} \label{eq:eijm-h3}
\int_1^{q^{-1}}\int_1^{q^{-1}} \frac{1}{xy} \frac{1}{q}\{C(qx,qy,q) - C(1,qy,q) - C(qx,1,q) + C(1,1,q)\}dxdy
\\
= \int_1^{q^{-1}}\int_1^{q^{-1}} \frac{R(x,y,1) - R(\infty,y,1) - R(x,\infty,1) + R(\infty,\infty,1)}{xy} dxdy
+ (\log q)^2 O(q^{\gamma\xi/(\gamma+\xi+1)}).
\end{multline}
Furthermore,~{\eqref{eq:xgammaR3} implies that 
\begin{align*}
\sup_{x,y \geq 1} x^\gamma \Big|R(x,y,1) - R(\infty,y,1) - R(x,\infty,1) + R(\infty,\infty,1) \Big| & < \infty 
\\
\sup_{x,y \geq 1} y^\gamma \Big|R(x,y,1) - R(\infty,y,1) - R(x,\infty,1) + R(\infty,\infty,1) \Big| & < \infty 
\end{align*}
to see this apply the triangle inequality and use~{\eqref{eq:xgammaR3} on both resulting parts noting that $y = \infty$ is explicitly allowed in~{\eqref{eq:xgammaR3}. 
Thus
\begin{multline} \label{eq:eijm-h4}
\Big(\int_1^\infty \int_1^\infty - \int_1^{q^{-1}} \int_1^{q^{-1}}\Big) \frac{\Big|R(x,y,1) - R(\infty,y,1) - R(x,\infty,1) + R(\infty,\infty,1) \Big|}{xy} dx dy
\\ 
\leq K \Big(\int_1^\infty \int_1^\infty - \int_1^{q^{-1}} \int_1^{q^{-1}}\Big) \frac{ x^{-\gamma} \wedge y^{-\gamma}}{xy} dx dy = o(1)
\end{multline}
Thus combining~\eqref{eq:eijm-h1}-~\eqref{eq:eijm-h4} with~\eqref{eq:emixalt-jm} and~\eqref{eq:emixalt-jm-q} the claim follows. \hfill $\Box$

\subsection{Proof of Theorem~\ref{th:Gammacons}}\label{sec:prGammacons}

{Throughout this subsection, all convergences and $o(\cdot), O(\cdot), o_\bP(\cdot), P_\bP(\cdot)$ terms will be as $n \to \infty$ unless stated otherwise}. A close look at the proof that follows shows that it continues to hold under the following high-level condition: there exists a $\beta >0$ such that for any $I \subset V$ with $|I| \leq 3$ such that
\begin{equation}\label{eq:convratetailproc}
\sup_{\g x \in [0,1]^{|I|}} \frac{n}{k}\Big| \hat C_I(k \g x/n) - C_I(k \g x/n) \Big| = O_\bP(k^{-\beta}).
\end{equation}
For independent observations this is true with $\beta = 1/2$ as we will establish below. Under suitable short-range dependence such as $\alpha$-mixing with sufficiently fast decay of the mixing coefficients or conditions on physical dependence measures this type of result can be established by the usual chaining arguments. Note that process convergence is explicitly not required and the rate can be slower that the typical $k^{-1/2}$ rate that is expected when process convergence does hold. We omit details for the sake of brevity.

We will show later that there exists a $\psi > 0$ such that for all $I \subset V$ with $2\leq |I| \leq 3$ and any fixed $0 \leq T < \infty$ 
\begin{equation}\label{eq:unifC}
\sup_{\g x \in [0,n/k]^{|I|-1}\times[0,T]} \Big| \widehat R_I(\g x) - \frac{n}{k} C_I(k \g x/n) \Big| = O_\bP(n^{-\psi}) .
\end{equation}
Note that this differs from common results on convergence of estimators of $R$ because some arguments are now allowed to vary over growing sets.

Now recall the representation for $e_{i}^{(m),\ell}(q)$ given in~\eqref{eq:epowalt-q} and apply it with $q = k/n$ to obtain
\begin{multline*} 
e_{i}^{(m),\ell}(k/n) = \int_{0}^{1} \frac{(n/k) C_{i,m}(kx/n,k/n)\ell(-\log x)^{\ell-1}}{x} dx 
\\
+ \int_{1}^{n/k} \frac{((n/k)C_{i,m}(kx/n,k/n) - 1)\ell( - \log x)^{\ell-1}}{x} dx.
\end{multline*}
Further note that by the bound $(n/k)C_{i,m}(kx/n,k/n) \leq x$ we also have 
\begin{multline*} 
e_{i}^{(m),\ell}(k/n) = \int_{1/k}^{1} \frac{(n/k) C_{i,m}(kx/n,k/n)\ell(-\log x)^{\ell-1}}{x} dx
\\
+ \int_{1}^{n/k} \frac{((n/k)C_{i,m}(kx/n,k/n) - 1)\ell( - \log x)^{\ell-1}}{x} dx + o(1).
\end{multline*}
Combining this with~\eqref{eq:hatepowalt} and~\eqref{eq:unifC} we find that
\begin{align*}
\Big| \hat e_{i}^{(m),\ell} - e_{i}^{(m),\ell}(k/n) \Big| &\leq \int_{1/k}^{n/k} \frac{\ell|\log x|^{\ell-1}}{x} dx \sup_{x \in [0,n/k]} \Big|\hat R_{i,m}(x,1) - \frac{n}{k}C_{i,m}(kx/n,k/n)\Big| + o(1)
\\
&=  O_\bP(n^{-\psi} (\log n)^\ell) + o(1) = o_{\bP}(1).
\end{align*}
It now remains to show that for $k/n\to q \in [0,1)$ we have $e_{i}^{(m),\ell}(k/n) \to e_{i}^{(m),\ell}(q)$. For $q>0$ this statement follows by uniform continuity of $C_{i,m}$ combined with the dominated convergence theorem after noting that $|(n/k) C_{i,m}(kx/n,k/n)| \leq x$ and after noting that in this case the integration range $[1,n/k]$ remains bounded. For $q = 0$ this statement was established in the proof of Proposition~\ref{prop:consvar} (there we considered a general $q\to 0$, replace that $q$ by $k/n$). In summary, we proved that for $k/n \to q \in [0,1)$ 
\begin{equation}\label{eq:h1}
\hat e_{i}^{(m),\ell} = e_{i}^{(m),\ell}(q) + o_{\bP}(1).    
\end{equation}
The proof of $\hat e_{ij}^{(m)} = e_{ij}^{(m)}(k/n) + o_{\bP}(1)$ is similar, and for the sake of brevity we only treat the more complicated case $i,j,m$ all different. Apply the representation in~\eqref{eq:emixalt-jm-q} with $q = k/n$ and note that
\begin{align*}
e_{i,j}^{(m)}(\frac{k}{n})
=~ & \int_0^1\int_0^1 \frac{n}{k}\frac{C_{i,j,m}(\frac{kx}{n},\frac{ky}{n},\frac{k}{n})}{xy} dxdy 
\\
&~+  \int_0^1\int_1^{n/k} \frac{n}{k}\frac{C_{i,j,m}(\frac{kx}{n},\frac{ky}{n},\frac{k}{n}) - C_{i,j,m}(1,\frac{ky}{n},\frac{k}{n})}{xy} dxdy 
\\
&~+ \int_1^{n/k}\int_0^1 \frac{n}{k}\frac{C_{i,j,m}(\frac{kx}{n},\frac{ky}{n},\frac{k}{n}) - C_{i,j,m}(\frac{kx}{n},1,\frac{k}{n})}{(n/k)xy} dxdy 
\\
&~ +  \int_1^{n/k}\int_1^{n/k} \frac{n}{k}\frac{C_{i,j,m}(\frac{kx}{n},\frac{ky}{n},\frac{k}{n}) - C_{i,j,m}(1,\frac{ky}{n},\frac{k}{n}) - C_{i,j,m}(\frac{kx}{n},1,\frac{k}{n}) + C_{i,j,m}(1,1,\frac{k}{n})}{xy} dxdy
\\
=~ & \int_{1/k}^1\int_{1/k}^1 \frac{n}{k}\frac{C_{i,j,m}(\frac{kx}{n},\frac{ky}{n},\frac{k}{n})}{xy} dxdy 
\\
&~+  \int_{1/k}^1\int_1^{n/k} \frac{n}{k}\frac{C_{i,j,m}(\frac{kx}{n},\frac{ky}{n},\frac{k}{n}) - C_{i,j,m}(1,\frac{ky}{n},\frac{k}{n})}{xy} dxdy 
\\
&~+ \int_1^{n/k}\int_{1/k}^1 \frac{n}{k}\frac{C_{i,j,m}(\frac{kx}{n},\frac{ky}{n},\frac{k}{n}) - C_{i,j,m}(\frac{kx}{n},1,\frac{k}{n})}{xy} dxdy 
\\
&~ +  \int_1^{n/k}\int_1^{n/k} \frac{n}{k}\frac{C_{i,j,m}(\frac{kx}{n},\frac{ky}{n},\frac{k}{n}) - C_{i,j,m}(1,\frac{ky}{n},\frac{k}{n}) - C_{i,j,m}(\frac{kx}{n},1,\frac{k}{n}) + C_{i,j,m}(1,1,\frac{k}{n})}{xy} dxdy
\\ 
&~ + o(1) 
\end{align*}
where the equality follows from the bounds $C_{i,j,m}(qx,qy,q)/q \leq x \wedge y$. Combining this with~\eqref{eq:hatemixalt2} and~\eqref{eq:unifC} we find that
\begin{align*}
&\Big|\hat e_{i,j}^{(m)} - e_{i,j}^{(m)}(k/n) \Big| 
\\
&\leq \int_{1/k}^{n/k} \int_{1/k}^{n/k} \frac{4}{xy} dxdy \sup_{x,y \in [0,n/k]} \Big|\hat R_{i,j,m}(x,y,1) - \frac{n}{k}C_{i,j,m}(kx/n,ky/n,k/n)\Big| + o(1)
\\
& = O((\log n)^2)O_\bP(n^{-\psi}) + o(1) = o_\bP(1). 
\end{align*} 
Now continuity of $C_{i,j,m}$ together with the dominated convergence theorem imply that for $k/n \to q \in (0,1)$ we also have $e_{i,j}^{(m)}(k/n) \to e_{i,j}^{(m)}(q)$, while for $k/n \to 0$ this follows from the arguments given in the proof of Proposition~\ref{prop:consvar}. In summary, we have established that for $k/n \to q \in [0,1)$ also $\hat e_{i,j}^{(m)} = e_{i,j}^{(m)}(q) + o_\bP(1)$. Combining this with the representations~\eqref{eq:repgamma}, \eqref{eq:repgammahat} and~\eqref{eq:h1} this shows that $\hat \Gamma^{(m)} = \Gamma^{(m)}(q) + o_\bP(1)$. To complete the proof it thus remains to prove~\eqref{eq:unifC}.

\bigskip
\bigskip

\noindent	
\textbf{Proof of~(\ref{eq:unifC})}
We begin with a proof of the following result: for independent observations we have for any $I \subset V, |I| \leq 3$
\begin{equation} \label{eq:unifratehatC}
\sup_{\g x \in [0,1]^{|I|}} \frac{n}{k}\Big| \hat C_I(k \g x/n) - C_I(k \g x/n) \Big| = O_\bP(k^{-1/2}).
\end{equation}
By the results in \cite{CH87} we have for $j \in V$
\[
\sup_{x \in [0,1]} |(n/k)\hat F_j^-(kx/n) - x| = O_\bP(k^{-1/2}).
\]
Thus we have with probability tending to one $\widehat{\g F}^-([0,k/n]^d) \subset [0,2k/n]^d$, which together with Lipschitz continuity of $C$ with Lipschitz constant $1$ implies that with probability tending to one
\begin{align*}
&\sup_{\g x \in [0,1]^d} \frac{n}{k}\Big| \hat C(k \g x/n) - C(k \g x/n) \Big|
\\
\leq & \sup_{\g x \in [0,1]^d}\frac{n}{k}\Big| \hat C^\circ(\widehat{\g F}^-(k\g x/n)) - C(\widehat{\g F}^-(k\g x/n)) \Big| + \sup_{\g x \in [0,1]^d}\frac{n}{k}\Big| C(\widehat{\g F}^-(k\g x/n)) - C(k \g x/n) \Big|
\\
\leq & \sup_{\g x \in [0,2]^d}\frac{n}{k}\Big| \hat C^\circ(k\g x/n) - C(k\g x/n) \Big| + \sup_{\g x \in [0,1]^d} \sum_{j=1}^d \Big|\frac{n}{k}\hat F_j^-(kx_j/n) - x_j\Big|
\\
= &  \sup_{\g x \in [0,2]^d}\frac{n}{k}\Big| \hat C^\circ(k\g x/n) - C(k\g x/n) \Big| + O_\bP(k^{-1/2})
\end{align*}
where we recall that the notation $\hat C^\circ$ was introduced in~\eqref{eq:Ccirc}. Now if $k/n \to q >0$, it follows by standard results about the empirical process indexed by rectangles that 
\[
\sup_{\g x \in [0,2]^d}\frac{n}{k}\Big| \hat C^\circ(k\g x/n) - C(k\g x/n) \Big| = O_\bP(n^{-1/2}) = O_\bP(k^{-1/2}).
\]
When $k/n \to 0$ the bound
\[
\sup_{\g x \in [0,2]^d}\frac{n}{k}\Big| \hat C^\circ(k\g x/n) - C(k\g x/n) \Big| = O_\bP(k^{-1/2})
\]
follows from corresponding results on the tail empirical process. This completes the proof of~\eqref{eq:unifratehatC} and we now continue with the proof of~\eqref{eq:unifC}. Observe that for $v \geq 1$
\[
\frac{n}{k}\{ \hat C(k \g x/n) - C(k \g x/n) \} = v \frac{n}{kv} \{ \hat C(kv (\g x/v)/n) - C(kv (\g x/v)/n) \}. 
\] 
Thus, setting $v = (n/k)^{-\alpha}$, we have for any $1 \geq \alpha > 0$ and any $I \subset V$ 
\begin{equation} \label{eq:hatCCOP}
\sup_{\g x \in [0,(n/k)^\alpha]^d} \frac{n}{k}\Big| \hat C_I(k \g x_I/n) - C_I(k \g x_I/n) \Big| = O_\bP(k^{-\beta}(n/k)^\alpha).
\end{equation}
Note in particular that when $ k/n^{(1+\kappa)/(1+\beta)}$ is bounded away from zero for some $\beta >\kappa > 0$ we can directly set $\alpha = 1$ and obtain
\[
\sup_{\g x \in [0,n/k]^d} \frac{n}{k}\Big| \hat C_I(k \g x_I/n) - C_I(k \g x_I/n) \Big| = O_\bP(n^{-\kappa}).
\]
In particular, this implies that~\eqref{eq:unifC} holds in the case $k/n^{(1+\beta/2)/(1+\beta)}$ bounded away from zero in which case we can set $\psi = \beta/2$.
	
The case $k = o(n^{(1+\beta/2)/(1+\beta)})$ will be discussed next. Observe that for any two functions $f,g: \R^d \to \R$ which are non-decreasing in every coordinate we have for $\g a \leq \g b$ (inequalities are interpreted coordinate-wise) 
\begin{equation}\label{eq:monb}
\sup_{\g x \in [\g a, \g b]} |f(\g x) - g(\g x)| \leq |g(\g b) - g(\g a)| + |g(\g b) - f(\g b)| + |g(\g a) - f(\g a)|.
\end{equation}  
This follows from a combination of the bounds
\begin{align*}
f(\g x) - g(\g x) &\leq f(\g b) - g(\g a) = f(\g b) - g(\g b) + g(\g b) - g(\g a)
\\
f(\g x) - g(\g x) &\geq f(\g a) - g(\g b) = f(\g a) - g(\g a) + g(\g a) - g(\g b).
\end{align*}
For any $\g x \in (\R^+)^{|I|}$ define the vectors $\g v_{k,n}(\g x)$ with entries $(\g v_{k,n}(\g x))_i := n/k$ if $x_i \geq (n/k)^\alpha$ and $x_i$ otherwise and $\g w_{k,n}(\g x)$ with entries $\infty$ if $x_i \geq (n/k)^\alpha$ and $x_i$ otherwise. With this notation we have for $I = (i_1,i_2,i_3)$, uniformly on $[(n/k)^{\alpha},n/k]\times[0,n/k]\times[0,T]$,
\begin{align*}
\frac{n}{k}C_I(k \g v_{k,n}(\g x)/n) &\geq \frac{n}{k}C_I(k \g x/n) 
\geq \frac{n}{k}C_I(k (\g x \wedge (n/k)^\alpha)/n) 
\\
& \stackrel{(a)}{=} R_I(\g x \wedge (n/k)^\alpha) + O((k/n)^{(1-\alpha)\xi - \alpha})
\\
& \stackrel{(b)}{=} R_I(\g w_{k,n}(\g x)) + O((k/n)^{(1-\alpha)\xi - \alpha}+(k/n)^{\gamma\alpha}) 
\\
& \stackrel{(c)}{=} \frac{n}{k}C_I(k \g v_{k,n}(\g x)/n) + O((k/n)^{(1-\alpha)\xi - \alpha}+(k/n)^{\gamma\alpha}).
\end{align*}
Here $(a)$ follows by~\eqref{eq:cbound1}applied with $\delta =\alpha, q = n/k$, $(b)$ follows by~\eqref{eq:tailR3} applied with $\delta =\alpha, q = n/k$ and $(c)$ follows by~\eqref{eq:cbound1} applied with $\delta =\alpha, q = n/k$ and $I = (i_2,i_3)$ when $x_{i_2} < (n/k)^\alpha$ and holds trivially when $x_{i_2} \geq (n/k)^\alpha$ since in that case $R_I(\g w_{k,n}(\g x)) = x_3 = \frac{n}{k}C_I(k \g v_{k,n}(\g x)/n) $.  
	
In summary, we have proved
\begin{multline} \label{eq:uniftailC}
\sup_{|I| = 3} \sup_{\g x\in [(n/k)^{\alpha},n/k]\times[0,n/k]\times[0,T]} \Big|\frac{n}{k}C_I(k \g v_{k,n}(\g x)/n) - \frac{n}{k}C_I(k (\g x \wedge (k/n)^\alpha)/n) \Big|
\\
= O((k/n)^{(1-\alpha)\xi-\alpha} + (k/n)^{\alpha\gamma})
\end{multline} 

Now for any $\g x \in \big([0,n/k]^{2}\backslash [0,(n/k)^\alpha]^{2}\big)\times[0,T]$ apply~\eqref{eq:monb} with $f = \hat C_I, g = C_I, \g a = k (\g x \wedge (n/k)^\alpha)/n, \g b = k \g v_{k,n}(\g x)/n $ to obtain for any $|I| = 3$
\begin{align*}
& \frac{n}{k}\Big| \hat C_I(k \g x/n) - C_I(k \g x/n) \Big|
\\
\leq &\frac{n}{k}\Big|C_I(k \g v_{k,n}(\g x)/n) - C_I(k (\g x \wedge (k/n)^\alpha)/n) \Big| + \frac{n}{k}\Big| \hat C_I(k \g v_{k,n}(\g x)/n) - C_I(k \g v_{k,n}(\g x)/n) \Big|
\\
& \quad + \frac{n}{k}\Big| \hat C_I(k (\g x \wedge (n/k)^\alpha)/n) - C_I(k (\g x \wedge (n/k)^\alpha)/n) \Big|.
\end{align*}
Now by~\eqref{eq:uniftailC} we have (note the supremum in~\eqref{eq:uniftailC} is over all $I$ with $|I| = 3$, so the first two coordinates can be interchanged)
\begin{align*}
&\sup_{|I| =3} \sup_{\g x \in \{[0,n/k]^{2}\backslash [0,(n/k)^\alpha]^{2}\}\times[0,T]} \frac{n}{k}\Big|C_I(k \g v_{k,n}(\g x)/n) - C_I(k (\g x \wedge (k/n)^\alpha)/n) \Big|
\\
= &O((k/n)^{(1-\alpha)\xi-\alpha} + (k/n)^{\alpha\gamma}).
\end{align*}
Next, note that by the definition of $\g v_{k,n}(\g x)$ we have 
\[
\frac{n}{k} \hat C_I(k\g v_{k,n}(\g x)/n) = x_3 + O_\bP(k^{-\beta}) = \frac{n}{k}C_I(k \g v_{k,n}(\g x)/n) + O_\bP(k^{-\beta})
\] 
uniformly in $\g x \in [(n/k)^\alpha,n/k]^2 \times[0,T]$. Moreover, if $I = (i_1,i_2,i_3)$ then for $\g x = (x_1,x_2,x_3) \in [(n/k)^\alpha,n/k]\times[0,(n/k)^\alpha)\times[0,T]$ we have $C_I(\g x) = C_{(i_2,i_3)}(x_2,x_3)$ and the same is true for $\hat C_I$. Thus by~\eqref{eq:hatCCOP}
\begin{align*}
&\sup_{|I| =3} \sup_{\g x \in [(n/k)^\alpha,n/k]\times[0,(n/k)^\alpha)\times[0,T]} \frac{n}{k}\Big| \hat C_I(k \g v_{k,n}(\g x)/n) - C_I(k \g v_{k,n}(\g x)/n) \Big|
\\
\leq&\sup_{|J| = 2} \sup_{\g y \in [0,(n/k)^\alpha]\times[0,T]} \frac{n}{k}\Big| \hat C_J(k \g y/n) - C_J(k \g y/n) \Big|
\\
= &~O_{\bP}(k^{-\beta}(n/k)^\alpha).
\end{align*}
Combining the arguments above we find 
\[
\sup_{|I| =3} \sup_{\g x \in \{[0,n/k]^{2}\backslash [0,(n/k)^\alpha]^{2}\}\times[0,T]} \frac{n}{k}\Big| \hat C_I(k \g v_{k,n}(\g x)/n) - C_I(k \g v_{k,n}(\g x)/n) \Big| = O_{\bP}(k^{-\beta}(n/k)^\alpha + k^{-\beta})
\]
Finally, again by~\eqref{eq:hatCCOP}
\begin{align*}
\sup_{|I| =3} \sup_{\g x \in \{[0,n/k]^{2}\backslash [0,(n/k)^\alpha]^{2}\}\times[0,T]} \frac{n}{k}\Big| \hat C_I(k (\g x \wedge (n/k)^\alpha)/n) - C_I(k (\g x \wedge (n/k)^\alpha)/n) \Big|
= O_\bP(k^{-\beta}(n/k)^\alpha)
\end{align*}
Combining the bounds above we obtain
\[
\sup_{\g x \in [0,n/k]^{|I|-1}\times[0,T]} \Big| \frac{n}{k}\hat C_I(k \g x/n) - \frac{n}{k} C_I(k \g x/n) \Big| = O_\bP(k^{-\beta}(n/k)^\alpha) + O((k/n)^{(1-\alpha)\xi-\alpha} + (k/n)^{\alpha\gamma} + k^{-\beta})
\]
Now recall that we are in the case
\[
n^{\theta} \leq k \leq n^{(1+\beta/2)/(1+\beta)}
\]
where $\theta > 0$ is from the assumptions. Under this assumption we can make $\alpha$ sufficiently small to obtain
\[
O_\bP(k^{-\beta}(n/k)^\alpha) + O((k/n)^{(1-\alpha)\xi-\alpha} + (k/n)^{\alpha\gamma} + k^{-\beta}) = o_\bP(n^{-\zeta})
\]
for some $\zeta > 0$. This completes the proof of~\eqref{eq:unifC}. \hfill $\Box$

\subsection{Proofs of alternative representations}\label{sec:altrepr}

\textbf{Proof of~\eqref{eq:epowalt} and~\eqref{eq:epowalt-q}} Recall the following representation for the expected value of a non-negative random variable $X$
\[
\E[X] = \int_{[0,\infty)} \bP(X > x) dx.
\]
The claim in~\eqref{eq:epowalt} follows by applying this representation to the non-negative random variables $(\log Y_i^m )^\ell \einsfun\{Y_i^m > 1\}$ and 
$
(-\log(Y_i^m))^\ell \einsfun\{Y_i^m \leq 1\}
$ 
and collecting terms. For example
\begin{align*}
\E[ (\log(Y_i^m))^\ell \einsfun\{Y_i^m > 1\} ] &= \int_{[0,\infty)} \bP\Big((\log(Y_i^m))^\ell \einsfun\{Y_i^m > 1\} > x\Big) dx 
\\
& = \int_{[0,\infty)} \bP\Big(Y_i^m\einsfun\{Y_i^m > 1\} > \exp(x^{1/\ell})\Big) dx
\\
& = \int_{[0,\infty)} \bP\Big(Y_i^m > \exp(x^{1/\ell})\Big) dx
\\
& = \int_{[0,\infty)} R_{i,m}(\exp(-x^{1/\ell}),1) dx
\\
& = \int_{(0,1]} \frac{R_{i,m}(x,1)\ell(- \log x)^{\ell-1}}{x} dx
\end{align*}
where the last equality follows with the substitution $u = \exp(-x^{1/\ell})$. Similar arguments show that
\begin{align*}
\E[ (- \log(Y_i^m))^\ell \einsfun\{Y_i^m \leq 1\} ] &= \int_{[0,\infty)} \bP\Big((-\log(Y_i^m))^\ell \einsfun\{Y_i^m \leq 1\} > x\Big) dx 
\\
& = \int_{[0,\infty)} \bP\Big(- \log Y_i^m \einsfun\{Y_i^m \leq 1\} > x^{1/\ell}\Big) dx
\\
& = \int_{[0,\infty)} \bP\Big( \log Y_i^m \einsfun\{Y_i^m \leq 1\} < - x^{1/\ell}\Big) dx
\\
& = \int_{[0,\infty)} \bP\Big( \log Y_i^m < - x^{1/\ell}\Big) dx
\\
& = - \int_{[0,\infty)} \bP\Big( \log Y_i^m \geq - x^{1/\ell} \Big) - 1 dx
\\
& = - \int_{[0,\infty)} R_{i,m}(\exp(x^{1/\ell}),1) - 1 dx
\\
& = - \int_{[1,\infty)} \frac{(R_{i,m}(x,1)-1)\ell(\log x)^{\ell-1}}{x} dx.
\end{align*}
Finally note that
\[
\E[(\log Y_i^m)^\ell] = \E[(\log Y_i^m)^\ell\einsfun\{Y_i^m > 1\}] + (-1)^{\ell}\E[(-\log Y_i^m)^\ell \einsfun\{Y_i^m \leq 1\}]
\]
and~\eqref{eq:epowalt} follows by collecting terms. The claim in~\eqref{eq:epowalt-q} follows by similar arguments and details are omitted for the sake of brevity. \hfill $\Box$

\bigskip
\noindent
\textbf{Proof of~\eqref{eq:emixalt-jj},~\eqref{eq:emixalt-jj-q},~\eqref{eq:emixalt-jm} and~\eqref{eq:emixalt-jm-q}}
Since the proofs of all statements are similar we only outline the proof of~\eqref{eq:emixalt-jm}. To this end observe that for non-negative random variables $X,Y$ we have
\[
\E[XY] = \int_{[0,\infty)^2} \bP(X>x,Y>y) dxdy.
\]
For a proof, note that
\begin{align*}
\E[XY] &= \int_{\Omega} X(\omega)Y(\omega) d\bP(\omega) = \int_{\Omega} \int_{[0,\infty)^2} \einsfun_{(0,X(\omega))\times(0,Y(\omega))}(x,y) dxdy d\bP(\omega)
\\
& =  \int_{[0,\infty)^2} \int_{\Omega} \einsfun_{(0,X(\omega))\times(0,Y(\omega))}(x,y) d\bP(\omega) dxdy = \int_{[0,\infty)^2} \bP(X>x,Y>y) dxdy
\end{align*}
where the order of integration can be interchanged by the Tonelli theorem since the integrand is non-negative. Using this representation and similar computations as in the proof of~\eqref{eq:epowalt} show that
\begin{align*}
&\E[(-\log Y_i^m)(-\log Y_j^m) \einsfun\{Y_i^m \leq 1, Y_j^m \leq 1\}] = \int_0^1\int_0^1 \frac{R_{i,j,m}(x,y,1)}{xy} dxdy 
\\
&\E[(\log Y_i^m)(-\log Y_j^m) \einsfun\{Y_i^m > 1, Y_j^m \leq 1\}]= - \int_0^1\int_1^{\infty} \frac{R_{i,j,m}(x,y,1) - R_{i,j,m}(\infty,y,1)}{xy} dxdy
\\
&\E[(-\log Y_i^m)(\log Y_j^m) \einsfun\{Y_i^m \leq 1, Y_j^m > 1\}]= - \int_1^{\infty}\int_0^1 \frac{R_{i,j,m}(x,y,1) - R_{i,j,m}(x,\infty,1)}{xy} dxdy
\end{align*}
and
\begin{multline*}
\E[\log Y_i^m \log Y_j^m \einsfun\{Y_i^m > 1, Y_j^m > 1\}]
\\
=  \int_1^{\infty}\int_1^{\infty} \frac{R_{i,j,m}(x,y,1) - R_{i,j,m}(\infty,y,1) - R_{i,j,m}(x,\infty,1) + R_{i,j,m}(\infty,\infty,1)}{xy} dxdy.
\end{multline*}
Combining those expressions we obtain~\eqref{eq:emixalt-jm}. \hfill $\Box$

\bigskip
\bigskip

\noindent
\textbf{Proof of~(\ref{eq:hatemixalt}),~(\ref{eq:hatepowalt}) and~(\ref{eq:hatemixalt2})}
Begin by defining for $I = (i_1,\dots,i_j)$
\[
\check R_{I}(\g x) := \frac{1}{k} \sum_{t=1}^n I\Big\{ \hat F_{i_1}(U_{ti_1}) \leq kx_1/n,\ldots, \hat F_{i_j}(U_{ti_j}) \leq kx_j/n \Big\}.
\]
We have almost surely, {as $n \to \infty$,} 
\begin{equation} \label{eq:R-Rcheck}
\sup_{\g x \in [0,n/k]^{|I|}} \Big| \widehat R_{I}(\g x) - \check R_{I}(\g x) \Big| = O(1/k),
\end{equation}
this follows for instance from equation (3) in \cite{RWZ17} and the following discussion.

Next consider any integrable function $g$ with anti-derivative $G$ such that $G(1) = 0$. Then, defining $\hat U_{ti} := \hat F_{i}(U_{ti})$ and noting that by definition $1 \geq \hat U_{ti} \geq 1/n$,
\begin{align*}
\int_{1/k}^1 I\Big\{ \hat U_{ti} \leq kx/n\Big\} g(x) dx &= \int_{[n\hat U_{ti}/k,1 \vee (n\hat U_{ti}/k)]} g(x) dx
= - G(n\hat U_{ti}/k)I\Big\{ \hat U_{ti} \leq k/n\Big\}, 
\\
\int_{1}^{n/k} I\Big\{ \hat U_{ti} > kx/n\Big\} g(x) dx &= \int_{[1 \wedge (n\hat U_{ti}/k),n\hat U_{ti}/k)} g(x) dx =  G(n\hat U_{ti}/k)I\Big\{ \hat U_{ti} > k/n\Big\}.
\end{align*} 
Combining the above we find
\begin{equation}
\int_{1/k}^1 I\Big\{ \hat U_{ti} \leq kx/n\Big\} g(x) dx + \int_{1}^{n/k} \Big( I\Big\{ \hat U_{ti} \leq kx/n\Big\} -1 \Big) g(x) dx = - G(n\hat U_{ti}/k). \label{eq:reprint} 
\end{equation}
To obtain~\eqref{eq:hatepowalt} apply this result with $G(x) = (-\log x)^\ell$ to find that
\begin{align*}
&\frac{1}{k} \sum_{t=1}^n G\Big(\frac{n \hat F_{i}(U_{ti})}{k}\Big) I\{\hat F_j(U_{tj}) \leq k/n \}
\\
= &- \frac{1}{k} \sum_{t=1}^n \Big(\int_{1/k}^1 I\Big\{ \hat U_{ti} \leq kx/n\Big\} g(x) dx + \int_{1}^{n/k} \Big( I\Big\{ \hat U_{ti} \leq kx/n\Big\} -1 \Big) g(x) dx \Big)I\{\hat F_j(U_{tj}) \leq k/n \}
\\
= &- \int_{1/k}^{1} \check R_{ij}(x,1) g(x) dx - \int_{1}^{n/k} (\check R_{ij}(x,1) - 1)g(x) dx
\end{align*} 
where we used the equality $\sum_{t=1}^n I\{\hat F_j(U_{tj}) \leq k/n \} = k$ in the last line (note that by independence across $t$ all $U_{tj}, t=1,\dots,n$ take different values with probability one). Apply the above equality with $G(x) = (-\log x)^\ell$ to obtain
\[
\hat e_{i}^{(m),\ell} = \int_{1/k}^{1} \frac{\check R_{ij}(x,1) \ell (- \log x)^{\ell-1}}{x} dx + \int_{1}^{n/k} \frac{(\check R_{ij}(x,1) - 1)\ell (- \log x)^{\ell-1}}{x} dx.
\]
Now~\eqref{eq:hatepowalt} follows by an application of~\eqref{eq:R-Rcheck}. The proofs of~\eqref{eq:hatemixalt} and~\eqref{eq:hatemixalt2} follow by very similar arguments using the function $G(x) = - \log x$ in~\eqref{eq:reprint} and details are omitted for the sake of brevity. \hfill $\Box$

\section{Proof of Theorem~\ref{th:consMST-hd} and Theorem~\ref{th:consMSTGamma-hd}.}

\subsection{Preliminaries for the proofs of Theorem~\ref{th:consMST-hd} and Theorem~\ref{th:consMSTGamma-hd}.} 

Let $\rho_{ij}: i \neq j \in V$ be arbitrary numbers satisfying $\rho_{ij}  = \rho_{ji}$. Consider trees $(V,E') = T' \neq T = (V,E)$. Define
\begin{equation}\label{eq:defmurho}
\mu_\rho := \min_{(h,l) \notin E} \min_{(i,j) \in \ph(hl;T)} (\rho_{hl} - \rho_{ij} ).  
\end{equation}
Then
\begin{equation}\label{eq:lowboundrhodiff}
\sum_{(i,j)\in E'} \rho_{ij} - \sum_{(i,j)\in E} \rho_{ij} \geq |E \setminus E'| \mu_\rho.
\end{equation}  
To see this recall that in the proof of Proposition~\ref{prop:chimst} we construct a bijective map $\tau: E \to E'$ such that for each $(h,l) = \tau((i,j))$ it holds that $(i,j) \in \ph(hl;T)$. By the definition of $\mu_\rho$ we have for $\tau(i,j) \notin E$
\[
\rho_{\tau(i,j)} - \rho_{ij} \geq \mu_\rho.
\]
Since $\tau$ is bijective we can write
\[
\sum_{(i,j)\in E'} \rho_{ij} - \sum_{(i,j)\in E} \rho_{ij}
= \sum_{(i,j)\in E} \rho_{\tau(i,j)} - \sum_{(i,j)\in E} \rho_{ij} \geq |E' \setminus E| \mu_\rho,
\]
where the inequality follows since there are exactly $|E' \setminus E|$ terms in the first sum where $\tau(i,j) \notin E$. This completes the proof of~\eqref{eq:lowboundrhodiff}.

Next assume that $\hat \rho_{ij}$ are estimators for $\rho_{ij}$, that $\hat T_\rho = (V, \hat E)$ denotes the minimal spanning tree with respect to $\hat \rho_{ij}$ and that $T = (V,E)$ is the minimal spanning tree with respect to $\rho_{ij}$. Then 
\begin{equation}\label{eq:rhomstprob} 
\bP\big(\hat T_\rho \neq T\big) \leq \bP\Big( \max_{(i,j): i\neq j, i,j \in V } |\hat\rho_{ij} - \rho_{ij}| \geq \frac{\mu_\rho}{2} \Big).
\end{equation}
For a proof assume that $\hat T_\rho \neq T$. Since $\hat T_\rho$ is the minimal spanning tree with weights $\hat \rho_{ij}$ we must have
\[
\sum_{(i,j) \in \hat E} \hat\rho_{ij} - \sum_{(i,j) \in  E} \hat\rho_{ij} \leq 0.
\]
This can be rewritten as
\begin{align*}
0 \geq & \sum_{(i,j) \in \hat E} \hat\rho_{ij} - \sum_{(i,j) \in  E} \hat\rho_{ij}\notag
\\
= & \sum_{(i,j) \in \hat E} \{\hat\rho_{ij} - \rho_{ij}\} - \sum_{(i,j) \in E } \{ \hat\rho_{ij} - \rho_{ij}\} + \sum_{(i,j) \in \hat E} \rho_{ij} - \sum_{(i,j) \in E } \rho_{ij}\notag
\\
= & \sum_{(i,j) \in \hat E\setminus E} \{\hat\rho_{ij} - \rho_{ij}\} - \sum_{(i,j) \in E \setminus \hat E} \{ \hat\rho_{ij} - \rho_{ij}\} + \sum_{(i,j) \in \hat E} \rho_{ij} - \sum_{(i,j) \in E} \rho_{ij}\notag
\\
\geq & - 2 |\hat E\setminus E| \max_{(i,j): i\neq j, i,j \in V } |\hat\rho_{ij} - \rho_{ij}| + |\hat E\setminus E| \mu_\rho\notag
\\
= & |\hat E\setminus E|\Big( - 2\max_{(i,j): i\neq j, i,j \in V } |\hat\rho_{ij} - \rho_{ij}| + \mu_\rho  \Big). 
\end{align*}
Here we used \eqref{eq:lowboundrhodiff} in the second line from below. Rearrange terms to obtain
\[
\max_{(i,j): i\neq j, i,j \in V } |\hat\rho_{ij} - \rho_{ij}| \geq \frac{\mu_\rho}{2}.
\]
This shows~\eqref{eq:rhomstprob}.

\subsection{Proofs of Theorem~\ref{th:consMST-hd}.}

A crucial ingredient in the proof is the following concentration bound for $\hat \chi_{ij}$.

\begin{proposition}\label{prop:chiconc}
	There exists a universal constant $K$ such that for all $s >0$ we have
	\[
	\bP\Big( |\hat \chi_{ij} - \chi_{ij}(k/n)| > s \Big) \leq 5 \exp\Big(-\frac{3k}{10}\Big\{\frac{s^2}{K^2}\wedge 1 \Big\} \Big).
	\]
\end{proposition} 

This result is proved separately further below. We now move on to the main result. Recall that Prim's algorithm returns the minimal spanning tree and that this algorithm depends only on the relative order of $\rho_{ij}$ and not on their exact values. Thus the minimal spanning trees $\hat T_\chi$ corresponding to $\hat\rho_{ij} = -\log \hat\chi_{ij}$ and $\tilde T_\chi$ corresponding to $\tilde \rho_{ij} = - \hat \chi_{ij}$ are the same. 
{We distinguish two cases. If $\mu_\chi^Y = 0$, the bound in the theorem is trivial and there is nothing to prove. Hence assume $\mu_\chi^Y > 0$. In this case the inequality in~\eqref{eq:orderchi2} is strict and $T$ is the unique minimal spanning tree with respect to the distances $-\chi_{ij}^Y$.} Hence also
\[
\bP(\hat T_\chi \neq T) = \bP(\tilde T_\chi \neq T).
\]
Apply~\eqref{eq:rhomstprob} with $\tilde \rho_{ij} = -\hat\chi_{ij}, \rho_{ij} = -\chi_{ij}^Y$ to obtain
\begin{align*}
\bP\big(\tilde T_\chi \neq T\big) &\leq \bP\Big( \max_{(i,j): i\neq j, i,j \in V } |\hat\chi_{ij} - \chi_{ij}^Y| \geq \frac{\mu_\chi^Y}{2} \Big)
\\
&\leq \bP\Big( \max_{(i,j): i\neq j, i,j \in V } |\hat\chi_{ij} - \chi_{ij}(k/n)| \geq \frac{\mu_\chi^Y}{2} - \delta_{k/n}\Big).
\end{align*}
Combined with Proposition~\ref{prop:chiconc} this yields
\[
\bP\big(\hat T_\chi \neq T\big) \leq 5d^2 \exp\Big(-\frac{3k}{10}\Big\{\Big(\frac{\mu_\chi^Y - 2\delta_{k/n}}{2K}\Big)_+^2\wedge 1 \Big\} \Big).
\]
This completes the proof. \hfill $\Box$

\subsection{Proof of Theorem~\ref{th:consMSTGamma-hd} }\label{sec:proofth:consMSTGamma-hd}

We will give a detailed proof for $\hat T_\Gamma^{(m)}$ and only provide an outline for $\hat T_\Gamma^{w}$ since both arguments are very similar. Let $\rho_{ij} = \Gamma_{ij}^{(m)}, \hat \rho_{ij} = \hat \Gamma_{ij}^{(m)}$. {If $\min_{(i,j) \in E} \Gamma_{ij}^{(m)} = 0$ the statement of the theorem is trivial and there is nothing to prove. Hence, in what follows assume that $\min_{(i,j) \in E} \Gamma_{ij}^{(m)} > 0$, in which case T is the unique minimal spanning tree with respect to $\rho_{ij} = \Gamma_{ij}^{(m)}$.} Recall the definition of $\mu_\rho$ in~\eqref{eq:defmurho} and apply Proposition~\ref{prop_tree_gamma} to derive that
\[
\mu_\rho \geq \min_{i\neq j} \Gamma_{ij}^{(m)}. 
\]
Apply~\eqref{eq:rhomstprob} to obtain
\begin{equation}\label{eq:gammaconsequence}
\bP\big(\hat T_\Gamma^{(m)} \neq T\big) \leq \bP\Big( \max_{(i,j): i\neq j, i,j \in V } |\hat\Gamma^{(m)}_{ij} - \Gamma^{(m)}_{ij}| \geq \frac{\min_{(i,j) \in E} \Gamma_{ij}^{(m)}}{2} \Big).
\end{equation}

The next ingredient for the proof are non-asymptotic bounds for $|\hat \Gamma_{ij}^{(m)} - \Gamma_{ij}^{(m)}|$ which are established in \cite{ELV2021}. {More precisely, the latter paper makes the following assumptions (the numbering below corresponds to the numbering in \cite{ELV2021}): for each ordered $I \subset V$ with $|I| \in \{2,3\}$ there exist functions $R_I: [0,\infty)^{|I|} \to [0,\infty)$ with the properties below. For simplicity of notation and to be consistent with the notation in \cite{ELV2021}, we write $R_{ij}$ for $I = (i,j)$.

\medskip

Assumption 2 (Bounded densities) For each $i,j \in V, i\neq j$ the functions $R_{ij}$ have mixed partial derivatives $r_{ij}$ satisfying
\begin{equation} \label{eq:densityR_new}
r_{ij}(x, y) := \frac{\partial^2}{\partial x \partial y} R_{ij}(x, y) \leq \frac{K(\beta)}{x^\beta y^{1-\beta}}, \quad (x, y) \in (0, \infty)^2,
\end{equation}
for constants $K(\beta)$ and every $\beta \in [-\eps, 1+\eps]$, for some some $\eps>0$.

\medskip	
	
Assumption 3 (Second order) The marginal distributions $F_1,\dots,F_d$ are continuous and there exist positive constants $K', \xi'$ such that for all $J \subset V$, $|J| \in \{2,3\}$ and $q \in (0, 1]$,
\begin{equation}\label{eq:RJstand}
\sup_{\g x \in [0,1]^{|J|}} \Big|q^{-1} \bP(F_J(\g X_J) > 1 - q\g x)- R_J(\g x) \Big| \leq K'q^{\xi'}.
\end{equation}

\medskip

Assumption 4 (Tail) There exist positive constants $K_T, \xi_T$ such that for all $i \neq j \in V$ and $q \in (0, 1]$,
\begin{equation}\label{eq:tailR}
1 - R_{ij}(q^{-1}, 1) \leq K_T q^{\xi_T}.
\end{equation}

}

{
We will now show that under conditions (B), (T) the vector $\g X$  satisfies Assumption~3 and Assumption~4 above with $R_{I}(\g x)$ defined through~\eqref{eq:defR} on $[0,1]^{|I|}$ and extended to $[0,\infty)^{|I|}$ through $R_I(c \g x) = c R_I(\g x)$. We will further show that those $R_I$ satisfy~\eqref{eq:densityR_new} in Assumption 2 under assumption (D).   

Assumption 3 directly follows from~\eqref{eq:bounRnkR'} with $K',\xi'$ from Assumption 3 satisfying $K' = K_R, \xi' = \xi$. Note that~\eqref{eq:bounRnkR'} further follows from (B) with $\gamma_1 = \xi$ and the limits $R_I$ equal the $R_I$ defined in the present paper. The latter also implies that Assumption 4 is a direct consequence of~\eqref{eq:tailRij'}, and thus (T).

To see that~\eqref{eq:densityR_new} in Assumption 2 follows from (D), observe that }
\[
R_I(\g x) = \ell_I(1)\mathbb P({\g Y_{(I)}} \geq \g x).
\]
Hence if ${\g Y_{(I)}}$ has density $f_I$ then $R_I$ has density
\[
r_I(x,y) = \frac{f_I(1/x,1/y)}{\ell_I(1)x^2y^2}.
\] 
Note that by homogeneity of ${\g Y_{(I)}}$ the density $f_I$ satisfies 
\begin{equation}\label{eq:fIhom}
f_I(cx,cy) = c^{-3} f_I(x,y)
\end{equation} 
for all $c,x,y$ such that $(x,y), (cx,cy) \in \mathcal L$. Thus $r_I(x,y) = xy f_I(y,x)$. Hence the assumption in (D) on $r_I$ is equivalent to requiring  
\begin{equation}\label{eq:densequiv}
f_I(x,y) \leq \ell_I(1) K(\beta) \frac{1}{y^{1+\beta}x^{2+\beta}}.
\end{equation}
This shows the equivalence between (D) and~\eqref{eq:densityR_new}. 

{ The discussion above combined with an application of Proposition~1 and Theorem~3 in~\cite{ELV2021} implies the following result. } 

\begin{theorem}[\cite{ELV2021}]\label{thm:elv}
Let (B), (T) hold and $\zeta \in (0, 1]$ be arbitrary. Let $\kappa = \gamma\xi/(1+\gamma+\xi)$. There exist positive constants $C$, $c$ and $M$ only depending on $K$, $\kappa$ and $\zeta$ such that for any $k \geq n^\zeta$ and $\lambda \leq \sqrt{k}/(\log n)^4$,
\[
\mathbb P\bigg( \max_{i,j,m \in V} | \hat \Gamma_{i,j}^{(m)} - \Gamma_{i,j}^{(m)} | > C \Big\{ \Big(\frac{k}{n}\Big)^\kappa (\log(n/k))^2 + \frac{(\log(n/k))^2(1 + \lambda)}{\sqrt{k}} \Big\} \bigg) \leq M d^3 e^{-c\lambda^2}.
\]
If in addition (D) holds, there exists a positive constant $\bar C$ only depending on $K$, $\kappa$, $\zeta$, $\eps$ and $K(\beta)$ such that for any $k$ and $\lambda$ as above,
\[
\mathbb P \bigg( \max_{i,j,m \in V} | \hat \Gamma_{i,j}^{(m)} - \Gamma_{i,j}^{(m)} | > \bar C \Big\{ \Big(\frac{k}{n}\Big)^\kappa (\log(n/k))^2 + \frac{1 + \lambda}{\sqrt{k}} \Big\} \bigg) \leq M d^3 e^{-c\lambda^2}.
\]
	
\end{theorem}

To lighten notation let $b_{n,k} := (k/n)^\kappa (\log(n/k))^2$ and assume $\lambda > 1$ so that $1+\lambda$ can be replaced by $\lambda$ at the cost of possibly enlarging the constant $C$ by a factor of at most $2$. Then the bound in the second part can be reformulated as follows: for all $C(k^{-1/2}+ b_{n,k}) \leq t \leq C[(\log n)^{-4}+b_{n,k}]$  
\[
\mathbb P \Big( \max_{i,j,m \in V} | \hat \Gamma_{i,j}^{(m)} - \Gamma_{i,j}^{(m)} | > t \Big) \leq M d^3 \exp(-ck[t/C-b_{n,k}]^2).
\]
Thus for all $t \geq C(k^{-1/2}+ b_{n,k})$
\[
\mathbb P \Big( \max_{i,j,m \in V} | \hat \Gamma_{i,j}^{(m)} - \Gamma_{i,j}^{(m)} | > t \Big) \leq M d^3\Big\{ \exp(-ck[t/C-b_{n,k}]^2) \wedge \exp(-ck/(\log n)^8) \Big\}.
\]
Combine this with~\eqref{eq:gammaconsequence} to obtain that under the conditions of the second part
\[
\bP\big(\hat T_\Gamma^{(m)} \neq T\big)
\leq  
M d^3\Big\{ \exp(-ck[\min_{(i,j) \in E} \Gamma_{ij}^{(m)}/2C-b_{n,k}]^2) \wedge \exp(-ck/(\log n)^8) \Big\}
\]
and under (A1) and (A2) 
\[
\bP\big(\hat T_\Gamma^{(m)} \neq T\big) \leq \Big\{d\exp\Big(- k\frac{(\min_{(i,j) \in E} \Gamma_{ij}^{(m)})^2}{4 \tilde C^2}  \Big)\Big\} \vee \Big\{7d^4\exp\Big(- \sqrt{k}\Big)\Big\}.
\]
This completes the proof for $\hat T_\Gamma^{(m)}$. 

The proof for $\hat T_\Gamma^w$ proceeds similarly and we only provide an outline. Let $\rho_{ij}^w := \sum_{m=1}^d w_m \Gamma^{(m)}_{ij}$ and $\hat \rho_{ij}^w := \sum_{m=1}^d w_m \hat \Gamma^{(m)}_{ij}$. Then
\[
\rho_{ij}^w = \sum_{m=1}^d w_m \Gamma^{(m)}_{ij} = \sum_{m=1}^d w_m \sum_{(s,t) \in \ph(ij;T)} \Gamma^{(m)}_{st} = \sum_{(s,t) \in \ph(ij;T)} \sum_{m=1}^d w_m \Gamma^{(m)}_{st} = \sum_{(s,t) \in \ph(ij;T)} \rho_{st}^w. 
\] 
This shows that for $\mu_\rho$ defined in~\eqref{eq:defmurho} we have
\[
\mu_\rho \geq \min_{(i,j) \in E} \sum_{m=1}^d w_m \Gamma^{(m)}_{ij}.
\]
Apply~\eqref{eq:rhomstprob} to obtain
\begin{equation}\label{eq:gammawconsequence}
\bP\big(\hat T_\Gamma^w \neq T\big) \leq \bP\Big( \max_{(i,j): i\neq j, i,j \in V } |\hat\rho^w_{ij} - \rho^w_{ij}| \geq \frac{1}{2} \min_{(i,j) \in E} \sum_{m=1}^d w_m \Gamma^{(m)}_{ij}\Big).
\end{equation}
Moreover, note that under the assumption $w_m \geq 0, \sum_{m=1}^d w_m = 1$ we have
\[
\max_{(i,j): i\neq j, i,j \in V } |\hat\rho^w_{ij} - \rho^w_{ij}| \leq \max_{(i,j),m: i\neq j, i,j,m \in V }\Big| \hat\Gamma_{ij}^{(m)} - \Gamma_{ij}^{(m)} \Big| \sum_{m=1}^d w_m =  \max_{(i,j),m: i\neq j, i,j,m \in V }\Big| \hat\Gamma_{ij}^{(m)} - \Gamma_{ij}^{(m)} \Big|. 
\]
Thus 
\[
\bP\big(\hat T_\Gamma^w \neq T\big) \leq \bP\Big( \max_{(i,j),m: i\neq j, i,j,m \in V }\Big| \hat\Gamma_{ij}^{(m)} - \Gamma_{ij}^{(m)} \Big| \geq \frac{1}{2}\min_{(i,j) \in E} \sum_{m=1}^d w_m \Gamma^{(m)}_{ij} \Big).
\]
Noting that the bounds from Theorem~\ref{thm:elv} are uniform in $i,j,m$ completes the proof by exactly the same arguments as for $\hat T_\Gamma^{(m)}$. 
\hfill $\Box$

\subsection{Proof of Proposition~\ref{prop:chiconc}}

We begin by introducing some useful notation. Define the random variables $ U_i := 1 - F_i(X_i)$. Denote the joint distribution of $\g U := (U_1,\dots,U_d)$ by $C$ and the joint distribution of $(U_i,U_j)$ by $C_{ij}$.
Next define the random variables $U_{ti} := 1 - F_i(X_{ti})$ (here $X_{ti}$ denotes the $i$'th entry of the vector $\g X_t$) and the vectors $\g U_t := (U_{t1},\dots,U_{td})^\top$. Denote by $\hat F_i$ the empirical distribution function $U_{1i},\dots,U_{ni}$. Define the vector $ \widehat{\g F}_{ij}^-(\g x) := (\hat F_i^-(x_1), \hat F_j^-(x_2))$, the function
\begin{equation} \label{eq:Ccirc}
\hat C^\circ(\g x) := \frac{1}{n} \sum_{t=1}^n I\{U_{ti} \leq x_1,U_{tj} \leq x_2\}
\end{equation}
and $\hat C_{ij}(k \g x/n) := \hat C^\circ_{ij}(\widehat{\g F}_{ij}^-(k\g x/n))$. With this notation we have the decomposition
\begin{align}
\Big|\hat \chi_{ij} - \chi_{ij}(k/n)\Big| \notag 
=~& \frac{n}{k} \sum_{t = 1}^n \einsfun\{\tilde F_i(X_{ti}) >1- k/n , \tilde F_{j}(X_{tj}) > 1 - k/n\} - \chi_{ij}(k/n) \notag
\\
=~& \frac{n}{k} \hat C_{ij}(k/n,k/n) - \frac{n}{k} C_{ij}(k/n,k/n) \notag
\\
\leq~& \frac{n}{k}\Big|\hat C^\circ_{ij}(\widehat F_i^-(k/n),\widehat F_j^-(k/n)) - C_{ij}(\widehat F_i^-(k/n),\widehat F_j^-(k/n))\Big| \notag
\\
& + \frac{n}{k} \Big|C_{ij}(\widehat F_i^-(k/n),\widehat F_j^-(k/n)) - C_{ij}(k/n,k/n) \Big| \notag
\\
\leq~& \frac{n}{k}\Big|\hat C^\circ_{ij}(\widehat F_i^-(k/n),\widehat F_j^-(k/n)) - C_{ij}(\widehat F_i^-(k/n),\widehat F_j^-(k/n))\Big| \notag
\\
& + \frac{n}{k} \Big|\widehat F_i^-(k/n) - k/n\Big| + \frac{n}{k}\Big|\widehat F_j^-(k/n)) - k/n\Big| \label{eq:decompchiij}
\end{align}
where we used Lipschitz continuity of $C_{ij}$ in the last line. Observe that 
\[
\widehat F_i^-(k/n) = U_{n:k,i}
\]
where $U_{n:k,i}$ denotes the $k$'th order statistic of the sample $U_{1i},\dots,U_{ni}$. 
Now by Inequality 1 in Chapter 11.3 of~\cite{shorack2009empirical} we find that 
\begin{align*}
\bP\Big(\frac{n}{k}\Big|U_{n:k,i} - \frac{k}{n+1}\Big| \geq tk^{-1/2}  \Big)
= \bP\Big(\sqrt{n}\Big|U_{n:k,i} - \frac{k}{n+1}\Big| \geq t\sqrt{k/n}  \Big)
\leq 2\exp\Big(-\frac{t^2}{2}\tilde\psi(t/\sqrt{k})\Big).
\end{align*}
Here 
\[
\tilde \psi(x) := \frac{2}{x^2}[x - \log(1+x)] 
\]
satisfies $\tilde\psi(0) = 1$, moreover $\tilde\psi$ is non-negative and decreasing on $[0,\infty)$ by Proposition 1 in Chapter 11.3 of~\cite{shorack2009empirical}. Hence with probability at least $1-4\exp(-\tilde\psi(1)\lambda^2/2)$ we have for $0 \leq \lambda k^{-1/2} \leq 1$
\[
\frac{n}{k} \Big|\widehat F_i^-(k/n) - k/n\Big| + \frac{n}{k}\Big|\widehat F_j^-(k/n)) - k/n\Big| \leq 2\lambda k^{-1/2} + 2 \frac{n}{k}\Big(\frac{k}{n} - \frac{k}{n+1} \Big) \leq 2\lambda k^{-1/2} + 2n^{-1}.
\] 
Defining the event 
\[
\Omega_1(\lambda) := \Big\{ \frac{n}{k} \Big|\widehat F_i^-(k/n) - k/n\Big| + \frac{n}{k}\Big|\widehat F_j^-(k/n)) - k/n\Big| \leq 4\lambda k^{-1/2} \Big\}
\]
we find that for $1 \leq \lambda \leq k^{1/2}$
\begin{equation}\label{eq:expboundquant}
\bP\Big( \Omega_1(\lambda) \Big) \geq 1-4\exp(-\tilde\psi(1)\lambda^2/2) \geq 1 - 4\exp(-3\lambda^2/10). 
\end{equation}
Further note that on $\Omega_1(\lambda)$ we have 
\[
\max(\widehat F_i^-(k/n), \widehat F_j^-(k/n)) \leq 5 k/n
\]
provided that $1 \leq \lambda \leq k^{1/2}$. Under this condition we obtain on $\Omega_1(\lambda)$
\[
\frac{n}{k}\Big|\hat C^\circ_{ij}(\widehat F_i^-(k/n),\widehat F_j^-(k/n)) - C_{ij}(\widehat F_i^-(k/n),\widehat F_j^-(k/n))\Big| \leq \sup_{x,y \in [0,5k/n]^2} \frac{n}{k}\Big|\hat C^\circ_{ij}(x,y) - C_{ij}(x,y)\Big|.
\]
Note that the latter supremum can be rewritten as
\[
\sup_{x,y \in [0,5k/n]^2} \frac{n}{k}\Big|\hat C^\circ_{ij}(x,y) - C_{ij}(x,y)\Big| = \sup_{f \in \mathcal{F}_{n,k}} |\bP_n f - \bP f|
\]
where $\bP_n$ denotes the empirical measure of the sample $(U_{ti}, U_{tj})_{t=1,\dots,n}$, $\bP f := \E[f(U_i,U_j)]$ for $(U_i,U_j) \sim C_{ij}$, $\bP_n f := n^{-1}\sum_{i=1}^n f(U_{ti}, U_{tj})$, and
\[
\mathcal{F}_{n,k} := \Big\{f: [0,1]^2 \to \R: (u,v) \mapsto \frac{n}{k}\einsfun\{u \leq kx/n, v \leq ky/n\} \Big| x,y \in [0,5k/n]^2\Big\}.
\]
The function class $\mathcal{F}_{n,k}$ is VC-subgraph (see section 2.6 in \cite{VW96}) and by Theorem 2.6.7 in \cite{VW96} we obtain
\begin{equation} \label{eq:entr}
N(\eps,\mathcal{F}_{n,k},L_2(\bP_n)) \leq \Big(\frac{A \|F\|_{L^2(\bP_n)}}{\eps} \Big)^V
\end{equation}
a.s. for constants $A,V$ independent of $n,k$. Moreover any $f \in \mathcal{F}_{n,k}$ satisfies the bound
\begin{equation} \label{eq:envelope}
|f(u,v)| \leq F(u,v) := \frac{n}{k}\einsfun\{u \leq 5k/n, v \leq 5k/n \} \leq \frac{n}{k}
\end{equation}
and we have $\|F\|_{L^2(\bP_n)} \leq \sqrt{5n/k}$ as well as
\begin{equation} \label{eq:var}
\sigma^2_{n,k} := \sup_{f \in \mathcal{F}_{n,k}} \bP f^2 = (n/k)^2 \E[\einsfun\{U_i \leq 5k/n, U_j \leq 5k/n \}] = \E[F(U_i,U_j)^2] = \|F\|^2_{L^2(\bP)}
\end{equation}
Apply the symmetrization inequality (see the second paragraph in section~2.2 of \cite{K06}) and inequality (2.2) from \cite{K06} to obtain for a universal constant $c_0$ and a constant $K_1$ that does not depend on $i,j,k,n$ 
\begin{align} 
\E\Big[\sup_{f \in \mathcal{F}_{n,k}} |\bP_n f - \bP f|\Big] \leq  c_0 \Big[\sigma_{n,k} \Big(\frac{V}{n} \log \frac{A \|F\|_{L^2(\bP)}}{\sigma^2_{n,k}}\Big)^{1/2} + \frac{Vn/k}{n} \log \frac{A \|F\|_{L^2(\bP)}}{\sigma^2_{n,k}}  \Big] 
\leq  K_1 k^{-1/2} \label{eq:Gexpect}
\end{align}
Next we will use the following refined version of Talagrand's concentration inequality, which states that for any countable class of measurable functions $\mathcal{F}$ with elements mapping into $[-M,M]$,
\begin{equation} \label{eq:Gcontract}
\bP\Big( \sup_{f \in \mathcal{F}}|\bP_n f - \bP f| \geq 2 \E\Big[\sup_{f \in \mathcal{F}}|\bP_n f - \bP f|\Big] + c_1 n^{-1/2 }\Big(\sup_{f \in \mathcal{F}} Pf^2\Big)^{1/2} \sqrt{t} + n^{-1}c_2 M t \Big) \leq e^{-t},
\end{equation}
for all $t>0$ and some universal constants $c_1,c_2 >0$. This is a special case of Theorem 3 in \cite{mass2000} (in the notation of that paper, set $\eps = 1$). By~\eqref{eq:envelope} we can set $M = n/k$ and combining this with~\eqref{eq:var}--\eqref{eq:Gexpect} we obtain for any $t \geq 1$ and a constant $K_2$ independent of $n,k,i,j,t$ 
\[
\bP\Big( \sup_{f \in \mathcal{F}_{n,k}} |\bP_n f - \bP f| \geq K_2\{(t/k)^{1/2} + t/k\} \Big) \geq 1 - e^{-t}.
\]
In other words, we have established that 
\begin{equation}\label{eq:boundomega2}
\bP\Big(\Omega_2(t)\Big) \geq 1 - e^{-t};
\end{equation}
where
\[
\Omega_2(t) := \Big\{\sup_{f \in \mathcal{F}_{n,k}} |\bP_n f - \bP f| \leq K_2\Big((t/k)^{1/2} + t/k\Big) \Big\}.
\]
Observe that on $\Omega_1(\sqrt{t}) \cap \Omega_2(t)$ we have for $1 \leq \sqrt{t} \leq \sqrt{k}$
\begin{align*}
&\frac{n}{k}\Big|\hat C^\circ_{ij}(\widehat F_i^-(k/n),\widehat F_j^-(k/n)) - C_{ij}(\widehat F_i^-(k/n),\widehat F_j^-(k/n))\Big|
\\
& + \frac{n}{k} \Big|\widehat F_i^-(k/n) - k/n\Big| + \frac{n}{k}\Big|\widehat F_j^-(k/n)) - k/n\Big|
\\
\leq & K_2\Big((t/k)^{1/2} + t/k\Big) + 4\sqrt{t} k^{-1/2}
\\
\leq & K_3 (t/k)^{1/2}
\end{align*}
for a constant $K_3$ independent of $i,j,n,k,t$.
and that by~\eqref{eq:expboundquant} applied with $\lambda = \sqrt{t}$ and~\eqref{eq:boundomega2}  
\[
\bP\Big( \Omega_1(\sqrt{t}) \cap \Omega_2(t) \Big) \geq 1 - \bP\Big( \Omega_1(\sqrt{t})^C\Big) - \bP\Big(\Omega_2(t)^C \Big) \geq 1 - 5 \exp(-3t/10).
\]
Combined with~\eqref{eq:decompchiij} this shows that for any $1 \leq i,j \leq d$ and any $1 \leq t \leq k$ we have 
\[
\bP\Big( |\hat \chi_{ij} - \chi_{ij}(k/n)| > k^{-1/2}t^{1/2}K_3 \Big) \leq 5 \exp(-3t/10).
\]
Note that the restriction $t \geq 1$ can be dropped since the bound becomes trivial for $t < 1$. This implies: there exists universal constant $K > 0$ such that for $s \leq K$ 
\[
\bP\Big( |\hat \chi_{ij} - \chi_{ij}(k/n)| > s \Big) \leq 5 \exp(-3ks^2/10K^2).
\]
Since for $s \geq K$ we have
\[
\bP\Big( |\hat \chi_{ij} - \chi_{ij}(k/n)| > s \Big) \leq \bP\Big( |\hat \chi_{ij} - \chi_{ij}(k/n)| > K \Big) \leq 5 \exp(-3k/10)
\]
this completes the proof. \hfill $\Box$

\bibliographystyle{Chicago}
\bibliography{references}

\end{document}